\documentclass[a4paper,fleqn,usenatbib]{mnras}

\usepackage[T1]{fontenc}
\usepackage{times}

\usepackage{graphicx}	
\usepackage{amsmath}	
\usepackage{amssymb}	

\newcommand{\ra}{r_\mathrm{a}}

\newcommand{\md}{\mathrm{d}}

\newcommand{\rp}{r_\mathrm{p}}

\newcommand{\epsGR}{\epsilon_\mathrm{GR}}
\newcommand{\epsstr}{\epsilon_\mathrm{strong}}
\newcommand{\epsweak}{\epsilon_\mathrm{weak}}
\newcommand{\epspibytwo}{\epsilon_{\pi/2}}

\newcommand{\jmin}{j_\mathrm{min}}
\newcommand{\emax}{e_\mathrm{max}}

\newcommand{\nn}{\nonumber}

\def\ba{\begin{eqnarray}}
\def\ea{\end{eqnarray}}

 \title[Secular dynamics of binaries - III. GR precession]{Secular dynamics of binaries in stellar clusters - III. doubly-averaged dynamics in the presence of general relativistic precession}

\author[C. Hamilton \& R. R. Rafikov]{
  Chris Hamilton$^{1}$\thanks{E-mail: ch783@cam.ac.uk} and Roman R. Rafikov$^{1,2}$\thanks{John N. Bahcall Fellow at the Institute for Advanced Study}\\
$^1$Department of Applied Mathematics and Theoretical Physics, University of
Cambridge, Wilberforce Road, Cambridge CB3 0WA, UK\\
$^2$Institute for Advanced Study, Einstein Drive, Princeton, NJ 08540, USA}

\date{Accepted XXX. Received YYY; in original form ZZZ}

\pubyear{2020}

\begin{document}
\label{firstpage}
\pagerange{\pageref{firstpage}--\pageref{lastpage}}
\maketitle

\begin{abstract}
Secular evolution of binaries driven by an external (tidal) potential is a classic astrophysical problem.  Tidal perturbations can arise due to an external point mass, as in the Lidov-Kozai (LK) theory of hierarchical triples, or due to an extended stellar system (e.g. galaxy or globular cluster) in which the binary resides.
For many applications, general-relativistic (GR) apsidal precession is important, and has been accounted for in some LK calculations. Here we generalise and extend these studies 
by exploring in detail the effect of GR precession on (quadrupole-level) tidal evolution of binaries orbiting in  arbitrary axisymmetric potentials (which includes LK theory as a special case).  We study the (doubly-averaged) orbital dynamics for arbitrary strengths of GR and binary initial conditions and uncover entirely new phase space morphologies with important implications for the binary orbital evolution. We also explore how GR precession affects secular evolution of binary orbital elements when the binary reaches high eccentricity ($e\to 1$) and delineate several different dynamical regimes. 
Our results are applicable to a variety of astrophysical systems. In particular, they can be used to understand the high-eccentricity behaviour of (cluster) tide-driven compact object mergers --- i.e. LIGO/Virgo gravitational wave sources --- for which GR effects are crucial. 
\end{abstract}

\begin{keywords}
gravitation -- celestial mechanics -- stars: kinematics and
dynamics -- galaxies: star clusters: general -- binaries: general -- gravitational wave sources
\end{keywords}



\section{Introduction}

The problem of the relative motion of two bound point masses has formed the basis of celestial mechanics since it was first successfully tackled mathematically by Newton in 1687. In 1915 Einstein updated the solution, showing that the lowest order correction to Newton's elliptical orbit (in the small parameter $G(m_1+m_2)/ac^2$, with $m_i$ the constituent masses and $a$ the binary semimajor axis) was simply an extra prograde apsidal precession at a rate
\begin{align}
    \dot{\omega}_\mathrm{GR} = \frac{\dot{\omega}_\mathrm{GR}|_{e=0}}{1-e^2}~~~{\rm with}~~~\dot{\omega}_\mathrm{GR}|_{e=0}=\frac{3[G(m_1+m_2)]^{3/2}}{a^{5/2}c^2},
    \label{eqn:omega_dot_GR}
\end{align}
where $e$ is the orbital eccentricity, and $\dot{\omega}_\mathrm{GR}|_{e=0}$ is the GR precession rate for a circular orbit.  Einstein's solution is now known as the first post-Newtonian (1PN) approximation to the two-body problem.

A century on, the LIGO/Virgo Collaboration has detected, and continues to detect, dozens of merging compact object (black hole or neutron star) binaries \citep{LIGO2018,LIGO2020}.  These discoveries certainly warrant an astrophysical explanation, which is complicated by the fact that the timescale for an isolated compact object binary to merge via gravitational wave (GW) emission is often much longer than the age of the Universe.  For instance, an isolated circular black hole binary with $m_1=m_2=30M_\odot$ will only merge within $10^{10}$yr if its initial semimajor axis $a$ is $\lesssim 0.2 \mathrm{AU}$.  Thus nature must have a way of forcing these relativistic binaries to such small separations.  


One way to achieve this outcome is by driving an initially wide binary to a very high eccentricity, $e\to 1$, so that for a given semimajor axis, a binary's pericentre distance $p\equiv a(1-e)$ is greatly diminished. In this case the repeated close approaches of the binary components allow significant energy and angular momentum to be dissipated in bursts of GWs, efficiently shrinking the binary orbit and accelerating the merger. Thus, in recent years much effort has gone into searching for mechanisms by which binaries might achieve very high eccentricity. One broad category of proposed mechanisms consists of secular eccentricity excitation of binaries by some perturbing tidal\footnote{Throughout this paper, unless explicitly stated otherwise, the word `tidal' refers to the tidal gravitational force acting upon a binary due to an external companion (star, stellar cluster, etc), and not to e.g. the internal fluid tides of a star.} potential. This could be the tidal potential due to a tertiary point mass (e.g. a star) that is gravitationally bound to the binary, in which case the dynamics are described by the Lidov-Kozai (LK) theory \citep{Lidov1962,Kozai1962}, or simply the mean field potential of the star cluster or galaxy in which the binary resides \citep{Heisler1986,Brasser2006,Hamilton2019a,Hamilton2019b,Bub2019}.   

However, when investigating such merger channels it is almost always necessary to account for the effect of (1PN) GR precession of the binary's pericentre angle. This is because GW emission primarily occurs during close pericentre passages when the binary is highly eccentric, and this is precisely the regime in which GR precession is most important (equation \eqref{eqn:omega_dot_GR}). For similar reasons it is often necessary to include GR precession (as well as other short-range precession effects, such as those arising from rotational or tidal bulges, see e.g. \citealt{Liu2015,Munoz2016}) in studies of LK secular evolution, which rely on tidal dissipation (inside one or both binary components) to shrink the binary orbit \citep{Fabrycky2007,Antonini2016}. Tidal dissipation is strongest when $e\to 1$, meaning that GR precession is important as well. 

GR precession is routinely accounted for in LK population synthesis calculations (e.g. \citealt{Antonini2014,Rodriguez2015,Liu2015,Liu2018,Hamers2018_VRR,Samsing2019}). Its effect is understood as increasing a binary's prograde apsidal precession rate as $e\to 1$, preventing the perturber from coherently torquing the binary and effectively stopping the reduction of the binary's angular momentum. A result of this so-called `relativistic quenching' effect is a reduction in the maximum eccentricity that the binary can reach, even if the initial inclination between binary and perturber orbits is favourable \citep{Fabrycky2007}. Some authors have derived approximations to this maximum eccentricity in the limit where GR precession can be treated as a small perturbation to the LK evolution \citep{Miller2002,Blaes2002,Wen2003,Veras2010,Liu2015,Anderson2017,Grishin2018}. Also, \citet{Iwasa2016} looked at the modification of the phase space portrait of the LK problem in the presence of GR precession, although their study was far from exhaustive. However, so far nobody has studied carefully the impact of the GR precession for binaries perturbed by general tidal potentials \citep{Brasser2006,Hamilton2019c,Bub2019} where we expect similar considerations to apply. 

The main purpose of this paper is to explore systematically the effect of GR precession on the underlying phase space dynamics and eccentricity evolution of a tidally perturbed binary. We will focus exclusively upon the `doubly-averaged' (hereafter DA)\footnote{`Double-averaging' here refers to averaging first over the binary's fast `inner' orbital motion and second over `outer' orbital motion of the binary's barycentre relative to its perturber --- see \citet{Hamilton2019a}.} dynamics of binaries perturbed by quadrupole-order tidal potentials.  We will also make the test-particle approximation, i.e. assume that the binary's outer orbital motion relative to its perturber contains much more angular momentum than its internal Keplerian orbit \citep{Naoz2016}. The quadrupolar and test-particle approximations are very good ones for the applications we have in mind (such as compact object binaries perturbed by globular cluster tides), but they can be relaxed, see \S\ref{sec:limitations}. The DA approximation will be relaxed in an upcoming paper (Hamilton \& Rafikov, in prep.).
The present study complements our investigation of DA cluster tide-driven dynamics of binaries begun in \citet{Hamilton2019a,Hamilton2019b}, hereafter `Paper I' and `Paper II' respectively. The DA theory developed in Papers I and II includes the test-particle quadrupole LK problem as a limiting case, but is more general and dynamically more rich, particularly when GR precession is included, as we show here. 

In \S\ref{sec:dynamical_framework} we write down the doubly-averaged perturbing Hamiltonian and establish the notation that we will use for the rest of the paper. In particular we introduce the key parameter $\epsGR$ which measures the strength of GR precession relative to tidal torques. In \S\ref{sec:phase} we explore the phase space behaviour as $\epsGR$ is varied; the quantitative results that we quote in this section are derived in Appendix \ref{sec:mathematical_details_Gamma_positive}.
In \S\ref{sec:High_ecc_behaviour} we investigate very high eccentricity behaviour in the presence of weak or moderate GR precession. In particular we explore how finite $\epsGR$ modifies both the maximum eccentricity reached and the timescale of high eccentricity episodes. In \S\ref{sec:discussion} we discuss our results in the light of the existing literature, and comment on the limitations of our study. We summarise in \S\ref{sec:summary}. Lastly, in Appendix \ref{sec:analytic} we provide for the first time an explicit, analytical solution to the DA equations of motion for all orbital elements in the high eccentricity limit.
We also check the accuracy of this solution against direct numerical integration of the DA equations of motion. 



\section{Dynamical framework}
\label{sec:dynamical_framework}
  
We consider a binary orbiting in an arbitrary time-independent, axisymmetric external potential $\Phi$. For the remainder of the paper we will refer to this as the `cluster' potential, though it can in reality be due to an axisymmetric galaxy, point mass, or whatever.  We refer to the binary's barycentric motion around the cluster --- which is assumed to follow the trajectory of a test particle in the potential $\Phi$, and need not be circular --- as the `outer' orbit.  The binary's `inner' orbit is described by the usual orbital elements: semimajor axis $a$, eccentricity $e$, inclination $i$, argument of pericentre $\omega$, longitude of the ascending node $\Omega$ and mean anomaly $M$.  Inclination is measured relative to the plane perpendicular to the cluster's symmetry ($Z$) axis or, in the case of a spherical potential, relative to the outer orbital plane. The angle to the line of nodes is measured relative to an arbitrary fixed axis in this plane.  

Then the dynamical evolution of the binary's inner orbital elements is governed by the secular `doubly-averaged' perturbing Hamiltonian (Paper I)\footnote{For simplicity we have replaced the notation $\overline{\langle H_1 \rangle}_M, \, \langle H_\mathrm{GR} \rangle_M$ from Papers I and II with $H_1, H_\mathrm{GR}$.}
\begin{align}  
H = CH^* \equiv C( H_1^* + H_\mathrm{GR}^*), \,\,\,\,\,\,\,\, \mathrm{where} \,\,\,\,\,\,\, C \equiv Aa^2/8. 
\label{eqn:DA_Hamiltonian} 
\end{align} 
Here $A$ is a constant with units of $($frequency$)^2$ which measures the strength of the tidal torque and sets the timescale of secular evolution. It is completely determined by stipulating the form of the cluster potential $\Phi$ and the outer orbit of the binary; to order of magnitude, $A^{-1/2}$ is comparable to the period of the binary's outer orbit.  For reference, we provide the value of $A$ for the LK problem, i.e. when $\Phi$ is the Keplerian potential (Paper I, Appendix B):
\begin{align}
\label{eqn:A_LK}
A_\mathrm{LK} =\frac{G\mathcal{M}}{2a_\mathrm{g}^3(1-e_\mathrm{g}^2)^{3/2}},
\end{align}
where $\mathcal{M}$ is the tertiary perturber's mass and $a_\mathrm{g}$, $e_\mathrm{g}$ are respectively the semimajor axis and eccentricity of the binary's outer orbit relative to the tertiary perturber.

Next, $H_1^*$ and $H_\mathrm{GR}^*$ are the dimensionless Hamiltonians accounting for quadrupole-order cluster tides and GR pericentre precession, respectively:
\begin{align} 
   & H_1^* = (2+3e^2)(1-3\Gamma\cos^2i)-15\Gamma e^2 \sin^2 i \cos 2\omega, 
\label{H1Star}
\\
& H_\mathrm{GR}^* =  -\epsilon_\mathrm{GR}(1-e^2)^{-1/2}. \label{HGRStar}
\end{align}
The crucial quantity $\Gamma$ in \eqref{H1Star} is a dimensionless parameter which is fully determined (like $A$) by stipulating $\Phi$ and the outer orbit --- see \S\ref{sec:note_on_Gamma} for discussion. The relative strength of GR precession is measured in equation \eqref{HGRStar} by the crucial parameter
\begin{align} 
\epsilon_\mathrm{GR} &\equiv \frac{24G^2(m_1+m_2)^2}{c^2Aa^4} \sim \frac{n_\mathrm{K}^2}{A}\left(\frac{v}{c} \right)^2
\sim  \dot{\omega}_\mathrm{GR}|_{e=0} t_\mathrm{sec}
\label{eq:epsGR} \\
&=  0.258 \times \left( \frac{A^*}{0.5}\right)^{-1}\left( \frac{\mathcal{M}}{10^5M_\odot}\right)^{-1}\left( \frac{b}{\mathrm{pc}}\right)^{3} \nn \\ & \times \left( \frac{m_1+m_2}{M_\odot}\right)^{2}  \left( \frac{a}{20 \, \mathrm{AU}}\right)^{-4}. 
\label{eq:epsGRnumerical}
\end{align} 
Here  $n_{\rm K}=\sqrt{G(m_1+m_2)/a^3}$ and $v \sim \sqrt{G(m_1+m_2)/a}$ are the Keplerian mean motion and typical orbital speed of the inner orbit of the binary, respectively, while $t_\mathrm{sec}$ is the timescale of secular eccentricity oscillations in the non-GR limit (Paper II, equations (33)-(34)). In the numerical estimate \eqref{eq:epsGRnumerical} we have assumed that the binary is orbiting a spherical cluster with scale radius $b$ and total mass $\mathcal{M}$. Typical values of the dimensionless parameter $A^* \equiv A/(G\mathcal{M}/b^3)$ are mapped out in \S6 of Paper I. 
   
As in Paper I we introduce Delaunay variables (actions) $L=\sqrt{G(m_1+m_2) a}, J=L\sqrt{1-e^2}$, and $J_z = J\cos i$, their corresponding angles being $M$, $\omega$ and $\Omega$ respectively. Since \eqref{eqn:DA_Hamiltonian} is independent of the mean anomaly $M$, the action $L$ is conserved, so we can choose to work with the following dimensionless variables:
\begin{align}  
\Theta \equiv J_z^2/L^2 = (1-e^2)\cos^2 i, \,\,\,\,\,\,\,\,\, 
j\equiv J/L = \sqrt{1-e^2}.
\label{eq:AMdefs}
\end{align} 
Obviously $j$ is just the dimensionless angular momentum. The definitions (\ref{eq:AMdefs}) imply that $e$ and $j$ must obey
\begin{align}    
0\leq e\leq e_\mathrm{lim} \equiv \sqrt{1-\Theta},~~~~~~~~~~~\Theta^{1/2}\leq j\leq 1,
\label{eq:AMconstr}
\end{align}
to be physically meaningful for a given $\Theta$.
With this notation established, we can rewrite the dimensionless Hamiltonians \eqref{H1Star}-\eqref{HGRStar} in the form
\begin{align}   
H_1^* =& \left[ (j^2 - 3\Gamma \Theta)( 5-3j^2) \right. \nn \\ & \left. - 15\Gamma(j^2-\Theta)(1-j^2) \cos 2\omega \right]j^{-2}
\label{eqn:H1star_omega_j},
\\
 H_\mathrm{GR}^* &= - \epsGR j^{-1}. 
\label{eqn:HGR_omega_j}
\end{align} 
Since both Hamiltonians are independent of $\Omega$, the dimensionless quantity $\Theta$ is an integral of motion. The total dimensionless Hamiltonian $H^*= H_1^*+H^*_\mathrm{GR}$ can be taken as the other integral of motion. The equations of motion fully describing the evolution of the dimensionless variables $\omega$, $j$ are
\begin{align}
\frac{\md \omega}{\md t} &= \frac{C}{L} \frac{\partial H^*}{\partial j} = \frac{C}{L} \frac{\partial}{\partial j} (H_1^* + H_\mathrm{GR}^*) \nn \\ &= \frac{6C}{L}\frac{[5\Gamma\Theta - j^4 +5\Gamma(j^4-\Theta)\cos2\omega + \epsGR j/6]}{j^3},
\label{eom1} \\ 
\frac{\md j}{\md t} &=  -\frac{C}{L} \frac{\partial H^*}{\partial \omega}
=
-\frac{C}{L} \frac{\partial H_1^*}{\partial \omega}
 \nn \\
 &= -\frac{30\Gamma C}{L} \frac{(j^2-\Theta)(1-j^2)}{j^2}\sin 2\omega.
\label{eom2} 
\end{align} 
Since $\omega, j$ are decoupled from $\Omega$, the evolution of the nodal angle $\Omega$ can be explored separately using the equation of motion  
\begin{align} 
      \frac{\md \Omega}{\md t} &= C\frac{\partial H^*}{\partial J_z}= C\frac{\partial H^*_1}{\partial J_z} \nn \\ &= -\frac{6C\Gamma}{L}\Theta^{1/2} \frac{5-3j^2-5\cos 2\omega (1-j^2)}{j^2}.
      \label{eqn:dOmegadt_DA}
\end{align}
Obviously, the equation of motion for $J_z$ is trivial, $\md J_z/\md t=-\partial H/\partial \Omega=0$.

Given that $H^*(\omega,j)$ is a constant we can use equations (\ref{eqn:H1star_omega_j})-(\ref{eqn:HGR_omega_j}) to eliminate $\omega$ from equation \eqref{eom2}. Following a derivation analogous to that of equation (30) in  Paper II, and without making any approximations, we find
\begin{align}  
\frac{\md j}{\md t} =
\pm \frac{6C}{Lj^2} & \Bigg\{
 (25\Gamma^2-1)\left[(j_+^2-j^2)(j^2-j_-^2)-\frac{\epsGR}{3(1+5\Gamma)}j\right] \nn \\ &\times 
\left[j^2(j_0^2-j^2)+\frac{\epsGR}{3(5\Gamma-1)}j\right] \Bigg\}^{1/2},
\label{eq:djdtGR}
\end{align}
where
\begin{align}  
j_\pm^2 &\equiv \frac{\Sigma \pm \sqrt{\Sigma^2-10\Gamma\Theta\left(1+ 5\Gamma\right)}}{1+5\Gamma}, 
\label{eqn:jpm}
\\ 
    j_0^2 &\equiv 1-D,
    \label{eqn:j0}
\end{align}
with
\begin{align} 
\label{eqn:def_Sigma}
\Sigma &\equiv \frac{1+5\Gamma}{2} +5\Gamma\Theta + \left(\frac{5\Gamma-1}{2} \right)D,
\\
 D &\equiv  \frac{H^*/3 - 2/3 +2\Gamma\Theta}{1-5\Gamma}
 \nn \\ &= e^2\left(1+\frac{10\Gamma}{1-5\Gamma}\sin^2i\sin^2\omega\right) - \frac{\epsGR}{3(1-5\Gamma)\sqrt{1-e^2}}.
 \label{eqn:def_D}
\end{align} 
Note that the definitions of $j_\pm^2$, $j_0^2$, $\Sigma$, and $D$ are equivalent to those given in Paper II (equations (18), (17), (19) and (15) respectively) except that we have replaced $H_1^*$ in equation (\ref{eqn:def_D}) by $H^* = H_1^* + H_\mathrm{GR}^*$, i.e. we have used the value of the Hamiltonian that includes GR precession. 
Therefore in the limit $\epsGR = 0$, equation \eqref{eq:djdtGR} reduces to equation (30) of Paper II. Note also that $j_\pm^2$, $j_0^2$ are not necessarily positive. We will use equation \eqref{eq:djdtGR} extensively when we study high-eccentricity behaviour in \S\ref{sec:High_ecc_behaviour}.
\\
\\
Some shorthand notation will be necessary as we proceed. In particular, several different values of $e$ and $j$ will come with distinct subscripts.  We provide a summary of our notation in Table \ref{tab:notation}.
\begin{table*}
	\centering
	\caption{Key to different variables.}
	\label{tab:notation}
	\begin{tabular}{llc} 
		\hline
		Symbol & Description & Defining equation(s)\\
		
		\hline
		
			$\epsGR$ & Parameter determining strength of GR precession. & \eqref{eq:epsGR} \\
			
        $\epspibytwo$ & Critical value of $\epsGR$ dictating the behaviour of fixed points at $\omega = \pm\pi/2$. & \eqref{eqn:epspibytwo} \\
						
		$\epsstr$ & Critical value of $\epsGR$ for the onset of `strong' GR precession, $\epsstr = 3(1+5\Gamma)$. & \eqref{eqn:epsstrong} \\
		
		$\epsweak$ & Critical value of $\epsGR$ defining the upper bound of the `weak' GR regime. & \eqref{eqn:epsweak} \\

		$e, \, j$ & Binary eccentricity, dimensionless angular momentum $j=\sqrt{1-e^2}$. & \eqref{eq:AMdefs} \\

		$e_\mathrm{f}, \,j_\mathrm{f}$ & $e, j$ values of fixed points at $\omega = \pm \pi/2$ when $\epsGR=0$. & \eqref{eqn:jf} \\
		
		$e_\mathrm{f,\pi/2},\, j_\mathrm{f,\pi/2}$ & $e, j$ values of fixed points at $\omega = \pm \pi/2$ when $\epsGR\neq0$. & \eqref{eqn:FPs_at_omega_piby2}\\
		
		$e_\mathrm{f,0}, \,j_\mathrm{f,0}$ & $e, j$ values of fixed points at $\omega = 0$, possible only when $\epsGR\neq 0$. & \eqref{eqn:jf0} \\
				
		$e_\mathrm{max},\, j_\mathrm{min}$  & Maximum $e$, minimum $j$ values. & \\
		
		$e_\mathrm{lim}$ & Upper limit on possible values of eccentricity,  $e_\mathrm{lim} = \sqrt{1-\Theta}$. & \eqref{eq:AMconstr}\\
	
		$e_0, i_0, \omega_0$ & Initial values of $e,i,\omega$. \\
		
		$j_\pm, j_0$. & Important functions of $e_0, i_0, \omega_0, \Gamma, \epsGR$
		 (note $j_0$ is \textit{not} the initial $j$ value).
		& \eqref{eqn:jpm}, \eqref{eqn:j0}\\
    		
		\hline
	\end{tabular}
\end{table*}


\subsection{A note on \texorpdfstring{$\Gamma$}{Gamma}}
\label{sec:note_on_Gamma}


The dimensionless quantity $\Gamma$ is crucial for our investigation because it determines the morphology of the phase space, and therefore sets fundamental constraints on the allowed dynamical behaviour (Paper II).  Its value is computed by time-averaging the outer orbit in the potential $\Phi$ (Paper I).  Typically it has to be computed numerically but in special cases (e.g. spherically symmetric potentials) it can be calculated (semi-)analytically.  For example, the test-particle quadrupole LK problem (e.g. \citealt{Antognini2015,Naoz2016}) corresponds exactly to the limit $\Gamma = 1$, while the problem of binaries orbiting in the midplane of a thin disk (e.g. \citealt{Heisler1986}) corresponds to $\Gamma = 1/3$. Binaries orbiting inside the potential generated by a homogenous sphere (similar to the inner regions of a globular cluster) effectively have $\Gamma \to 0$. For a binary on a circular outer orbit of radius $R$ in a Plummer potential with a scale radius $b$, we get $\Gamma = (1+4\zeta^2)^{-1}$, where $\zeta \equiv b/R$ (this special case was considered by \citealt{Brasser2006}; compare their disturbing function (A.5) with our equation \eqref{H1Star}).

In general, $\Gamma$ can take any value. Moreover, we showed in Paper II (for $\epsGR=0$) that at critical $\Gamma$ values of $\pm1/5, 0$, bifurcations occur in dynamics, meaning that we need to explore separately four regimes:
\begin{align}
    & \Gamma > 1/5, \label{eqn:Gamma_I} \\ 
    & 0 < \Gamma \leq 1/5, \label{eqn:Gamma_II}\\
    & -1/5 < \Gamma \leq 0, \label{eqn:Gamma_III}\\
    & \Gamma \leq -1/5. \label{eqn:Gamma_IV}
\end{align}
However, one can show that for sensible spherical potentials only $0 < \Gamma \leq 1$ is possible (Paper I, Appendix D). 
If one lets the binary population in a cluster simply trace the underlying stellar density profile then it turns out that cored clusters (those with a flat density profile $\rho(r) \to$ const. as $r \to 0$, like the Plummer profile) have a significant fraction of their binaries in the $0 < \Gamma \leq 1/5$ regime \eqref{eqn:Gamma_II}, and the rest in the $\Gamma > 1/5$ regime \eqref{eqn:Gamma_I}. On the contrary, cusped clusters (those with $\rho(r) \propto r^{-p}$ as $r\to 0$ for some $p>0$, like the Hernquist profile) host a much larger fraction of their binaries in the regime $\Gamma > 1/5$ and relatively few in the regime $0 < \Gamma \leq 1/5$.  This has strong implications for the dynamical evolution of binaries in these various types of cluster (\S\ref{sec:phase}).

To get negative $\Gamma$ values typically requires a highly inclined outer orbit in a sufficiently non-spherical potential (Paper I).  Since our applications are mostly concerned with spherical or near-spherical potentials such as those of globular clusters, for which negative $\Gamma$ values are very rare, we concentrate on the regimes \eqref{eqn:Gamma_I}-\eqref{eqn:Gamma_II} in the main body of the paper. Discussion of the regimes \eqref{eqn:Gamma_III}-\eqref{eqn:Gamma_IV} can be found in Appendix \ref{sec:Gamma_negative} .


\section{Phase space behaviour}
\label{sec:phase}

  
To gain a qualitative understanding of the dynamics driven by the Hamiltonian equations \eqref{eom1}-\eqref{eom2}, one can fix the values of $\Gamma,\Theta, \epsGR$ and then plot $(\omega,e)$ phase space trajectories\footnote{When referring to motion in phase space we use the terms `trajectory' and `orbit' interchangeably.}, i.e. contours of constant $H^*$ in the $(\omega,e)$ plane.  We call such a plot a `phase portrait'.
   
In the particular case $\epsGR = 0$ these are simply contours of constant $H_1^*$ --- see Figures 4, 5, 6 \& 7 of Paper II. In that (non-GR) case, one finds that two distinct phase space orbit families are possible for $\Gamma>0$: circulating orbits, which run over all $\omega \in (-\pi, \pi)$, and librating orbits, which loop around fixed points located at $(\omega=\pm\pi/2, e=e_\mathrm{f})$, where $e_\mathrm{f} \equiv (1-j_\mathrm{f}^2)^{1/2}$ and
\begin{align} 
j_\mathrm{f} = \left(\frac{10\Gamma\Theta}{1+5\Gamma}\right)^{1/4},
\label{eqn:jf}
\end{align}
see Paper II, equation (12).  These fixed points correspond to non-trivial solutions to the system of equations $\md j/\md t = 0$, $\md \omega/\md t = 0$. For $\Gamma > 0$ --- i.e. in the important regimes \eqref{eqn:Gamma_I}-\eqref{eqn:Gamma_II} --- fixed points \textit{always} exist in the phase portrait provided $\Theta$ is small enough (which can be achieved e.g. by starting with sufficiently large inclination). Note that this is not generally true for $\epsGR \neq 0$, as we will see below. The precise requirement for fixed points to exist when $\epsGR=0$ is (Paper II, \S2.2)
\begin{align}   
\Theta<\min\left(\Lambda,\,\Lambda^{-1}\right),~~~~~~
\mathrm{where}
~~~~~~
\Lambda(\Gamma)\equiv \frac{5\Gamma+1}{10\Gamma}. 
\label{eq:Theta_constr}
\end{align} 

\begin{figure*}
\centering
\includegraphics[width=0.9\linewidth,trim={0.9cm 0.3cm 0cm 0.2cm},clip]{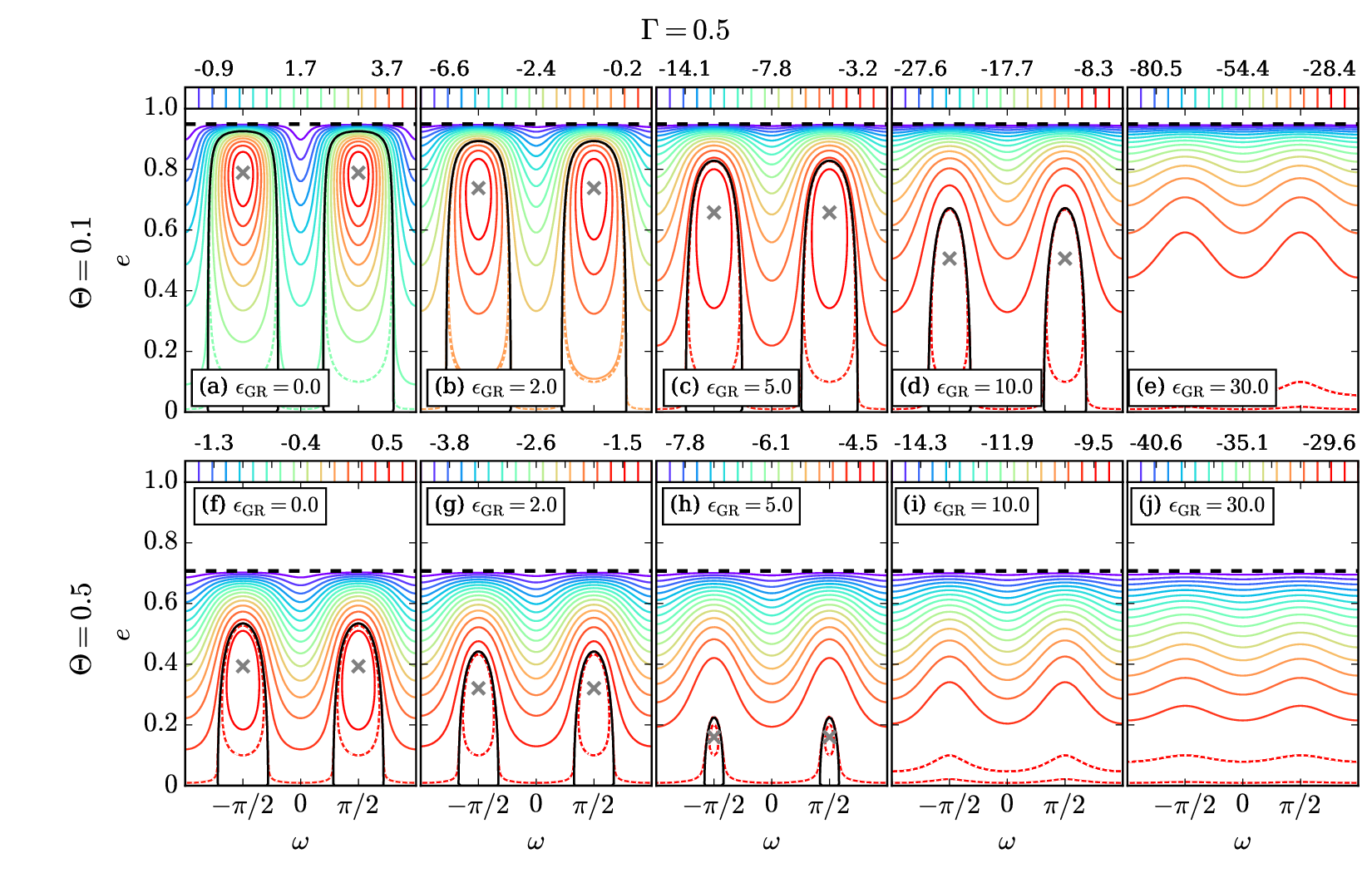}
\caption{Contour plots of constant Hamiltonian $H^* \equiv H_1^* + H^*_\mathrm{GR}$ in the $(\omega,e)$ plane for $\Gamma=0.5$.  In the top (bottom) row we fix $\Theta=0.1$ ($\Theta=0.5$). We increase $\epsGR$ from left to right, using the values $\epsGR=0, 2, 5, 10, 30$ indicated in each panel.  Contours are spaced linearly from the minimum (blue) to maximum (red) value of $H^*$ --- see the colour bar at the top of each panel. We have also added by hand dashed contours passing through $(\omega,e)=(0,0.01)$ and $(\omega,e)=(\pm\pi/2,0.1)$ in each panel. Dashed black horizontal lines indicate the limiting possible eccentricity $e_\mathrm{lim} = \sqrt{1-\Theta}$, while fixed points are shown with grey crosses should they exist.  Black separatrices illustrate the boundary between families of librating and circulating phase space trajectories.}
\label{fig:HStar_Contours_Gamma_pt5}
\end{figure*}

   \begin{figure*}
\centering
\includegraphics[width=0.9\linewidth,trim={0.9cm 0.3cm 0cm 0.2cm},clip]{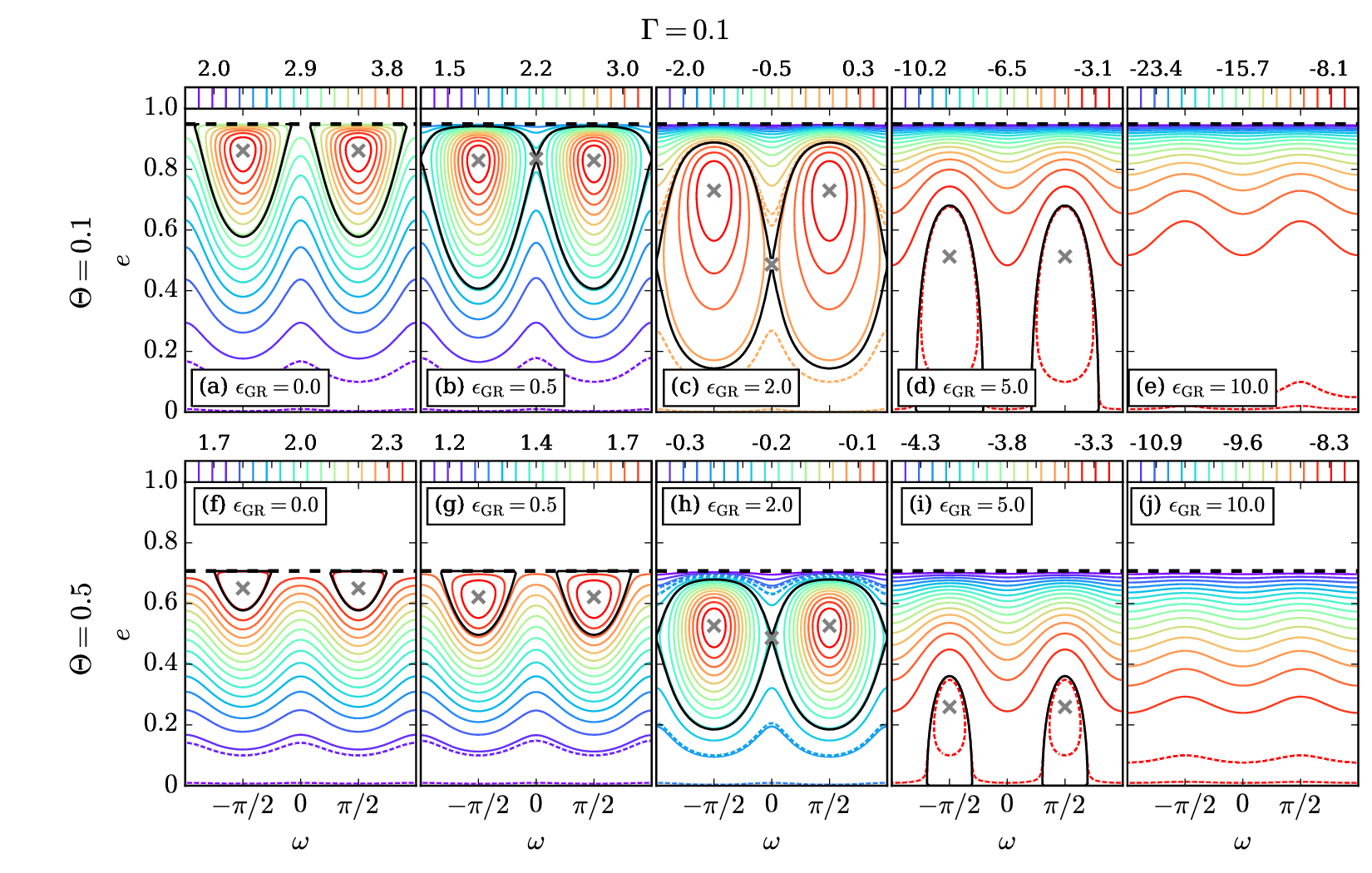}
\caption{Same as Figure \ref{fig:HStar_Contours_Gamma_pt5} except for $\Gamma=0.1$, and we have taken different values of $\epsGR$ to better demonstrate the new phase space behaviour. Note that dashed contours above the saddle points have the same $H^*$ value as the low-$e$ dashed contours. See text for details.}
\label{fig:HStar_Contours_Gamma_pt1}
\end{figure*}

In \S\ref{sec:High_ecc_behaviour} we will be interested exclusively in situations where some fraction of binaries can reach eccentricities very close to unity (i.e. $1-e \ll 1$).  Given that $\Theta = (1-e^2)\cos^2 i$ is conserved, a necessary condition for this is that $\Theta \ll 1$. 
Fixed points are crucial to this investigation because they may force an initially low-$e$ binary to very high maximum eccentricity $e_\mathrm{max}$.  Indeed, a key result of Paper II (again with $\epsGR=0$) was that for $\Gamma > 1/5$, whenever fixed points exist, $e_\mathrm{f}$ provides a \textit{lower bound} on $e_\mathrm{max}$. Equation \eqref{eqn:jf} implies that $e_\mathrm{f}$ is close to unity whenever $\Theta \ll 1$; high eccentricity excitation is then ubiquitous.  On the other hand, for $0 < \Gamma \leq 1/5$ the fixed points no longer provide a lower bound on circulating trajectories' $e_\mathrm{max}$ and so high-$e$ excitation is much rarer.  One consequence of this result is that cored clusters, which have a significant fraction of binaries in the $0 < \Gamma \leq 1/5$ regime, produce few tidally-driven compact object mergers compared to cusped clusters --- see \citet{Hamilton2019c}.  
\\
\\
In the rest of this section we explore how the phase space behaviour uncovered in Paper II (i.e. for $\epsGR=0$) is modified in the case of finite $\epsGR$, which we do separately for $\Gamma>1/5$  (\S\ref{sec:morphology_Gamma_Regime_I}) and for $0<\Gamma \leq 1/5$ (\S\ref{sec:morphology_Gamma_Regime_II}). We describe some properties of the fixed points in \S\ref{sect:fixed_points}, and then show how to calculate the maximum eccentricity of a given binary in \S\ref{sec:max_min_ecc_GR}. Details of the mathematical results that we quote throughout \S\S\ref{sec:morphology_Gamma_Regime_I}-\ref{sec:max_min_ecc_GR} are given in Appendix \ref{sec:mathematical_details_Gamma_positive}.
Note that the $\Gamma\leq 0$ regimes are treated in Appendix \ref{sec:Gamma_negative}.

           
\subsection{Phase space behaviour in the case \texorpdfstring{$\Gamma > 1/5$}{GamI}}
\label{sec:morphology_Gamma_Regime_I}
   

Figure \ref{fig:HStar_Contours_Gamma_pt5} shows phase portraits for the case $\Gamma=0.5 > 1/5$.  In the top (bottom) row we set $\Theta = 0.1\, (0.5)$. From left to right we vary $\epsGR$ taking $\epsGR=0, 2, 5, 10, 30$. In each panel a black horizontal dashed line shows the limiting possible eccentricity $e_\mathrm{lim} = \sqrt{1-\Theta}$ (equation \eqref{eq:AMconstr}). Contours are spaced linearly from the minimum (blue) to maximum (red) value of $H^*$ indicated by the colour bar at the top of each panel. Since the linearly sampled contours become too widely separated at low eccentricity, to illustrate the low-$e$ behaviour we have added dashed contours passing through $(\omega,e)=(0,0.01)$ and $(\omega,e)=(\pm\pi/2,0.1)$ in each panel.  Just like in Paper II, trajectories are split into librating and circulating families. We plot the separatrices between these families with solid black lines.  Fixed points are denoted with grey crosses.
   
In panels (a) and (f) we encounter the usual $\epsGR=0$ behaviour familiar from the LK problem: (I) there are fixed points\footnote{We have deliberately chosen $\Theta$ values such that the fixed points do exist for $\epsGR=0$, i.e. satisfying \eqref{eq:Theta_constr}.} at $\omega=\pm\pi/2$, each of which is surrounded by a region of librating orbits, (II) these librating islands are connected to $e=0$ line, (III) the family of circulating orbits runs `over the top' of the librating regions, and (IV) all phase space trajectories reach maximum eccentricity at $\omega=\pm\pi/2$.  

Inspecting the other panels, we see that the effect of increasing $\epsilon_\mathrm{GR}$ from zero is simply to push the fixed points at $\omega=\pm \pi/2$ to lower eccentricity. As a result, large amplitude eccentricity oscillations along a given secular trajectory are noticeably quenched as $\epsGR$ is increased, and the region of librating orbits is diminished in both area and vertical extent. Eventually the eccentricity of the fixed points reaches zero and so they vanish altogether, leaving only circulating orbits (panels (e), (i) and (j)).  

The phase space evolution for non-zero $\epsGR$ exhibited in Figure \ref{fig:HStar_Contours_Gamma_pt5} is characteristic of all systems with $\Gamma > 1/5$, including the LK case $\Gamma =1$, which has already been discussed to some degree by \citet{Iwasa2016} --- see \S\ref{sec:LK_literature}.  Of course, the precise characteristics, such as the eccentricity of the fixed points, the critical $\epsGR$ for fixed points to vanish, etc., do depend on the value of $\Gamma$, as we detail in \S\ref{sect:fixed_points}.

                     
\subsection{Phase space behaviour in the case  \texorpdfstring{$0 < \Gamma \leq 1/5$}{GamII}}
\label{sec:morphology_Gamma_Regime_II}


For $0 < \Gamma \leq 1/5$ (a regime typical of binaries orbiting the inner regions of a cored cluster), a similar but slightly more complex picture emerges. In Figure \ref{fig:HStar_Contours_Gamma_pt1} we show phase portraits similar to Figure \ref{fig:HStar_Contours_Gamma_pt5} except that we now take $\Gamma=0.1 < 1/5$, and pick some new values of $\epsGR$ to better demonstrate the modified phase space behaviour.  The strength of GR still increases from left to right. As in Figure \ref{fig:HStar_Contours_Gamma_pt5} we have added in dashed contours that take the values $H^*(\omega = 0, e=0.01)$ and $H^*(\omega = \pm\pi/2, e=0.1)$.

Starting with the non-GR case $\epsGR=0$, we immediately notice a qualitative difference between the phase space morphologies for $0<\Gamma \leq 1/5$ (Figures \ref{fig:HStar_Contours_Gamma_pt1}a,f) and  $\Gamma > 1/5$ (Figures \ref{fig:HStar_Contours_Gamma_pt5}a,f), discussed at length in Paper II. Although there are again fixed points at $\omega=\pm\pi/2$, the librating islands that surround them are now connected to $e= e_\mathrm{lim}$ (and not to $e=0$, like in the $\Gamma>1/5$ case).  As a result, circulating orbits run `underneath' librating orbits (rather than `over the top' as for $\Gamma > 1/5$) and the maximum  eccentricity of circulating orbits is found at $\omega = 0$ (rather than at $\omega = \pm\pi/2$). Crucially, unlike for $\Gamma > 1/5$, a binary that starts at low eccentricity does not necessarily reach a high eccentricity even if there are fixed points located near $e=1$.  This fact is responsible for the dearth of cluster-tide driven mergers in cored clusters, which host many binaries with $0<\Gamma\leq 1/5$ \citep{Hamilton2019c}.

As we increase $\epsGR$ from zero, the fixed points again get pushed to lower eccentricity (panels (b) and (g)). However, the effect of this for $\Gamma = 0.1$ is to initially \textit{increase}, rather than decrease, the fraction of the phase space area that is encompassed by the librating islands. Additionally, as the fixed points get pushed to lower eccentricity, a new family of high-eccentricity circulating orbits emerges once $\epsGR$ exceeds a threshold value which we determine in \S\ref{sec:fixed_points_0}. These phase space trajectories run `over the top' of the fixed points and have their eccentricity maxima at $\omega=\pm \pi/2$ (panels (b), (c) and (h)).  The qualitative change from $\epsGR=0$ is reflected in the fact that the librating island is now truly an \textit{island}, disconnected from both $e=0$ and $e = e_\mathrm{lim}$. This is different from the case $\Gamma > 1/5$, in which the lower portion of the librating regions {\it always} stretches down to $e= 0$ until $\epsGR$ becomes so large that fixed points cease to exist.

Physically these new features might have been anticipated.  First of all, in Paper II we saw that for $0 < \Gamma\leq 1/5$, the cluster-driven $\omega$ evolution of circulating trajectories is always retrograde (contrary to the $\Gamma>1/5$ case in which it is prograde). Since GR always promotes prograde precession, its effect in this $\Gamma$ regime is initially to slow down the overall precession rate $\dot{\omega}$, allowing for a more coherent torque compared to the case of $\epsilon_\mathrm{GR}=0$.  This leads to the appearance of new librating solutions. This is first true for binaries that previously lay just below the separatrix in the $(\omega,e)$ phase space, as these circulate the slowest (recall that the secular period diverges on the separatrix itself), giving $\dot{\omega} \sim 0$ for a relatively small value of $\epsGR$. On the other hand, at the highest binary eccentricities (near $e= e_\mathrm{lim}$) GR may dominate the dynamics, causing the binary's pericentre angle $\omega$ to precess rapidly, leading to the appearence of new high-$e$ circulating solutions.

Simultaneously with the new high-$e$ family of orbits, two saddle points (i.e. fixed points that are not local extrema of $H^*$) emerge at $\omega=0,\pi$. Passing through them are separatrices that isolate the distinct phase space orbital families in panels (b), (c) and (h). This is an entirely new phase space feature that is not found in LK theory, as it is only possible for $\Gamma \leq 1/5$ and only when GR is present, as we show in \S\ref{sec:fixed_points_0}. The `two-eyed' phase space structure of panels (b), (c) and (h) has therefore not been uncovered before. Note that for a system exhibiting this structure, a circulating trajectory `above' the saddle point can have the same $H^*$ value as a circulating trajectory `below' the saddle point.  In other words a single value of the Hamiltonian can correspond to two entirely different phase space trajectories. This can be seen in Figures \ref{fig:HStar_Contours_Gamma_pt1}c,h where the dashed contours circulating \textit{above} the librating islands appear because they have the same $H^*$ values as the manually added dashed low-$e$ contours passing through $(\omega,e) = (\pm\pi/2, 0.1)$ and $(0,0.01)$.

As $\epsGR$ is increased further, the eccentricity of the saddle points diminishes, similar to the fixed points at $\omega=\pi/2$. It is interesting to note that these various types of fixed points move at different `speeds' down the phase portrait as $\epsGR$ grows.  In particular, panels (d) and (i) of Figure \ref{fig:HStar_Contours_Gamma_pt1} demonstrate that for $0<\Gamma\leq 1/5$ there is a range of $\epsGR$ values where the saddle point at $\omega=0$ has gone below $e=0$ and so no longer exists, but the $\omega=\pm\pi/2$ fixed points still do exist.  

Even these remaining fixed points get pushed to (and past) $e=0$ as $\epsGR$ is increased ever further, leaving the entire phase space filled with circulating trajectories that have their eccentricity maxima at $\omega = \pm \pi/2$, just as for $\Gamma > 1/5$ (Figure \ref{fig:HStar_Contours_Gamma_pt1}e,j).  The amplitude of eccentricity oscillations decreases correspondingly until cluster tides are completely negligible and only GR apsidal precession remains.

\begin{figure*}
\centering
\includegraphics[width=0.99\linewidth]{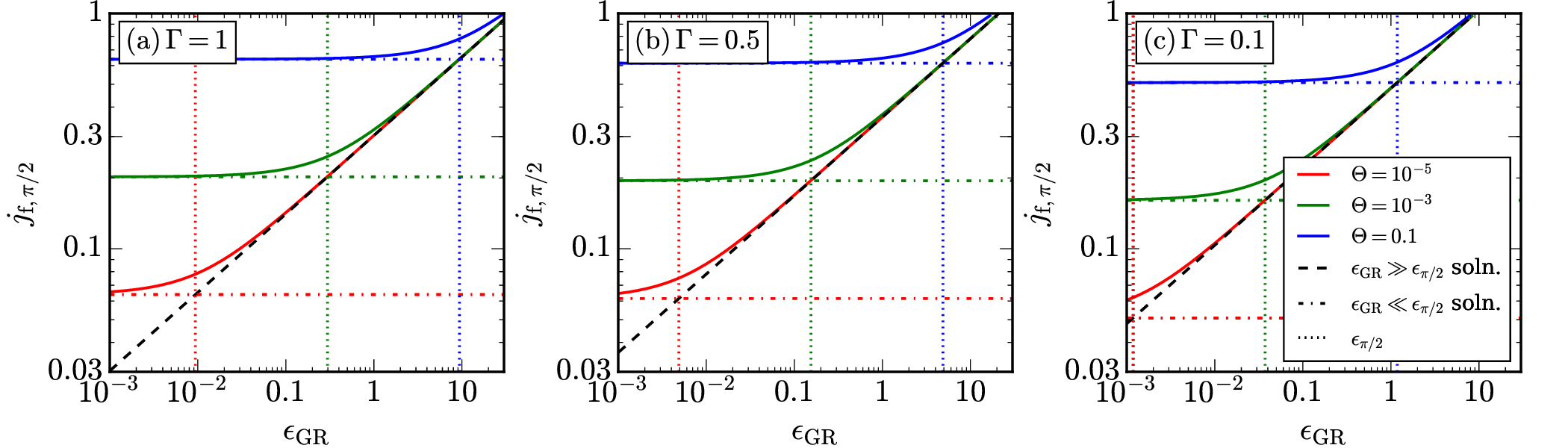}
\caption{Plots of $j_\mathrm{f,\pi/2}$ --- i.e. the values of $j$ for fixed points at $\omega=\pi/2$ --- for several values of $\Gamma$ and $\Theta$ (indicated on panels using labels and colours). Solid lines show the exact solution $j_{\mathrm{f},\pi/2}(\Gamma,\Theta,\epsGR)$ found by solving the quartic equation \eqref{eqn:FPs_at_omega_piby2}. Dot-dashed and dashed lines indicate the asymptotic solutions (\ref{eqn:jf_piby2_veryweakGR}) and (\ref{eqn:jf_piby2_weaktomodGR}) respectively, while vertical dotted lines indicate $\epspibytwo = \epspibytwo$ (see equation \ref{eqn:epspibytwo}) where the two asymptotic solutions match.}
\label{fig:j_fpi2}
\end{figure*}
 
   \begin{figure}
\centering
\includegraphics[width=0.84\linewidth]{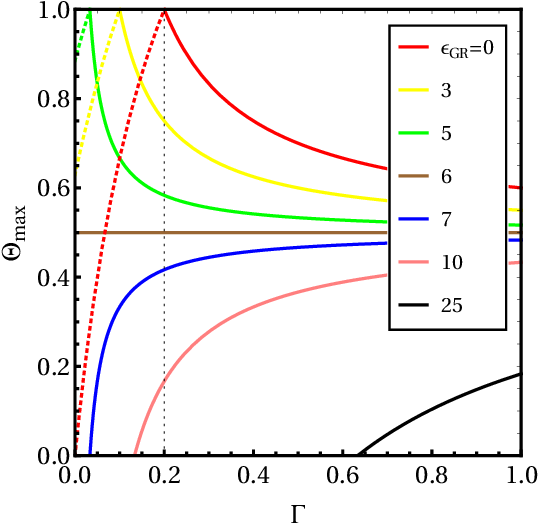}
\caption{Plots of $\Theta_\mathrm{max}$, the maximum value of $\Theta$ for which fixed points could exist at $\omega=\pm \pi/2$, defined by equation \eqref{eqn:Theta_constraint_fp_piby2}, as a function of $\Gamma$ for various values of $\epsGR$. Solid lines show $\Theta_\mathrm{max} = \Theta_1$ while dotted lines show $\Theta_\mathrm{max} = \Theta_2$.  The vertical dotted line corresponds to $\Gamma=1/5$. This figure is discussed in more detail after equation \eqref{eqn:Theta_crit}.}
\label{fig:Theta_max}
\end{figure}

           
\subsection{Fixed points}
\label{sect:fixed_points}
   

We now proceed to understand mathematically the nature of the various fixed points that we found in the phase portraits in \S\S\ref{sec:morphology_Gamma_Regime_I}-\ref{sec:morphology_Gamma_Regime_II}. By setting $\md j/\md t = 0$ in equation \eqref{eom2}, we see that all possible non-trivial fixed points\footnote{i.e. not corresponding to $j^2=\Theta$ or $j^2=1$.} are located on (i) the lines $\omega = \pm \pi/2$, as in Paper II, and/or (ii) the lines $\omega = 0, \pm \pi$, consistent with Figures \ref{fig:HStar_Contours_Gamma_pt5} and \ref{fig:HStar_Contours_Gamma_pt1}.  Finding the $j$ values of the fixed points requires plugging these $\omega$ values into $\md \omega/\md t = 0$, given by equation \eqref{eom1}, and solving the resulting algebraic equation for $j$. We do this next for each of the fixed points.

           
\subsubsection{Fixed points at $\omega = \pm \pi/2$}
\label{sec:fixed_points_pi_over_2}

In \S\ref{sec:fixed_points_omega_piby2}, we show how to calculate the $j$ value of the fixed points at $\omega=\pm\pi/2$, which we call $j_{\mathrm{f},\pi/2}$, for arbitrary $\epsGR$ and for any $\Gamma>0$, i.e. for both types of phase portraits shown in Figures \ref{fig:HStar_Contours_Gamma_pt5}, \ref{fig:HStar_Contours_Gamma_pt1}. The values of  $j_{\mathrm{f},\pi/2}$ are found as solutions to the quartic polynomial \eqref{eqn:FPs_at_omega_piby2}, and we illustrate their behaviour in Figure \ref{fig:j_fpi2} for several values of $\Gamma$ and $\Theta$. One can see that $j_{\mathrm{f},\pi/2}$ always {\it increases} with $\epsGR$ (see \eqref{eqn:djfdepsGR}), explaining why in Figures \ref{fig:HStar_Contours_Gamma_pt5},\ref{fig:HStar_Contours_Gamma_pt1} the fixed points at $\omega = \pm \pi/2$ always get pushed to lower $e$ as $\epsGR$ is gradually increased from zero. 

While the explicit expressions for $j_{\mathrm{f},\pi/2}$ are too complicated to be shown here, we can gain important insights by considering two limiting cases, namely when $\epsGR$ is much smaller/larger than a particular critical value:
\begin{align}
       \epspibytwo \equiv 6(10\Gamma\Theta)^{3/4}(1+5\Gamma)^{1/4}.
       \label{eqn:epspibytwo}
\end{align}
(In the top and bottom rows of Figure \ref{fig:HStar_Contours_Gamma_pt5}, $\epspibytwo$ takes values around $4.9$ and $16.3$ respectively). In the limit $\epsGR \ll \epspibytwo$, which we will call `very weak GR' regime, the $\epsGR$ term in \eqref{eqn:FPs_at_omega_piby2} is small and we find to lowest order in $\epsGR/\epspibytwo$
\begin{align}
j_\mathrm{f,\pi/2} \approx j_\mathrm{f}\left(1 + \frac{\epsGR}{4\epspibytwo}\right).
\label{eqn:jf_piby2_veryweakGR}
\end{align}
In the opposite limit $\epsGR \gg \epspibytwo$ the right hand side in \eqref{eqn:FPs_at_omega_piby2} becomes small and we find to lowest order in $\epspibytwo/\epsGR$
   \begin{align}
    j_\mathrm{f,\pi/2} \approx \left[ \frac{\epsGR}{6(1+5\Gamma)} \right]^{1/3}\left[1 + \frac{1}{3}\left( \frac{\epspibytwo}{\epsGR} \right)^{4/3} \right].
    \label{eqn:jf_piby2_weaktomodGR}
\end{align}
Figure \ref{fig:j_fpi2} shows that these asymptotic solutions match the actual $j_\mathrm{f,\pi/2}$ behaviour in the appropriate limits very well.

In \S\ref{sec:fixed_points_omega_piby2} we show also that for fixed points at $(\omega,j) = (\pm\pi/2,j_{\mathrm{f},\pi/2})$ to exist for a given $\Gamma>0$, the quantities $\Theta$ and $\epsGR$ must obey the inequalities
\begin{align}
    6\Theta^{1/2}[(1+5\Gamma)\Theta - 10\Gamma] < \epsGR < 6[1+5\Gamma - 10\Gamma\Theta],
        \label{eqn:epsGR_constraint_fp_piby2}
\end{align}
and 
   \begin{align}
&\Theta < \Theta_\mathrm{max} \equiv \begin{cases} \Theta_1,\,\,\,\,\,\,\,\,\,\,\,\,\,\,\,\,\,\,\,\,\,\,\,\,\,\,\,\,\,\, \Gamma > 1/5, \\ \mathrm{min}[\Theta_1
    , \Theta_2], \,\,\,\,\,\, 0<\Gamma \leq 1/5.
    \label{eqn:Theta_constraint_fp_piby2}
    \end{cases}
\end{align}
Here $\Theta_2$ is the smallest positive real solution to equation \eqref{eqn:Theta_crit}, while
\begin{align}
    \Theta_1 \equiv \frac{1+5\Gamma}{10\Gamma}\left(1-\frac{\epsGR}{2\epsstr}\right) = \frac{1+5\Gamma-\epsGR/6}{10\Gamma},
    \label{eqn:Theta_1}
\end{align}
and we have defined
\begin{align}
\epsstr \equiv 3(1+5\Gamma),
        \label{eqn:epsstrong}
\end{align}
a quantity that will appear repeatedly throughout this paper.  

In Figure \ref{fig:Theta_max} we plot $\Theta_\mathrm{max}$ (equation \eqref{eqn:Theta_constraint_fp_piby2}) as a function of $\Gamma$ for various values of $\epsGR$ (c.f. Figure 1 of Paper II). It is easy to check that the combinations of $\Gamma, \Theta$ and $\epsGR$ that give rise to $\omega=\pm\pi/2$ fixed points in Figure \ref{fig:HStar_Contours_Gamma_pt5} do obey the inequalities \eqref{eqn:epsGR_constraint_fp_piby2}-\eqref{eqn:Theta_constraint_fp_piby2}. 
Of course, in the limit $\epsGR \to 0$ equations (\ref{eqn:Theta_constraint_fp_piby2})-(\ref{eqn:epsstrong}) reduce to the non-GR constraint \eqref{eq:Theta_constr}. Finally we note that for sufficiently small $\Theta$, the conditions \eqref{eqn:epsGR_constraint_fp_piby2}, \eqref{eqn:Theta_constraint_fp_piby2} reduce simply to the requirement that
\begin{align}
\label{eqn:FPs_pi_by_2_criterion_small_Theta}
    \epsGR < 2\epsstr \,\,\,\,\,\, (\mathrm{for} \,\,\, \Theta \ll 1).
\end{align}
In other words, if \eqref{eqn:FPs_pi_by_2_criterion_small_Theta} is not satisfied then there are no fixed points even for initially orthogonal inner and outer orbits ($i_0=90^\circ$). 
This reflects what we see in  Figures \ref{fig:HStar_Contours_Gamma_pt5}d,e and \ref{fig:HStar_Contours_Gamma_pt1}d,e (see also \S\ref{sec:phase_small_Theta}).

           
\subsubsection{Fixed points at $\omega = 0,\pi$}
\label{sec:fixed_points_0}

Fixed points at $\omega = 0,\pi$ are unique to the $0<\Gamma\leq 1/5$ regime (for $\Gamma>0$). They are always saddle points, and we explore their properties mathematically in \S\ref{sec:fixed_points_omega_0}. From now on, for brevity we will simply refer to them as being located at $\omega=0$ rather than $\omega=0, \pm\pi$, because phase space locations separated in $\omega$ by multiples of $\pi$ are equivalent (see equation \eqref{eqn:H1star_omega_j}). 

As we demonstrate in \S\ref{sec:fixed_points_omega_0}, these saddle points are always located at $(\omega ,j) =(0, j_\mathrm{f,0})$ where
\begin{align}
    j_\mathrm{f,0} \equiv \left[ \frac{\epsGR}{6(1-5\Gamma)} \right]^{1/3}.
    \label{eqn:jf0}
\end{align}
Note that $j_\mathrm{f,0}$ is independent of $\Theta$, which can is reflected in Figure \ref{fig:HStar_Contours_Gamma_pt1}c,h. Also, the constraint \eqref{eq:AMconstr} implies that fixed points exist at $(\omega ,j) =(0, j_\mathrm{f,0})$ if and only if
\begin{align}
    \Theta^{3/2} < \frac{\epsGR}{6(1-5\Gamma)} < 1.
    \label{eqn:fp_omega_0_inequality}
\end{align}
Obviously $\epsGR$ must be finite for the inequality \eqref{eqn:fp_omega_0_inequality} to hold even for very small $\Theta$ --- hence fixed points at $\omega = 0$ do not exist for $\epsGR = 0$, which is why they were not found in Paper II and do not exist in Figures \ref{fig:HStar_Contours_Gamma_pt5}a,f or \ref{fig:HStar_Contours_Gamma_pt1}a,f. 

In addition we learn from \eqref{eqn:fp_omega_0_inequality} that there are never any fixed points at $\omega=0$ for $\Gamma > 1/5$ regardless of $\epsGR$, which explains the phase space structure in Figure \ref{fig:HStar_Contours_Gamma_pt5}. In particular this implies that for the LK problem ($\Gamma=1$) the only possible fixed point locations are the standard ones at $\omega = \pm\pi/2$, regardless of the value of $\epsGR$. For positive $\Gamma$, fixed points at $\omega =0$ can be realised only for $\Gamma \leq 1/5$ and we see from \eqref{eqn:jf0}-\eqref{eqn:fp_omega_0_inequality} that when $\epsGR$ exceeds the threshold value $6(1-5\Gamma)\Theta^{3/2}$ (corresponding to $\epsGR=0.095$ and $1.06$ in the top and bottom rows of Figure \ref{fig:HStar_Contours_Gamma_pt1}, respectively), a fixed point appears at the limiting eccentricity $e_\mathrm{lim} = \sqrt{1-\Theta}$. Increasing $\epsGR$ at fixed $\Gamma$ always acts to increase $j_\mathrm{f,0}$, i.e. to decrease the eccentricity $e_\mathrm{f,0} \equiv (1-j_\mathrm{f,0}^2)^{1/2}$ of this particular fixed point. As we increase $\epsGR$ to the threshold value $\epsGR=6(1-5\Gamma)$ (which is independent of $\Theta$ and corresponds to $\epsGR=3$ in Figure \ref{fig:HStar_Contours_Gamma_pt1}), the saddle point vanishes through $e=0$, leaving only the fixed points at $\omega=\pm\pi/2$.

Beyond that threshold, as mentioned in \S\ref{sec:fixed_points_0}, there is a range of $\epsGR$ values for which the saddle point at $\omega=0$ is no longer present, but the $\omega=\pm\pi/2$ fixed points still do exist. Combining the constraints \eqref{eqn:epsGR_constraint_fp_piby2}, \eqref{eqn:Theta_constraint_fp_piby2} and \eqref{eqn:fp_omega_0_inequality} we see that for $\Theta \ll 1$ this range is given approximately by
\begin{align}
   6(1-5\Gamma) < \epsGR < 6(1+5\Gamma).
   \label{eqn:piby2_but_no_saddle}
\end{align} 
The lower limit here is exact, while the upper limit is correct to zeroth order in $\Theta$. Within this range the qualitative behaviour resembles the $\Gamma > 1/5$ behaviour we saw in Figure \ref{fig:HStar_Contours_Gamma_pt5}; in particular, the maximum eccentricity of all orbits is found at $\omega = \pm\pi/2$. The range \eqref{eqn:piby2_but_no_saddle} is important because it allows for eccentricity excitation of initially near-circular binaries, which is not possible in the $0<\Gamma \leq 1/5$ regime when $\epsGR=0$ (see \S\ref{sec:max_ecc_circular}).


\subsection{Determination of the maximum eccentricity of a given orbit}
\label{sec:max_min_ecc_GR}  

 
Our next goal is to calculate the maximum eccentricity $e_\mathrm{max}$ reached by a binary given the initial conditions $(\omega_0, e_0, \Theta, \Gamma, \epsGR)$.  In particular, we wish to know if a binary will reach $\emax \to 1$, since this is the regime in which dissipative effects (e.g. GW emission) can become important.
 
For $\Gamma > 1/5$, a binary's maximum eccentricity is always found at $\omega=\pi/2$ regardless of whether its phase space orbit librates or circulates (Figure \ref{fig:HStar_Contours_Gamma_pt1}). Plugging $\omega = \pi/2$ into $H^*(\omega,j)$ gives us a depressed quartic equation:
\begin{align}
    j^4 & + \left(  \frac{H^*-24\Gamma\Theta-5-15\Gamma}{3(1+5\Gamma)} \right) j^2 + \frac{\epsGR}{3(1+5\Gamma)}j \nn \\ &+ \frac{10\Gamma\Theta}{1+5\Gamma} = 0.
    \label{eqn:depressed_quartic}
\end{align}
We call real roots of equation \eqref{eqn:depressed_quartic} $j(\omega=\pi/2)$. In the limit $\epsGR=0$,  equaton \eqref{eqn:depressed_quartic} reduces to a quadratic for $j^2(\omega=\pi/2)$ and we recover the solution (18) of Paper II.  For $\epsGR\neq 0$ the real roots of \eqref{eqn:depressed_quartic} can still be written down analytically but they are too complicated to be worth presenting here. The minimum angular momentum $\jmin$ (corresponding to the maximum eccentricity $\emax \equiv \sqrt{1-\jmin^2}$) will then be given by the smallest physical root $j(\omega=\pi/2)$, i.e. the smallest root of \eqref{eqn:depressed_quartic} that satisfies $\sqrt{\Theta} < j(\omega=\pi/2) < 1$. 

The situation is slightly more complex for $0 < \Gamma \leq 1/5$. In this case we must first work out whether an orbit librates or circulates  (and if it circulates, to which circulating family it belongs, since it can be above or below the saddle point, as in Figures \ref{fig:HStar_Contours_Gamma_pt1}b,c,h).  To do so we use the procedure given in \S\ref{sec:lib_or_circ} to calculate $j(\omega=0)$, which is the solution to the depressed cubic equation \eqref{eqn:depressed_cubic} that results from plugging $\omega = 0$ into $H^*(\omega,j)$. If the orbit circulates `below' the librating regions and the saddle point then we have $\jmin = j(\omega=0)$. Otherwise $\jmin$ is found at $\omega=\pm\pi/2$ and we proceed as for $\Gamma > 1/5$ by solving equation \eqref{eqn:depressed_quartic}.

  
\subsubsection{Maximum eccentricity achieved by initially near-circular binaries}
\label{sec:max_ecc_circular}

We can gain further insight and connect to the results of previous LK studies by considering the simplified case of initially near-circular binaries, $e_0 \approx 0$.  Evaluating the integrals of motion $H^*$ and $\Theta$ with the initial condition $e_0=0$ we find
\begin{align}
    H^* = 2(1-3\Gamma\cos^2i_0)-\epsGR, \,\,\,\,\,\,\,\,\,\, \Theta = \cos^2i_0.
    \label{eqn:H_Theta_circular}
\end{align}
Note the lack of $\omega_0$ dependence in these constants.

Now, for $\Gamma > 1/5$ eccentricity is always maximised at $\omega = \pi/2$, so can be found by solving \eqref{eqn:depressed_quartic}. Plugging \eqref{eqn:H_Theta_circular} into \eqref{eqn:depressed_quartic} we find that $\jmin$ is the solution to the equation
\begin{align}
0 =& (j-1) \nn \\ &\times \left[j^3 + j^2 -  \frac{(10\Gamma\cos^2i_0 + \epsGR/3)}{1+5\Gamma} j - \frac{10\Gamma\cos^2i_0}{1+5\Gamma}
    \right].
    \label{eqn:quartic_circular}
\end{align}
In the LK limit of $\Gamma =1$, equation \eqref{eqn:quartic_circular} is equivalent to e.g. equation (34) of \citet{Fabrycky2007} or equation (50) of \citet{Liu2015}\footnote{Note that there is a typo in \citet{Liu2015}'s equation (50) --- the factor of $3/5$ on the right hand side should be $5/3$.}.  Note that $\jmin=1$ (i.e. $\emax =0$) is a solution to this equation.  It is the correct solution in the special case of a perfectly initially circular orbit, $e_0 \equiv 0$, which necessarily remains circular forever. This is because perfectly circular binaries feel no net torque from the external tide, which can be seen by plugging $j=1$ into equation (\ref{eom2}).

Meanwhile, an orbit that has $e_0$ infinitesimally larger than zero can have $\jmin$ corresponding to a non-trivial solution of \eqref{eqn:quartic_circular}. This will be the case if and only if the $\omega = \pm\pi/2$ fixed points have not yet disappeared below $e=0$ (panels (a)-(d) and (f)-(h) of Figure \ref{fig:HStar_Contours_Gamma_pt5}). Because of the constraint \eqref{eqn:epsGR_constraint_fp_piby2}, a necessary (and for $i_0 \to 90^\circ$, sufficient) requirement for this is $\epsGR < 6(1+5\Gamma)$. In that case the fixed points bound the maximum eccentricity from below, so $\emax > e_{\mathrm{f},\pi/2} \equiv (1-j^2_{\mathrm{f},\pi/2})^{1/2}$. On the other hand, if $\epsGR$ is large enough that the fixed points \textit{have} disappeared through $e=0$ then we simply have $\emax = 0$ (see panels (e), (i), (j) of Figure \ref{fig:HStar_Contours_Gamma_pt5}).

Next we turn to the regime $0<\Gamma \leq 1/5$. By consulting Figure \ref{fig:HStar_Contours_Gamma_pt1} one can see that a finite eccentricity is only achieved if $\epsGR$ is sufficiently large that the saddle point at $\omega =0$ has passed `down' the $(\omega,e)$ phase space and disappeared through $e=0$, but also sufficiently small that the $\omega=\pm\pi/2$ fixed points still exist (as in Figure \ref{fig:HStar_Contours_Gamma_pt1}d,i). A necessary requirement for this (which is again sufficient in the case $i_0 \to 90^\circ$) is that \eqref{eqn:piby2_but_no_saddle} be true.  Then $e$ is maximised at $\omega=\pm\pi/2$ and $\jmin$ is a non-trivial solution to equation \eqref{eqn:quartic_circular}. 
On the other hand, if \eqref{eqn:piby2_but_no_saddle} is not satisfied then a binary that starts at $e_0 \approx 0$ never increases its eccentricity\footnote{There is another solution at $\omega = 0$ given by equation \eqref{eqn:circular_omega_0_solution} which is unphysical for $\Gamma > 0$ but will become important for $\Gamma \leq 0$ --- see \S\ref{sec:max_ecc_circular_negative_Gamma}.} even for $i_0=90^\circ$.

   \begin{figure*}
\centering
\includegraphics[width=0.78\linewidth]{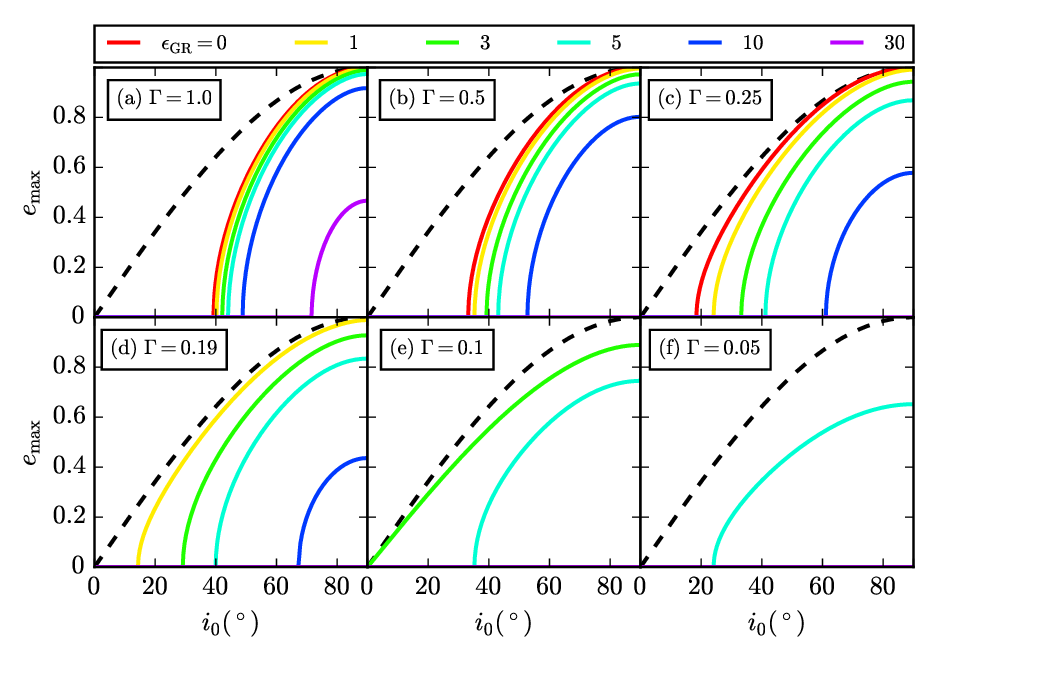}
\caption{Maximum eccentricity $\emax$ as a function of $i_0$ for initially near-circular binaries. Panels (a)-(c) are for $\Gamma > 1/5$ while panels (d)-(f) correspond to $0<\Gamma \leq 1/5$.  In each panel, different coloured lines represent the different values of $\epsGR$ (see legend). A dashed black line corresponds to $\emax = e_\mathrm{lim} = \sin i_0$. Note that for initially circular orbits to reach a non-zero $\emax$ we require fixed points to exist in the phase portrait at $\omega = \pm \pi/2$ but not at $\omega = 0$; for $i_0 \approx 90^\circ$ this corresponds to $6(1-5\Gamma) < \epsGR < 6(1+5\Gamma)$ --- see equation \eqref{eqn:piby2_but_no_saddle}. Note also that for $\Gamma < 1/5$, eccentricity excitation of near-circular binaries may be possible regardless of inclination, even when $i_0=0^\circ$, as for $\epsGR=3$} in panel (e).
\label{fig:emax_initially_circular}
\end{figure*}

Overall then, we see that for near-circular binaries to reach finite $\emax$ we require fixed points to exist in the phase portrait at $\omega = \pm \pi/2$ but not at $\omega = 0$, and this necessarily requires $\epsGR$ to satisfy \eqref{eqn:piby2_but_no_saddle}.  

In Figure \ref{fig:emax_initially_circular} we plot $\emax$ as a function of $i_0$ for initially near-circular binaries. Panels (a)-(c) are for $\Gamma > 1/5$ (c.f. Figure 3 of \citet{Fabrycky2007} and Figure 6 of \citet{Liu2015}) while panels (d)-(f) correspond to $0<\Gamma \leq 1/5$.  In each panel, different coloured solid lines represent the different values of $\epsGR$, while a dashed black line indicates the limiting eccentricity $e_\mathrm{lim} = \sqrt{1-\Theta} = \sin i_0$ (the highest possible $e$ for an initially near-circular binary, corresponding to $j=\cos i_0$).
We see that for $\Gamma > 1/5$, the effect of increasing $\epsGR$ at a fixed $i_0$ (and therefore a fixed $\Theta$) is always to decrease $\emax$. This is what we would expect by comparing the top and bottom rows of Figure \ref{fig:HStar_Contours_Gamma_pt5}. Moreover, if we consider the most favourable orbital inclination $i_0= 90^\circ$ then we can easily derive the exact solution to \eqref{eqn:quartic_circular}. We find that either $\jmin=1$ (so $\emax=0$), or that 
\begin{align}\jmin=\frac{1}{2}\left[\left(1+\frac{4\epsGR}{\epsstr}\right)^{1/2}-1\right],
\label{eqn:jmin_circular_i90}\end{align}
with $\epsstr$ defined in \eqref{eqn:epsstrong}; in the LK limit this result reduces to equation (35) of \citet{Fabrycky2007}. Expanding the solution \eqref{eqn:jmin_circular_i90} for $\epsGR/\epsstr \ll 1$ we find 
\begin{align}
    e_\mathrm{max} \approx 1-\frac{1}{2}\left(\frac{\epsGR}{\epsstr}\right)^2.
    \label{eq:emax_limit}
\end{align}
Thus we expect $\emax \to 1$ for these favourably inclined binaries when GR is negligible, but also that $\emax$ will deviate from $1$ considerably when $\epsGR$ starts approaching $\epsstr$, which is what we see in Figure \ref{fig:emax_initially_circular}a,b,c. Obviously this means that the smaller is $\Gamma$, the smaller $\epsGR$ needs to be to suppress the very highest eccentricities. Finally we note that there is no magenta curve --- corresponding to $\epsGR = 30$ --- in either panel (b) or panel (c).  This is because for these $\Gamma$ values the constraint \eqref{eqn:piby2_but_no_saddle} is violated for $\epsGR=30$, so the only possible solution to \eqref{eqn:quartic_circular} is $\emax=0$. 

Now consider the regime $0 < \Gamma \leq 1/5$ exhibited in panels (d)-(f).
The reader will notice the diminishing number of curves in these panels.  Indeed, there is not even a red curve corresponding to $\epsGR=0$.  This again is a consequence of the fact that for $\epsGR=0$, equation \eqref{eqn:piby2_but_no_saddle} cannot be satisfied, so that initially circular orbits achieve no eccentricity excitation ($\emax=0$). 

A related phenomenon is that in panel (e), the green ($\epsGR = 3$) curve asymptotes to the black dashed line $e=e_\mathrm{lim}$ as $i_0\to 0^\circ$.  This is also as expected: since $\Gamma=0.1$, equation \eqref{eqn:piby2_but_no_saddle} tells us $\epsGR = 3$ is precisely the lower bound on GR strength above which initially near-circular binaries can reach non-zero eccentricities at all $i_0$, since at this value of $\epsGR$ the saddle point crosses $e=0$ --- see Figure \ref{fig:HStar_Theta_pt001_logplot}i,j,k.

Note that this $0 < \Gamma \leq 1/5$ behaviour is completely different from that found for near-circular binaries in the $\Gamma > 1/5$ regime (and therefore to the known LK results).  For $\Gamma > 1/5$, taking  $i_0\approx 0^\circ$ inevitably leads to $\emax \approx 0$ --- in other words there is no eccentricity excitation for initially coplanar ($i_0=0$) orbits, regardless of $\epsGR$.  Moreover, for $\Gamma>1/5$ even if a binary can reach a finite maximum eccentricity for $\epsGR=0$, increasing $\epsGR$ always decreases this maximum eccentricity.  On the contrary, for $0 < \Gamma \leq 1/5$ reaching a finite $e_\mathrm{max}$ may be possible even for initially almost coplanar orbits, and a finite $\epsGR$ is actually \textit{necessary} to trigger the eccentricity excitation starting from a circular orbit.  Despite this, comparison of the top and bottom rows of Figure \ref{fig:emax_initially_circular} reinforces the idea that the $0 < \Gamma \leq 1/5$ regime admits far fewer high-eccentricity solutions than $\Gamma > 1/5$ as $\epsGR$ is varied.


\section{High eccentricity behaviour}
\label{sec:High_ecc_behaviour}


Our next goal is to understand the impact of GR precession on the time dependence of the binary orbital elements in the important limit of very high eccentricity, $e\to 1$.  This limit is relevant in a variety of astrophysical contexts. For example, the dramatic reduction of the binary pericentre distance that occurs when $e$ approaches unity can trigger short-range effects such as tidal dissipation (leading to hot Jupiter formation), GW emission (leading to compact object mergers), and so on.  Thus we wish to understand in detail how GR precession affects not only the maximum eccentricity  $e_\mathrm{max}$, but also the behaviour of $e(t)$ and other orbital elements in the vicinity of $e_\mathrm{max}$.

In \S\ref{sec:max_min_ecc_GR} we explained how to find $e_\mathrm{max}$ for arbitrary $\Gamma>0$, initial conditions $(e_0, i_0, \omega_0)$, and value of $\epsGR$.  Here we will examine the solutions quantitatively in the high eccentricity limit, and explore the time spent near highest eccentricity.  To this end we will make extensive use of equation \eqref{eq:djdtGR}, which tells us $\md j/\md t$ as a function of $j$. It is important to note that the solutions for extrema of $j$ at $\omega = 0$ and $\omega = \pm\pi/2$ are all contained within \eqref{eq:djdtGR}.  Indeed, setting the first square bracket inside the square root in \eqref{eq:djdtGR} to zero gives the depressed quartic equation \eqref{eqn:depressed_quartic} whose roots correspond to extrema of $j$ at $\omega = \pm\pi/2$, i.e. what we have so far called $j(\omega=\pm\pi/2)$.  Setting the other square bracket to zero gives the depressed cubic \eqref{eqn:depressed_cubic}-\eqref{eqn:q} which determines the roots at $\omega = 0$, i.e. what we called $j(\omega=0)$.

In this section we will focus on situations in which $e_\mathrm{max}$ is achieved at $\omega=\pm\pi/2$, since this is the most common prerequisite for $e\to 1$ (\S\S\ref{sec:morphology_Gamma_Regime_I}-\ref{sec:morphology_Gamma_Regime_II}). The rare cases in which $e$ approaches unity at $\omega=0$ are covered in Appendix \ref{sec:high_e_omega_0}. 

 
\subsection{Phase space behaviour for $\Theta \ll 1, \,\,\, \Gamma > 0$}
\label{sec:phase_small_Theta}

\begin{figure*}
\centering
\includegraphics[width=0.99\linewidth,trim={0.9cm 0.3cm 0cm 0.2cm},clip]{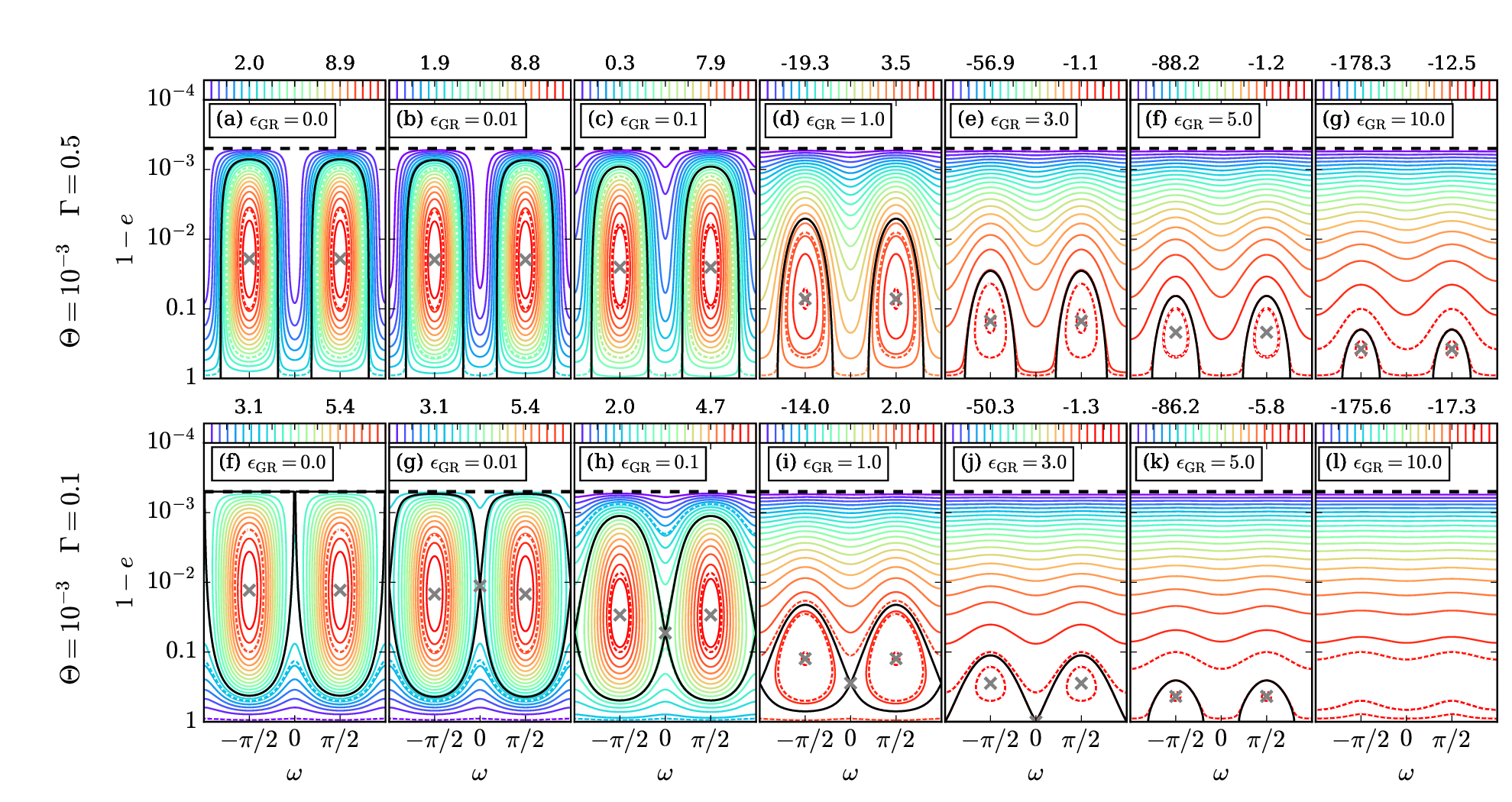}
\caption{As in Figures \ref{fig:HStar_Contours_Gamma_pt5} and \ref{fig:HStar_Contours_Gamma_pt1}, except we have (I) fixed $\Theta = 10^{-3}$ and used two values of $\Gamma$ (namely $0.5$ and $0.1$ for the top and bottom row respectively),  (II) plotted $1-e$ on the vertical axis using an inverted logarithmic scale (so that $e$ still increases vertically), (III) added by hand additional dashed contours with the value $H^*(\omega=\pm\pi/2,e=0.9)$, and (IV) used some new values of $\epsGR$.}
\label{fig:HStar_Theta_pt001_logplot}
\end{figure*}

We are interested in binaries that start with initial eccentricity $e_0$ not close to unity, and that are capable of reaching extremely high eccentricities $\emax\to 1$, i.e. $\jmin\to 0$. For this to be possible a necessary condition is that $\Theta \ll 1$, owing to the constraint \eqref{eq:AMconstr}. Hence it is important to understand the regime $\Theta \ll 1$ in detail.

In Figure \ref{fig:HStar_Theta_pt001_logplot} we show phase portraits for $\Gamma = 0.5$ (top row) and $\Gamma = 0.1$ (bottom row), this time fixing $\Theta = 10^{-3}$ in both cases, and adding in extra dashed contours\footnote{In addition to the dashed contours already included in Figures \ref{fig:HStar_Contours_Gamma_pt5} and \ref{fig:HStar_Contours_Gamma_pt1}.} with the value $H^*(\omega=\pm\pi/2,e=0.9)$. Note that on the vertical axis we now plot $1-e$ on a logarithmic scale, with eccentricity still increasing vertically as in Figures \ref{fig:HStar_Contours_Gamma_pt5}, \ref{fig:HStar_Contours_Gamma_pt1}. This allows us to see in detail how trajectories separate from $e\approx e_\mathrm{lim}$ as we increase $\epsGR$.

In these plots, $\Theta$ is sufficiently small that to a very good approximation the requirement for fixed points at $\omega = \pm\pi/2$ to exist is just $\epsGR < 2\epsstr$ (equation \eqref{eqn:FPs_pi_by_2_criterion_small_Theta}). This critical value is surpassed in panel (l), since in that case $\epsGR=10$ while $2\epsstr = 9$, which is why all fixed points have disappeared. Meanwhile the criterion for a saddle point to exist at $\omega=0$ (equation \eqref{eqn:fp_omega_0_inequality}) for $\Gamma = 0.1$ and $\Theta=10^{-3}$ is approximately $10^{-5} < \epsGR < 3$.  This is consistent with what we see in panels (f)-(l) --- note in particular the transitional point $\epsGR=3$ in panel (j).

Comparing the top and bottom rows of Figure \ref{fig:HStar_Theta_pt001_logplot}, one observes a striking difference between behaviour in the $\Gamma > 1/5$ and $0 < \Gamma \leq 1/5$ dynamical regimes.  For $\Gamma = 0.5 > 1/5$, an initially near-circular binary can be driven to very high eccentricity $(\gtrsim 0.99)$ even for $\epsGR = 1.0$ (panel (d)).  Conversely, for $\Gamma =0.1<1/5$ the phase space structure simply does not allow such behaviour (panels (f)-(i)). 
More precisely, for $0 < \Gamma \leq 1/5$, the eccentricity of the saddle point \eqref{eqn:jf0} acts as a hard boundary on the maximum eccentricity of low-$e$ orbits, and most of them do not get close even to that value. Even when $\epsGR$ is increased so that a new family of circulating orbits appears, and the librating region is significantly enlarged, the system admits very few solutions that start at low $e$ and achieve high $e$.
It is therefore unsurprising that one finds fewer cluster-tide driven compact object mergers from systems such as globular clusters that have a relatively high fraction of binaries in the $0 < \Gamma \leq 1/5$ regime \citep{Hamilton2019c}.


\subsection{High eccentricity behaviour for $\epsGR=0$}
\label{sect:high-e-eps-0}


Before embarking on a full study of high eccentricity evolution for arbitrary $\epsGR$, we first consider the case $\epsGR=0$. In that case the non-zero roots of the polynomial on the right hand side of \eqref{eq:djdtGR} are $j_\pm,j_0$, one of which will correspond to the minimum angular momentum $j_\mathrm{min}$. Then we can integrate \eqref{eq:djdtGR} with $\epsGR=0$ to find $t(j)$; the resulting expression involves an incomplete elliptical integral of the first kind (see \S2.6 of Paper II for the general $\Gamma$ case, and \citet{Vash1999, Kinoshita2007} in the LK case of $\Gamma =1$). 
Next, assuming that $j^2 \ll 1$, we can expand this elliptical integral to find\footnote{Note that one can get the same result simply by expanding the right hand side of \eqref{eq:djdtGR} for $j \ll 1$.} (see \S9.2 of Paper II)
\begin{align} 
j(t) = j_\mathrm{min}\sqrt{1+\left(\frac{t}{t_\mathrm{min}} \right)^2}, \,\,\,\,\,\,\, \mathrm{where} \,\,\,\,\,\,\,
t_\mathrm{min} &\equiv \frac{j_\mathrm{min}}{j_1 j_2}\tau,
\label{eq:t_min}
\end{align}
$j_1, j_2$ are the two roots \textit{not} corresponding to $j_\mathrm{min}$,
and $\tau$ is a characteristic secular timescale which is independent of $e_0, i_0, \omega_0$:
\begin{align}
\tau \equiv \frac{L}{6C\sqrt{\vert 25\Gamma^2-1 \vert}}.
\label{eq:tau}
\end{align}
Using the definitions of $C$ and $L$ one can show that $\tau$ is, up to constant factors, the same at $t_{\rm sec}$ defined after equation (\ref{eq:epsGRnumerical}). Note we have taken the origin of the time coordinate to coincide with $j=j_\mathrm{min}$. Clearly $t_{\rm min}$ is the characteristic evolution timescale in the vicinity of $j_{\rm min}$, i.e. the time it takes for $j$ to change from $j_{\rm min}$ to $\sqrt{2}j_{\rm min}$. 

Note that the solution (\ref{eq:t_min}) is quadratic in $t$ for $t\lesssim t_\mathrm{min}$ and linear when $t\gtrsim t_{\rm min}$, as long as $j$ remains $\ll 1$. It provides a better approximation to $j(t)$ over a wider interval of time near the peak eccentricity than the purely quadratic approximation adopted by \citet{Randall2018}, in their calculation of the GW energy emitted by a binary undergoing LK oscillations (\S\ref{sec:analytic_validity}).


\subsection{Modifications brought about by finite \texorpdfstring{$\epsGR$}{eGR}}
\label{sec:including_epsGR}


Before we proceed to examine the $j(t)$ behaviour, it is important to realise that including a finite $\epsGR$ affects the right hand side of \eqref{eq:djdtGR}, and therefore the value of $\jmin$, in two distinct ways.  First, there is the obvious explicit dependence on $\epsGR$ that appears twice in equation \eqref{eq:djdtGR}. Second, there is also an implicit dependence on $\epsGR$ in \eqref{eq:djdtGR} through the values of $j_\pm$ and $j_0$ (see equations \eqref{eqn:jpm}, \eqref{eqn:j0}). We will now discuss this implicit dependence, and then use the results to understand $j(t)$ behaviour in different asymptotic $\epsGR$ regimes. 

In the limit $\Theta \ll 1$, and assuming that $e_0$ is not too close to $1$ and $\Gamma$ is not too close to $1/5$, equations \eqref{eqn:def_Sigma}, \eqref{eqn:def_D} tell us that
\begin{align}
    \Sigma \approx
    (\epsstr+\epsGR)/6 +\mathcal{O}(e_o^2).
    \label{eqn:Sigma_small_Theta}
\end{align}
Equation \eqref{eqn:Sigma_small_Theta} implies that $\Sigma$, and hence $j_\pm^2$, will be modified significantly by GR only if $\epsGR \gtrsim \epsstr$, in agreement with what we saw in Figures \ref{fig:HStar_Contours_Gamma_pt5}, \ref{fig:HStar_Contours_Gamma_pt1} \& \ref{fig:HStar_Theta_pt001_logplot}. In this case, a perturbative approach around the non-GR solution will fail. We therefore say that any binary with $\epsGR \gtrsim \epsstr$ exists in the regime of `strong GR', which we explore in \S\ref{sec:very_strong_GR}. Conversely, if $\epsGR$ is in what we will call the `weak-to-moderate GR' regime:
\begin{align}
    \epsilon_\mathrm{GR} \ll \epsstr,
    \label{eqn:weak_to_moderate_GR_condition}
\end{align}
then $\Sigma \sim 1$, and $j_\pm$ will stay close to their non-GR values for $\Theta\ll \Sigma\sim 1$, namely (see equation (\ref{eqn:jpm}))
\begin{align} 
&j^2_+\approx \frac{2\Sigma}{1+5\Gamma} \sim 1,\,\,\,\,\,\,\,\,\,\,\,\,\,\mathrm{and}\,\,\,\,\,\,\,\,\,\,\,\,\, j^2_-\approx \frac{5\Gamma\Theta}{\Sigma} \sim \Theta \ll 1.
\label{eqn:j_plus_minus_small_Theta}
\end{align}
In other words the relative perturbations to $j_\pm$ induced by GR can be neglected.  Note that the weak-to-moderate GR regime \eqref{eqn:weak_to_moderate_GR_condition} already encompasses the very weak GR regime introduced in \S\ref{sec:fixed_points_pi_over_2}. In \S\S\ref{sec:weak_GR}-\ref{sec:moderate_GR} we will further delineate distinct `weak GR' and `moderate GR' regimes.

Also, using equations (\ref{eqn:j0}), (\ref{eqn:def_D}) it is easy to show that the {\it absolute} change to $j_0$ incurred by including GR will be small ($\ll 1$) whenever
\begin{align}
\epsGR \ll 3\vert 1-5\Gamma \vert \sqrt{1-e_0^2}.
\label{eqn:epsGR_ll_1_minus_5_Gamma}
\end{align}
Note that for $\Gamma$ not close to $1/5$ and $e_0$ not close to unity, the condition \eqref{eqn:epsGR_ll_1_minus_5_Gamma} is automatically guaranteed by the weak-to-moderate GR condition \eqref{eqn:weak_to_moderate_GR_condition}. In that case the relative perturbation to $j_0$ due to GR precession can be neglected (if $j_0\sim 1$).


\subsection{High-$e$ behaviour in the weak-to-moderate GR limit}
\label{sec:pert}


In the non-GR limit ($\epsGR=0$), for $\Gamma>0$ the vast majority of phase space trajectories that are capable of reaching very high eccentricities reach them at\footnote{The exception is for circulating orbits with $0 < \Gamma < 1/5$ that lie very close to the separatrix. These rare orbits are discussed in Appendix \ref{sec:high_e_omega_0}.} $\omega=\pm\pi/2$. 
As shown in Paper II, for $\Gamma>0$ the corresponding minimum angular momentum for these orbits is always $j_\mathrm{min}=j_- \ll 1$.  

We now want to see what happens to \eqref{eq:djdtGR} for finite $\epsGR \ll \epsstr$. From the discussion in \S\ref{sec:including_epsGR} we expect that we may neglect $j^2$ compared to $j_+^2$, $j_0^2$ in this regime. As a result we can write
\begin{align}  
\frac{\md j}{\md t} \approx
\pm \frac{6C}{Lj^{3/2}}
\sqrt{(25\Gamma^2-1)j_+^2 j_0^2\left[j^2-j_-^2-\gamma jj_-\right]\left[j+\sigma j_-\right]},
\label{eq:djdt_weak_to_moderate_GR}
\end{align}
where we defined the following dimensionless numbers:
\begin{align} \gamma &\equiv \frac{\epsGR}{3(1+5\Gamma)j_+^2j_-} =  \frac{2\epsGR}{\epsweak},
\label{eqn:dimensionless_number_1}
\\
\sigma &\equiv
\frac{\epsGR}{3(5\Gamma-1)j_0^2j_-} 
= \frac{2\epsGR}{\epsweak} \times  \frac{5\Gamma+1}{5\Gamma-1} \frac{j_+^2}{j_0^2},
\label{eqn:dimensionless_number_2}
\end{align}
with
\begin{align}
    \epsweak \equiv 6(1+5\Gamma)j_+^2j_- \approx \left(720\Gamma\Sigma\right)^{1/2}\Theta^{1/2}.
    \label{eqn:epsweak}
\end{align}
To get the second equality in \eqref{eqn:epsweak} we used the approximation \eqref{eqn:j_plus_minus_small_Theta}. Both $\epsweak$ and $\gamma$ are manifestly positive in the weak-to-moderate GR regime given $\Gamma>0$. Except in pathological cases, $\sigma$ is also positive for the regimes we are interested in here\footnote{This is true because $(5\Gamma-1)j_0^2$ is positive in the $\epsGR=0$ limit for all the cases we care about, namely any orbit with $\Gamma>1/5$ and librating orbits with $0<\Gamma \leq 1/5$. The inclusion of GR subtracts from $(5\Gamma-1)j_0^2$ by an amount $\epsGR/(3\sqrt{1-e_0^2})$.  For $j_0 \sim 1$ and $e_0^2 \ll 1$, this modification will not make $(5\Gamma-1)j_0^2$ negative as long as \eqref{eqn:epsGR_ll_1_minus_5_Gamma} is satisfied.}. 

To find the minimum $j$ at $\omega=\pm\pi/2$ we require the right hand side of \eqref{eq:djdt_weak_to_moderate_GR} to equal zero, which, as we mentioned earlier, means that the first square bracket inside the square root must vanish. This gives a quadratic equation for $j_\mathrm{min}$, the only meaningful (positive) solution to which is
\begin{align} 
j_{\rm min}&=\frac{\gamma j_-}{2}\left[ 1 + \sqrt{1+4\gamma^{-2}}\right]
\nn
\\
&= \frac{1}{2j_+^2\epsstr} \left[\epsGR + \sqrt{\epsGR^2+\epsweak^2}\right].
\label{eq:gen_sol_1}
\end{align}
Equations \eqref{eq:djdt_weak_to_moderate_GR} and \eqref{eq:gen_sol_1} work as long as $\Theta\ll 1$ and $\epsGR$ is in the weak-to-moderate GR regime, i.e. satisfies \eqref{eqn:weak_to_moderate_GR_condition} and \eqref{eqn:epsGR_ll_1_minus_5_Gamma}. 

Equation \eqref{eq:gen_sol_1} has been used by several authors in the LK limit of $\Gamma=1$ --- see \S\ref{sec:LK_literature}. Importantly, it allows us to write down a solution for the maximum eccentricity reached by initially near-circular binaries in the weak-to-moderate GR regime.  Indeed, let us put $\Theta = \cos^2 i_0$ and assume $\Theta \ll 1$ (i.e. $i_0 \approx 90^\circ$) so that the binary is capable of reaching very high eccentricity. Then $j_+ \approx 1$ and from \eqref{eq:gen_sol_1} we find
\begin{align} 
\label{eqn:jGR_highe_circular}
j_{\rm min} \approx \frac{1}{2}\left[\frac{\epsGR}{3(1+5\Gamma)}+\sqrt{\left[\frac{\epsGR}{3(1+5\Gamma)}\right]^2+\frac{40\Gamma\cos^2i_0}{1+5\Gamma}}\right].
\end{align}
Note that this result is consistent with what 
we found in \S\ref{sec:max_ecc_circular}, where we assumed near-circularity from the outset and made no (explicit) assumptions about $\jmin$ or $\epsGR$ other than \eqref{eqn:piby2_but_no_saddle}.  For instance: (I) we can alternatively derive \eqref{eqn:jGR_highe_circular} by solving equation \eqref{eqn:quartic_circular} in the limit $j\ll 1$; (II) if we take  $i_0 = 90^\circ$ in \eqref{eqn:jGR_highe_circular} then we get exactly the same result as if we expand \eqref{eqn:jmin_circular_i90} for $\epsGR\ll\epsstr$, namely equation \eqref{eq:emax_limit}.
Moreover, in the LK limit $\Gamma=1$ we recover from \eqref{eqn:jGR_highe_circular} a well-known result,
identical to\footnote{
Note that our definition of $\epsGR$ differs from what \citet{Miller2002} call $\theta_\mathrm{PN}$ and what \citet{Liu2015} call $\varepsilon_\mathrm{GR}$. Our $\epsGR$ is defined for any outer orbit in any axisymmetric potential, whereas their parameters are defined only in the Keplerian (LK) limit. In this limit, $\epsGR = 6\theta_\mathrm{PN} = 16\varepsilon_\mathrm{GR}$. } e.g. equation (8) of \citet{Miller2002} and equation (52) of \citet{Liu2015}.
\\
\\
It is now instructive to investigate separately the high-$e$ behaviour in the asymptotic regimes of {\it weak} and {\it moderate} GR precession (still assuming eccentricity is maximised at $\omega = \pm\pi/2$).


\subsubsection{Weak GR, $\epsGR \ll \epsweak$}
\label{sec:weak_GR}

\begin{figure}
    \centering
    \includegraphics[width=0.95\linewidth]{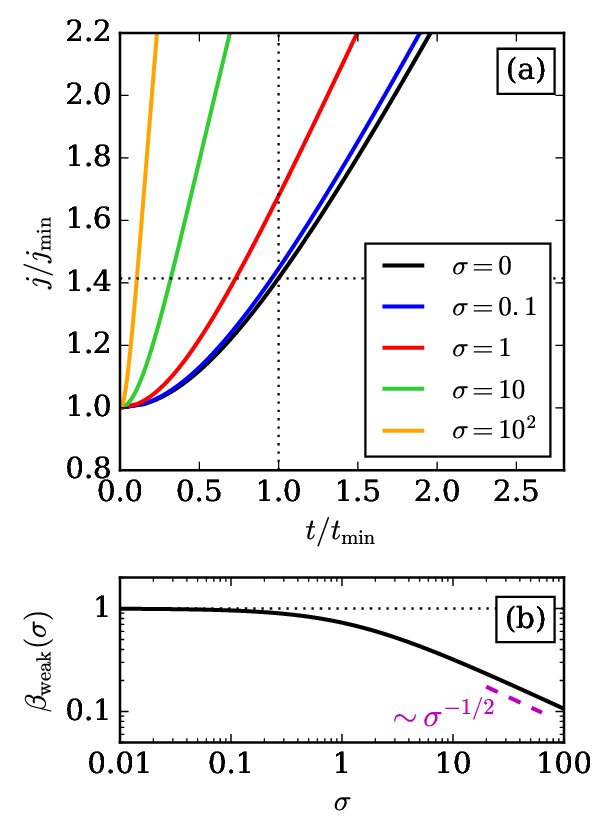}
    \caption{High-$e$ behaviour in the weak GR regime (\S\ref{sec:weak_GR}). Panel (a) shows the solution (\ref{eq:j_implicit_weakGR}) for $j(t)$ near the eccentricity peak for various values of $\sigma$ (equation (\ref{eqn:dimensionless_number_2})). The horizontal dotted line shows $j/j_\mathrm{min}=\sqrt{2}$ and the vertical dotted line shows $t=t_\mathrm{min}$. Panel (b) shows $\beta_\mathrm{weak}(\sigma)$, which is the time (in units of $t_{\rm min}$) over which the binary's $j/\jmin$ changes from $1$ to $\sqrt{2}$, defined by setting $j/\jmin=\sqrt{2}$ in the right hand side of \eqref{eq:j_implicit_weakGR}.  Note that both axes are on a logarithmic scale in this panel. A dashed magenta line shows the scaling $\beta_\mathrm{weak} \propto \sigma^{-1/2}$ for $\sigma \gg 1$.}
    \label{fig:j_of_t_weak}
\end{figure}

In the asymptotic regime of \textit{weak GR}, defined by $\epsGR \ll \epsweak$, the solution \eqref{eq:gen_sol_1} becomes approximately
\begin{align} 
j_{\rm min}&\approx j_-\left(1+\frac{\gamma}{2}\right)=
j_-\left(1+\frac{\epsGR}{\epsweak}\right).
\label{eq:weak_sol_1}
\end{align}
In other words, GR causes only a slight perturbation of $j_{\rm min}$ away from the non-GR value of $j_-$ at the relative level $\epsGR/\epsweak \ll 1$. 

To determine the time dependence of $j(t)$ in the vicinity of $\jmin$, we make use of the weak GR assumption to drop the $\gamma$ term in the first square bracket in \eqref{eq:djdt_weak_to_moderate_GR}.  The result is
\begin{align}  
\frac{\md j}{\md t} \approx
\pm \frac{6C}{Lj^{3/2}}
\sqrt{(25\Gamma^2-1)j_+^2 j_0^2\left[j^2-j_-^2\right]\left[j+\sigma j_-\right]}.
\label{eqn:djdt_weak_GR}
\end{align}
Integration of \eqref{eqn:djdt_weak_GR} gives an implicit solution for $j(t)$ in the form
\begin{align} 
\frac{t}{t_{\rm min}}=\int_1^{j/j_{\rm min}}\frac{\md x\, x^{3/2}}{\sqrt{(x^2-1)(x+\sigma)}},
\label{eq:j_implicit_weakGR}
\end{align}
where $t_\mathrm{min}$ is defined in equation \eqref{eq:t_min}.
In Figure \ref{fig:j_of_t_weak}a we plot the implicit solution for $j/j_\mathrm{min}$ as a function of $t/t_\mathrm{min}$ for various values of $\sigma$.

We can gain insight into $\sigma$ in the weak GR regime by using the fact that in this regime, $\epsGR \ll \jmin \approx j_- \sim \Theta^{1/2}$.  This allows us to simplify the expression \eqref{eqn:def_D} to 
\begin{align}
D \approx 1 + \frac{10\Gamma}{1-5\Gamma}\left(1-\frac{\Theta}{\jmin^2} \right)^{-1}.
\end{align}
Plugging this into \eqref{eqn:j0} and the resulting expression into \eqref{eqn:dimensionless_number_2} gives
\begin{align}
    \sigma \approx \frac{\epsGR}{30\Gamma\jmin}\left(1-\frac{\Theta}{\jmin^2} \right)^{-1} = \frac{\epsGR \chi}{30\Gamma\jmin},
\end{align}
where $\chi \geq 1$ is defined in equation \eqref{eqn:chi}. For typical values of $\chi \sim 1$, since $\epsGR \ll \jmin$ we expect $\sigma \ll 1$.  However, when $\chi$ greatly exceeds unity, $\sigma \gtrsim 1$ or even $\sigma \gg 1$ is also possible\footnote{Note that contrary to what a naive interpretation of \eqref{eqn:dimensionless_number_2} might suggest, the condition for $\sigma \gg 1$ is not that $\Gamma \to 1/5$.}.

In the case $\sigma \ll 1$, the term $\sigma j_-$ in the final square bracket in \eqref{eqn:djdt_weak_GR} can also be dropped compared to $j$.  Then equation \eqref{eqn:djdt_weak_GR} takes the same functional form as its non-GR analogue; integrating, we get a solution $j(t)$ in precisely the form (\ref{eq:t_min}) with\footnote{Note that $j_\pm, j_0$ depend on $\epsGR$ through \eqref{eqn:jpm}-\eqref{eqn:j0} only weakly, at the relative level $\mathcal{O}(\epsGR/\epsweak)$.}  
$j_\mathrm{min} \to j_-$ and $j_1, j_2 \to j_+, j_0$. This is reflected in Figure \ref{fig:j_of_t_weak}a, in which the black line ($\sigma =0$) is exactly the non-GR result from \eqref{eq:t_min}, and as expected $j=\sqrt{2}j_\mathrm{min}$ coincides with $t=t_\mathrm{min}$ in that case. 

However the assumption $\sigma \ll 1$ may not always be valid. Figure \ref{fig:j_of_t_weak}a shows that as we increase $\sigma$ the behaviour of $j(t)$ becomes more sharply peaked around $j_\mathrm{min}$ (when time is measured in units of $t_\mathrm{min}$), although against this trend one must remember that to change $\sigma$ is to change one or more of $\epsGR$, $\Gamma$, $j_0$ and $j_-$, any of which will modify $t_\mathrm{min}$. We are particularly interested in the value of $t_\mathrm{min}^\mathrm{weak}(\sigma)$, which is the time it takes for $j$ to go from $j_{\rm min}$ to $\sqrt{2}\jmin$ in the weak GR regime, to compare with the solution (\ref{eq:t_min}). By setting $j/\jmin = \sqrt{2}$ on the right hand side of \eqref{eq:j_implicit_weakGR} and $t=t_\mathrm{min}^\mathrm{weak}$ on the left, we find
\begin{align}
    t_\mathrm{min}^\mathrm{weak}(\sigma) = \beta_\mathrm{weak}(\sigma) t_\mathrm{min},
    \label{eqn:tmin_weak}
\end{align}
where $\beta_\mathrm{weak}(\sigma) $ is plotted as a function of $\sigma$ in Figure \ref{fig:j_of_t_weak}b. As expected $\beta_\mathrm{weak} \to 1$ for $\sigma \to 0$, i.e. in the limit of negligible GR precession. For finite GR, typical values of  $\beta_\mathrm{weak}$ are $\sim 1$ except for very large $\sigma \gtrsim 10$.  For $\sigma \gg 1$ we see that $\beta_\mathrm{weak}$ falls off like $\sim \sigma^{-1/2}$.

In Figures \ref{fig:Kep_a50}, \ref{fig:Kep_a30}, \ref{fig:Hern_a50} and \ref{fig:Hern_a40} we compare the weak GR solution for $j(t)$, namely equation \eqref{eq:j_implicit_weakGR}, to direct numerical integration of the DA equations of motion \eqref{eom1}, \eqref{eom2}, for binaries in different dynamical regimes.
Full details are given in \S\ref{sec:analytic_validity}; here we only note that the values of the key quantities $\Gamma$, $\epsGR$, $\epsweak$, $\sigma$, etc. are shown at the top of each figure. In every example, panel (a) shows $\log_{10}(1-e)$ behaviour in the vicinity of peak eccentricity, while panel (b) shows the same thing zoomed out over a much longer time interval\footnote{Note however that on the horizontal axis we plot time in units of $t_\mathrm{min}'$ (equation \eqref{eqn:tmin_DAsoln}) rather than $t_\mathrm{min}$, as explained in Appendix \ref{sec:analytic}.
}. 
The weak GR solution for $j(t)$ (equation \eqref{eq:j_implicit_weakGR}) is plotted in panels (a) and (b) with a dashed green line, while the
numerical solution is shown with a solid blue line.
We see that for $\epsGR \ll \epsweak$ (Figures \ref{fig:Kep_a50}, \ref{fig:Kep_a30}) this weak GR solution works very well, but that substantial errors begin to set in when $\epsGR$ approaches $\epsweak$ (Figures \ref{fig:Hern_a50}, \ref{fig:Hern_a40}).  
Finally, in each of these plots we also show with red dashed lines an `analytic' solution, equation \eqref{eqn:j_DA_analytic}, which coincides with \eqref{eq:t_min} provided $j_\mathrm{min}^4/\Theta \ll 1$.  As we have already stated, in the weak GR regime $j(t)$ takes the form \eqref{eq:t_min} provided that $\sigma \ll 1$, so it is unsurprising that in the plot with very small $\sigma$ (Figure \ref{fig:Kep_a50}) this analytic solution (equation \ref{eqn:j_DA_analytic}, denoted with red dashed lines) overlaps with the weak GR solution (equation \ref{eq:j_implicit_weakGR}, shown with green dashed lines).


\subsubsection{Moderate GR, $\epsweak \ll \epsGR \ll \epsstr$} 
\label{sec:moderate_GR}

Perhaps more interesting is the asymptotic regime of {\it moderate} GR, defined as $\epsweak\ll \epsGR \ll \epsstr$.  In this regime one finds from \eqref{eq:gen_sol_1} that
\begin{align} 
j_{\rm min}\approx\frac{\epsGR}{3(1+5\Gamma)j_+^2} = \gamma j_- =  \frac{2\epsGR }{\epsweak}j_- \gg j_-,
\label{eq:gen_sol_1p}
\end{align}
i.e. a significant perturbation of $j_{\rm min}$ away from $j_-$, resulting in a significantly reduced maximum eccentricity $e_{\rm max}$.

\begin{figure}
    \centering
    \includegraphics[width=0.95\linewidth]{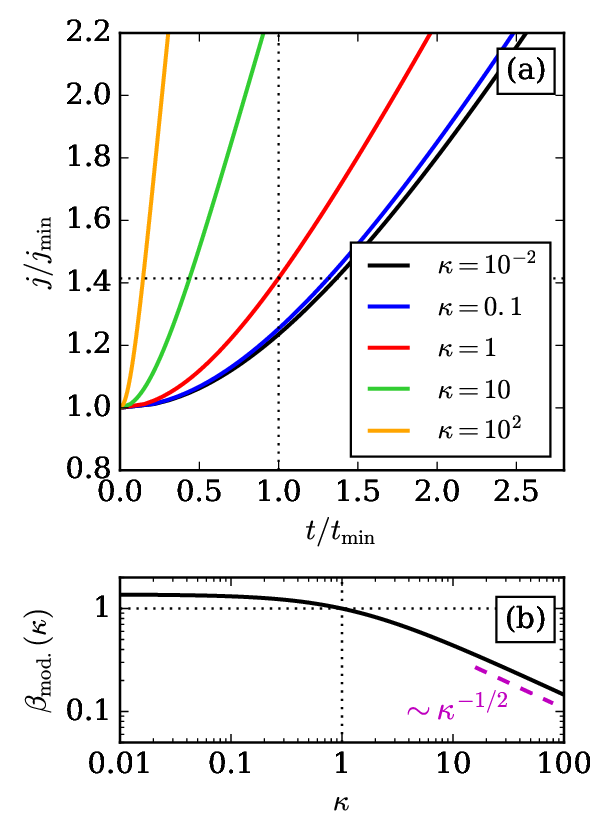}
    \caption{Similar to Figure \ref{fig:j_of_t_weak} except now for the moderate GR regime (\S\ref{sec:moderate_GR}). In panel (a) the solution for $j(t)$ is defined implicitly by equation \eqref{eq:j_implicit_modGR} and we plot it for various values of $\kappa$ (equation (\ref{eqn:moderateGR_ratio})). In panel (b) we show $\beta_\mathrm{mod}(\kappa)$, which is the time (in units of $t_\mathrm{min}$) over which $j$ changes from $j_\mathrm{min}$ to $\sqrt{2}j_\mathrm{min}$ in this regime. Dotted lines show $\kappa=1$ and $\beta_\mathrm{mod} = 1$, while a dashed magenta line shows the scaling $\beta_\mathrm{mod} \propto \kappa^{-1/2}$ for $\kappa \gg 1$.}
    \label{fig:j_of_t_mod}
\end{figure}

To determine the time dependence of $j(t)$ around $\jmin$ we neglect $j_-^2$ compared to $j^2$ in the first square bracket in \eqref{eq:djdt_weak_to_moderate_GR} and find
\begin{align}  
\frac{\md j}{\md t} &\approx
\pm \frac{6C}{Lj}
\sqrt{(25\Gamma^2-1)j_+^2 j_0^2\left[j-\gamma j_-\right]\left[j+\sigma j_-\right]}.
\label{eqn:djdt_moderate_GR}
\end{align}
Integration of \eqref{eqn:djdt_moderate_GR} gives an implicit solution for $j(t)$ in the form
\begin{align} 
\frac{t}{t_{\rm min}}=\int_1^{j/j_{\rm min}}\frac{x \,\md x}{\sqrt{(x-1)(x+\kappa)}},
\label{eq:j_implicit_modGR}
\end{align}
where
\begin{align}
\label{eqn:moderateGR_ratio}
\kappa \equiv \frac{\sigma}{\gamma} = \frac{j_+^2}{j_0^2} \frac{5\Gamma+1}{5\Gamma-1}.
\end{align}
Note that $t_\mathrm{min}$ in (\ref{eq:j_implicit_modGR}) is still defined by equation \eqref{eq:t_min} but taking $j_\mathrm{min}$ equal to its GR-modified value, namely $\gamma j_-$. In Figure \ref{fig:j_of_t_mod}a we plot this implicit solution for various values of $\kappa$.  Note also that $\kappa=1$ (red line) gives precisely the solution $j/j_\mathrm{min}$ in the form \eqref{eq:t_min}, and so unsurprisingly $j/j_\mathrm{min}=\sqrt{2}$ coincides with $t/t_\mathrm{min}=1$ in that case. As we increase $\kappa$ we see that the time spent near the minimum $j$ decreases (when measured in units of $t_\mathrm{min}$, which itself also depends on $\kappa$).

We can get a better feel for the quantity $\kappa$ in the moderate GR regime using the fact that in this regime, $\epsGR \sim \jmin \gg \Theta^{1/2}$. With this we can show from \eqref{eqn:def_Sigma} and \eqref{eqn:def_D} that
\begin{align}
\Sigma \approx \frac{\epsGR}{6\jmin}, \,\,\,\,\mathrm{and} \,\,\,\,
D \approx 1 + \frac{10\Gamma}{1-5\Gamma}\left(1-\frac{\epsGR}{30\Gamma \jmin} \right)^{-1}.
\end{align}
Plugging these results into \eqref{eqn:jpm} and \eqref{eqn:j0} and inserting the resulting expressions into \eqref{eqn:moderateGR_ratio}, we find
\begin{align}
    \kappa \approx \left(\frac{30\Gamma\jmin}{\epsGR}-1 \right)^{-1}.
\end{align}
Since $\jmin \sim \epsGR$ we typically expect the first term in the bracket to be $\gg 1$  resulting in $\kappa \ll 1$ .  However, as we will see in Appendix \ref{sec:analytic_validity}, much larger values of  $\kappa$ are also possible.

Analogous to \S\ref{sec:weak_GR}, by setting $j/\jmin = \sqrt{2}$ on the right hand side of \eqref{eq:j_implicit_modGR} and $t=t_\mathrm{min}^\mathrm{mod}$ on the left, we find that the time for $j$ to increase from $j_\mathrm{min}$ to $\sqrt{2}j_\mathrm{min}$ in the moderate GR regime is
\begin{align}
  t_\mathrm{min}^\mathrm{mod}(\kappa) = \beta_\mathrm{mod}(\kappa) t_\mathrm{min},
    \label{eqn:tmin_mod}
\end{align}
where $\beta_\mathrm{mod}(\kappa) $ is plotted as a function of $\kappa$ in Figure \ref{fig:j_of_t_mod}b. Clearly when $\kappa\sim 1$ (which is true for $\Gamma$ not too close to $1/5$) we have $\beta\sim 1$ and so $t_\mathrm{min}^\mathrm{GR} \sim t_{\rm min}$. 
But for $\kappa \gg 1$ the time spent in the high eccentricity state is somewhat reduced, with a scaling $t_\mathrm{min}^\mathrm{GR} \propto \kappa^{-1/2} t_{\rm min}$. However for this to be a significant effect requires rather extreme values of $\kappa \gg 1$.

Finally, in panels (a) and (b) of Figures \ref{fig:Kep_a15} and \ref{fig:Hern_a35} we compare the moderate GR solution \eqref{eq:j_implicit_modGR}, shown with dashed cyan lines, to direct numerical integration of the DA equations of motion, shown in solid blue. In both cases the moderate GR solution provides an excellent fit to the numerical result despite $\epsGR$ only being slightly larger than $\epsweak$.


\subsection{High-$e$ behaviour in the strong GR limit}
\label{sec:very_strong_GR}


The final asymptotic case to consider is that of strong GR, $\epsGR \gg \epsstr$. This regime is important for understanding the later stages of evolution of shrinking compact object binaries.  Indeed, GW emission  eventually brings any merging binary to a small enough semimajor axis to put it in this regime.

In this limit GR precession is the dominant effect, exceeding the secular effects of the external tide  --- see e.g. panels (e) and (j) of Figures \ref{fig:HStar_Contours_Gamma_pt5} and \ref{fig:HStar_Contours_Gamma_pt1}. Thus we anticipate that at high eccentricity the lowest order solution will be one of constant eccentricity and uniform prograde precession: 
\begin{align}
\label{eqn:high_e_strong_GR}
    j(t)=j(0), \,\,\,\,\,\,\, \omega = \omega(0) + \dot{\omega}_{\mathrm{GR}}(0)t,
\end{align}
where $\dot{\omega}_{\mathrm{GR}}(0) = (C/L)\epsGR/j^2(0)$ --- see equation \eqref{eqn:omega_dot_GR}. High eccentricity can therefore only be achieved if $j(0) \ll 1$ to start with.  This is actually a highly relevant scenario in practice because it is at very high eccentricity that GW emission, and hence the shrinkage of $a$ and the growth of $\epsGR$, is concentrated. Binaries periodically torqued to very high eccentricity by cluster tides eventually become trapped in a highly eccentric orbit as they enter the strong GR regime (Hamilton \& Rafikov, in prep.). Their phase space trajectories are then well described by \eqref{eqn:high_e_strong_GR}.

Interestingly, the minimum angular momentum $j_\mathrm{min}=j(0)$ predicted by the solution (\ref{eqn:high_e_strong_GR}) can still be described by the expression (\ref{eq:gen_sol_1}) in the limit $\epsGR \gg \epsstr$. 
To see this we note that for $\epsGR \gg \epsstr$ equation \eqref{eq:gen_sol_1} gives $j_{\min}\approx \epsGR/(j_+^2\epsstr)$.  
Then we take the expression \eqref{eqn:j_plus_minus_small_Theta} for $j_+^2$, and substitute into it the value of $\Sigma$ we get by taking $\epsGR \gg 1$ in (\ref{eqn:def_Sigma})-(\ref{eqn:def_D}), namely $\Sigma \approx \epsGR (1-e_0^2)^{-1/2}/6$. 
Putting these pieces together we find $j_\mathrm{min}\approx  \sqrt{1-e_0^2}=j(0)$. Thus the solution \eqref{eq:gen_sol_1} interpolates smoothly between the different asymptotic GR regimes.


\subsection{Evolution of $\omega(t)$ and $\Omega(t)$ as $e\to 1$}
\label{sect:om-Om}


So far we focused on understanding the behaviour of $j(t)$. Once this is determined one can understand the evolution of other orbital elements as well. In particular, since the Hamiltonian \eqref{eqn:DA_Hamiltonian} is conserved, we can use equations (\ref{eqn:H1star_omega_j})-(\ref{eqn:HGR_omega_j}) to express $\cos 2\omega$ entirely in terms of $j(t)$ and conserved quantities, leading to an explicit analytical expression for $\omega(t)$.  Finally one can plug this $\cos 2\omega(t)$ and $j(t)$ into the equation of motion \eqref{eqn:dOmegadt_DA} for $\Omega(t)$ and integrate the result.  Together with $J_z(t) = J_z(0) = \mathrm{const}$., this constitutes a complete solution to the DA, test-particle quadrupole problem with GR precession.

Unfortunately this proposed solution for $\omega(t)$ and $\Omega(t)$ is very messy for arbitrary values of $j$. Luckily, in the high eccentricity regime $j\ll 1$, one can make substantial progress by (i) making the additional (and often well-justified) assumption given by equation \eqref{eqn:j4_over_Theta_condition} and (ii) adopting the ansatz\footnote{Strictly speaking this ansatz is valid only for $\sigma \ll 1$, but is often a good approximation in the vicinity of $j_\mathrm{min}$ even for $\sigma \gg 1$ --- see \S\ref{sec:analytic_validity}.} (\ref{eq:t_min}) for $j(t)$. Then, as we show in Appendix \ref{sec:analytic}, one can derive relatively simple explicit analytical solutions for $\omega(t)$ and $\Omega(t)$. These solutions work very well as long as equation (\ref{eq:t_min}) is a good approximation to the $j(t)$ behaviour near the peak eccentricity, as we verify numerically in \S\ref{sec:analytic_validity}. To our knowledge, an explicit high-$e$ solution of this form accounting for GR precession has not been derived before even for the LK problem. It can be used for instance in order to explore the short-timescale (i.e. non-DA) effects near peak eccentricity, which are important for accurate calculation of the LK-driven merger rate \citep{Grishin2018}.


\section{Discussion}
\label{sec:discussion}


In this paper we have studied the impact of 1PN GR precession on secular evolution of binaries perturbed by cluster tides. A single dimensionless number $\Gamma$ effectively encompasses all information about the particular tidal potential and the binary's outer orbit within that potential. Meanwhile the relative strength of GR precession compared to external tides is characterised by the dimensionless number $\epsGR$ (equation \eqref{eq:epsGR}). 

In the main body of the paper we only discussed the systems with $\Gamma > 0$. Although the resulting dynamics are significantly complicated by bifurcations that occur at $\Gamma = \pm 1/5, 0$, for $\Gamma > 1/5$ our qualitative results are intuitive, falling in line with those gleaned from previous LK ($\Gamma=1$) studies that accounted for GR precession. However, for $0<\Gamma\leq 1/5$ we uncovered a completely new pattern of secular evolution, which we characterised in detail. Secular dynamics of binaries with negative $\Gamma$ (possible for binaries on highly inclined outer orbits in strongly non-spherical potentials, see Paper I) and non-zero $\epsGR$ is covered in Appendix \ref{sec:Gamma_negative}. As mentioned in \S\ref{sec:note_on_Gamma}, the $\Gamma\leq 0$ regime splits into two further regimes, namely $-1/5 < \Gamma \leq 0$ and $\Gamma \leq -1/5$. We found that in both of these regimes the resulting phase space structures and maximum eccentricity behaviour are considerably more complex and counter-intuitive than for $\Gamma > 0$.

Furthermore, we have explored the evolution of binary orbital elements in the limit of very high eccentricity (\S\ref{sec:High_ecc_behaviour}). This investigation revealed a number of distinct dynamical regimes that are classified according to the value of $\epsGR$. In \S\ref{sect:corollary} we summarise and systematise these regimes based on their physical characteristics. In \S\ref{sec:LK_literature} we compare our study to the existing LK literature, and in \S\ref{sec:limitations} we discuss its limitations.

\begin{figure}
\centering
\includegraphics[width=0.9\linewidth]{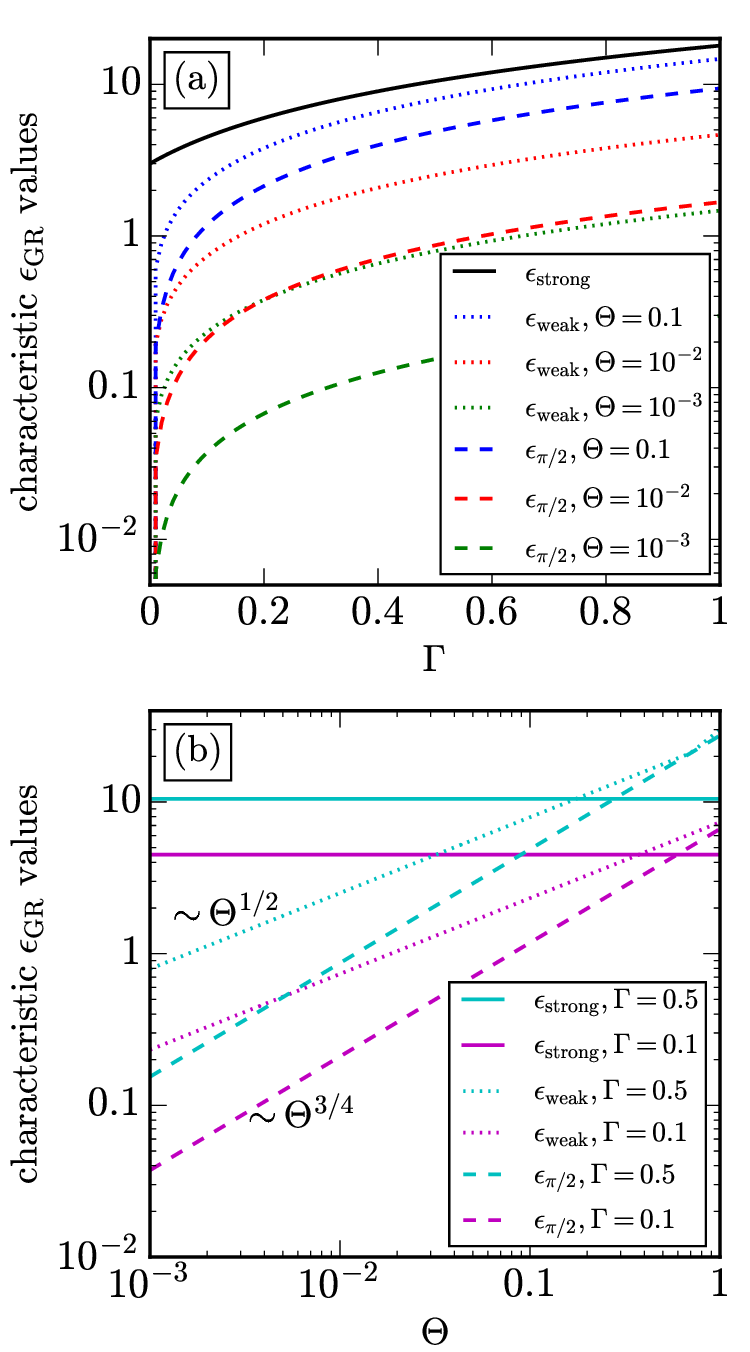}
\caption{(a) Plot showing several characteristic values of the GR strength $\epsGR$ as functions of $\Gamma>0$: $\epsstr$ (black solid line), $\epsweak$ (dotted lines) and $\epspibytwo$ (dashed lines).  The values of $\epsweak$ and $\epspibytwo$ depend on $\Theta$ so we show them for three $\Theta$ values, namely $0.1$ (blue), $10^{-2}$ (red) and $10^{-3}$ (green). (b) Same, but now as a function of $\Theta$, for $\Gamma=0.5$ (cyan) and $\Gamma=0.1$ (magenta).}
\label{fig:logeps}
\end{figure}


\subsection{Summary of $\epsGR$ regimes and their physical interpretation}
\label{sect:corollary}


In this study we introduced three characteristic values of $\epsGR$, namely $\epspibytwo$ (equation \eqref{eqn:epspibytwo}), 
$\epsweak$ (equation \eqref{eqn:epsweak}), and $\epsstr$ (equation \eqref{eqn:epsstrong}). The first two scale with $\Theta$ in such a way that in the high-$e$ limit, when $\Theta\ll 1$, one finds
\begin{align}
\epspibytwo\lesssim \epsweak\lesssim \epsstr.
\label{eq:hierarchy}
\end{align}
This hierarchy is illustrated in Figure \ref{fig:logeps}, in which we show how\footnote{to calculate $\epsweak$ for this figure we set $e_0=\epsGR=0$, so that $j_\pm^2$ is given by \eqref{eqn:jpm} with $\Sigma = (1+5\Gamma+10\Gamma\Theta)/2$} $\epspibytwo$, $\epsweak$, $\epsstr$ depend on both $\Gamma>0$ and $\Theta$.  
We see that decreasing $\Theta$ widens the gap between $\epsstr$ and $\epsweak$. This is as expected because small $\Theta$ tends to promote high-$e$, and since $\dot{\omega}_\mathrm{GR} \propto \epsGR/(1-e^2)$, the higher is $e$, the smaller is the critical value of $\epsGR$ at which GR effects become important. Note however that even for $\Theta = 10^{-3}$ the difference between $\epsstr$ and $\epsweak$ does not exceed $\sim 10^2$. Since $\epsGR \propto a^{-4}$ (equation (\ref{eq:epsGR})), a relatively small change in semimajor axis $a$ can easily shift $\epsGR$ from one asymptotic regime to another.

\begin{table*}
	\centering
	\caption{Key to references for different asymptotic GR regimes and corresponding high eccentricity results. }
	\label{tab:high_e}
	\begin{tabular}{lcccccc} 
		\hline
		 & Very weak GR &  Weak GR &  Moderate GR & Strong GR &\\
		
			  & $\epsGR \ll \epspibytwo$ & $\epspibytwo \ll \epsGR \ll \epsweak$ & $\epsweak \ll \epsGR \ll \epsstr$ & $\epsGR \gg \epsstr$\\
			
				\hline
				
        $\jmin$ [if found at $\omega = \pm\pi/2$] & \eqref{eq:weak_sol_1} & \eqref{eq:weak_sol_1} & \eqref{eq:gen_sol_1p}  & N/A \\
        
                $\jmin$ [if found at $\omega = 0$] & \eqref{eqn:j0_perturbative} if  \eqref{eqn:j0_expand_cond} true, else \S\S\ref{sec:lib_or_circ},\ref{sec:high_e_omega_0} & \S\S\ref{sec:lib_or_circ},\ref{sec:high_e_omega_0} & \S\S\ref{sec:lib_or_circ},\ref{sec:high_e_omega_0} & N/A \\
                
                $j(t)$ near $e\to 1$ & \eqref{eq:t_min} & \eqref{eq:j_implicit_weakGR} & \eqref{eq:j_implicit_modGR} & \eqref{eqn:high_e_strong_GR} \\
                
                $j_{\mathrm{f},\pi/2}$ & \eqref{eqn:jf_piby2_veryweakGR}  & \eqref{eqn:jf_piby2_weaktomodGR} & \eqref{eqn:jf_piby2_weaktomodGR} & none \\
                
                  $j_{\mathrm{f},0}$ & \eqref{eqn:jf0}  & \eqref{eqn:jf0} & \eqref{eqn:jf0} & none \\
						
		\hline
	\end{tabular}
\end{table*}

These characteristic values of $\epsGR$ allow us naturally to delineate four important regimes of secular dynamics: 
\begin{itemize}
    \item very weak GR: $\epsGR \lesssim \epspibytwo$,
    \item weak GR: $\epspibytwo \lesssim \epsGR \lesssim \epsweak$,
    \item moderate GR: $\epsweak \lesssim \epsGR \lesssim \epsstr$,
    \item strong GR: $\epsGR \gtrsim \epsstr$.
\end{itemize}
Based on our findings in \S\S\ref{sec:phase}-\ref{sec:High_ecc_behaviour} we now provide a description of the basic features of each regime. 
\\
{\bf Very weak GR}~~~In this limit (below the dashed curves in Figure \ref{fig:logeps}) GR precession has essentially no effect on the dynamics for $\Gamma>1/5$. More precisely, GR is too weak to affect either the locations of the fixed points at $\omega=\pm \pi/2$, which are given by equation (\ref{eqn:jf_piby2_veryweakGR}), or the maximum eccentricity reached by the binary at the same $\omega$ in the course of its tide-driven secular evolution, given by equation (\ref{eq:weak_sol_1}). Thus all $\Gamma>1/5$ results of Paper II, which were derived for $\epsGR=0$, are valid.  However, for $0<\Gamma\leq 1/5$ an important modification arises if $6(1-5\Gamma)\Theta^{3/2} < \epsGR \lesssim \epspibytwo$, which is that saddle points appear at $\omega=0,\pi$. These saddles do not exist for $\epsGR=0$ (see equation (\ref{eqn:fp_omega_0_inequality}) and \S\ref{sec:fixed_points_0}), but they do change the maximum eccentricity reached by the binary --- see Appendix \ref{sec:high_e_omega_0}.
\\
{\bf Weak GR}~~~In this regime (between the dashed and dotted curves in Figure \ref{fig:logeps}) GR precession starts to modify the $j$ locations of the fixed points at $\omega=\pm \pi/2$, which are now given by equation (\ref{eqn:jf_piby2_weaktomodGR}). At the same time, GR precession does not appreciably change $j_{\rm min}$ (or equivalently $e_{\rm max}$), which stays close to its $\epsGR=0$ value $j_-$ (see equation (\ref{eq:weak_sol_1})). If $\sigma \gg 1$ then GR also modifies the time spent in the high eccentricity state (equation \eqref{eqn:tmin_weak}). 
\\
{\bf Moderate GR}~~~In this regime (between the dotted and solid curves in Figure \ref{fig:logeps}) GR precession modifies not only the locations of the fixed points but also the values of $j_{\rm min}$ (and hence of $e_{\rm max}$), now given in equation (\ref{eq:gen_sol_1p}).  GR also modifies the time spent in the high-$e$ state (equation \eqref{eqn:tmin_mod}). In other words, in this regime GR precession presents an efficient barrier suppressing the maximum eccentricity reached by the binary in the course of its secular evolution.
\\ 
{\bf Strong GR}~~~In this limit (above the solid curves in Figure \ref{fig:logeps}) GR precession dominates the binary dynamics at all times; the quantities $j_\pm$ are significantly affected by GR precession (equation \eqref{eqn:Sigma_small_Theta}) and all fixed points in the phase portrait disappear (equations \eqref{eqn:FPs_pi_by_2_criterion_small_Theta}, \eqref{eqn:fp_omega_0_inequality}). Cluster tides drive only very small eccentricity oscillations on top of uniform GR precession, so that $e$ is roughly constant --- see equation (\ref{eqn:high_e_strong_GR}).

In Table \ref{tab:high_e} we summarise the main features of the asymptotic $\epsGR$ regimes that we have found in this and previous sections.
\\
\\

We may use this regime separation to shed light on the physical meaning of the characteristic $\epsGR$ values introduced in this work. To do so, we first note that the GR precession rate (\ref{eqn:omega_dot_GR}) can be written as $\dot{\omega}_\mathrm{GR}(j)=\dot{\omega}_\mathrm{GR}|_{e=0}j^{-2}\sim \epsGR t_{\rm sec}^{-1}j^{-2}$ (see the definition (\ref{eq:epsGR})). Next, consider some arbitrary cluster tide-driven process occurring on a characteristic timescale $t_{\rm ch}$. GR precession will affect this process if $\epsGR$ is such that
\begin{align}  
\dot{\omega}_\mathrm{GR}(j)t_{\rm ch}\sim 1, \,\,\,\,\,\,\,\,\,\, \mathrm{i.e.} \,\,\,\,\,\,\,\,\,\,\epsGR \frac{t_{\rm ch}}{t_{\rm sec}}j^{-2}\sim 1.
\label{eq:cond}
\end{align}
If $\epsGR$ satisfies \eqref{eq:cond}, or exceeds that value, then GR breaks the coherence of the tidal torque over the timescale $\sim t_\mathrm{ch}$, and so GR precession substantially interferes with the secular evolution. We now demonstrate how this simple physical argument leads one to the critical values $\epsstr$, $\epsweak$ and $\epspibytwo$.

First, in the strong GR regime we expect GR precession to dominate binary evolution at all times, even for near-circular orbits. Setting $t_{\rm ch}\sim t_{\rm sec}$ and $j\sim 1$ we obtain $\epsGR\sim 1$, which is consistent with the definition (\ref{eqn:epsstrong}) of $\epsstr$ up to a numerical coefficient.

Second, in the moderate GR regime, we anticipate that GR precession will present an effective barrier that stops the decrease of $j$ if $t_{\rm ch}$ is the characteristic timescale of secular evolution near the eccentricity peak. In \S\ref{sec:High_ecc_behaviour} we find quite generally this timescale to be $t_{\rm ch}\sim t_{\rm min}\sim j_{\rm min}t_{\rm sec}$ --- see e.g. equations (\ref{eq:t_min}) and (\ref{eqn:tmin_mod}). Plugging this into the condition (\ref{eq:cond}) and evaluating $\dot{\omega}_\mathrm{GR}$ at $j_{\rm min}$ we immediately find that $j_{\rm min}\sim \epsGR$, in agreement with equation (\ref{eq:gen_sol_1p}). When the GR barrier first emerges at the transition between weak and moderate regimes, $j_{\rm min}$ is still well approximated by the $\epsGR=0$ solution $j_-\sim \Theta^{1/2}$ (see equation \eqref{eqn:j_plus_minus_small_Theta}). As a result, the $\epsGR$ value corresponding to this transition is $\sim \Theta^{1/2}$, in agreement with the definition (\ref{eqn:epsweak}) of $\epsweak$.

Third, we expect fixed points in the phase portrait at $\omega=\pm\pi/2$ to be substantially displaced by GR precession when $\dot{\omega}_\mathrm{GR}(j_\mathrm{f})$ becomes comparable to the characteristic secular frequency $\dot\omega$ of libration around a fixed point. 
Since we are interested in the displacement of $j_\mathrm{f}$ by an amount $\sim j_\mathrm{f}$, we take this $\dot\omega$ from the $H^\star=$~const. contour centred on the $\omega=\pm\pi/2$ fixed point and with vertical extent $\sim j_\mathrm{f}$.
Plugging $j\sim j_\mathrm{f}$ into the equation (\ref{eom1}) and using the expression (\ref{eqn:jf}) for $j_\mathrm{f}$ we find $\dot\omega\sim \Theta^{1/4}t_{\rm sec}^{-1}$, so that in this case $t_{\rm ch}\sim \Theta^{-1/4}t_{\rm sec}$. 
Substituting this into the condition (\ref{eq:cond}) and again setting $j\sim j_\mathrm{f}\sim \Theta^{1/4}$ we find $\epsGR\sim \Theta^{3/4}$ for the transition between the weak and very weak GR regimes. This agrees with the definition of $\epspibytwo$ in equation (\ref{eqn:epspibytwo}).

Note that while these considerations allow us to understand the scalings of characteristic $\epsGR$ values with $\Theta$, one still needs the full analysis presented in \S\S\ref{sec:phase}-\ref{sec:High_ecc_behaviour} to obtain the numerical coefficients, which are actually quite important. Indeed, equations \eqref{eqn:epspibytwo}, \eqref{eqn:epsweak}, and \eqref{eqn:epsstrong} feature constant numerical factors which can substantially exceed unity, especially for the LK case of $\Gamma=1$.


\subsection{Relation to LK studies}
\label{sec:LK_literature}


Many authors who studied the LK mechanism and its applications have included 1PN GR precession in their calculations. The maximum eccentricity of an initially near-circular binary undergoing LK oscillations (i.e. the $\Gamma=1$ limit of \S\ref{sec:max_ecc_circular}) was derived by \citet{Miller2002,Blaes2002,Wen2003,Fabrycky2007,Liu2015}. Of these, \citet{Fabrycky2007} and \citet{Liu2015} also produced plots very similar to Figure \ref{fig:emax_initially_circular} that show how increasing $\epsGR$ decreases the maximum eccentricity achieved by initially near-circular binaries. Various authors have derived equations identical to, or very similar to, the quartic \eqref{eqn:depressed_quartic} and the weak-to-moderate maximum eccentricity solution \eqref{eq:gen_sol_1} in the LK limit --- see for instance equation (A7) of \citet{Blaes2002}, equation (8) of \citet{Wen2003}, equation (A6) of \citet{Veras2010}, and equations (64)-(65) of \citet{Grishin2018}.  Of course, because these studies only work with $\Gamma = 1$, the rather non-intuitive behaviour for $\Gamma <  1/5$ revealed in \S\ref{sec:max_ecc_circular} and Appendix \ref{sec:max_ecc_circular_negative_Gamma} has not been unveiled before. Moreover, to our knowledge no previous study has presented a clear classification of the different $\epsGR$ regimes (which we do in \S\ref{sect:corollary}), even in LK theory.

The quantitative results in the aforementioned papers have been employed in many practical calculations. Typically one simply adds the term \eqref{eqn:omega_dot_GR} to the singly- or doubly-averaged equations of motion along with any other short range forces or higher PN effects. In population synthesis calculations of compact object mergers  \citep{Antonini2012,Antonini2014,Silsbee2017,Liu2018} one often puts a sensible lower limit on the semimajor axis distribution below which GR is so strong that sufficient eccentricity excitation is impossible.
As explained in \S2.1 of \citet{Rodriguez2018} there are at least two ways to decide when GR dominates. One method is to take $\jmin$ corresponding to the pericentre distance that needs to be reached according to the problem at hand, and then equate $\dot{\omega}_\mathrm{GR}(\jmin)$ with the precession rate due to the tidal perturbations (see their equation (29)), which corresponds to equation (54) in Paper II. As discussed in \S\ref{sect:corollary}, this method would set a rough upper limit of $\epsGR \lesssim \jmin$; see equation (\ref{eq:gen_sol_1p}) for a more accurate expression. A second method is to demand that $\omega=\pm\pi/2$ fixed points do exist in the phase portrait \citep{Fabrycky2007} allowing for substantial eccentricity excitation to occur starting from the near-circular orbits, which is equivalent to $\epsGR \lesssim \epsstr$. However, this is not a very stringent requirement, and does not guarantee that the majority of systems with such $\epsGR$ would reach the required $\jmin$ --- many of them will be stopped by the GR barrier at eccentricities much lower than needed. The former method of setting an upper limit on $\epsGR$ is typically more stringent and allows more efficient selection of systems for Monte Carlo population synthesis (see Figure \ref{fig:logeps}). 

With regard to phase space structure, the only study we know of that resembles our \S\ref{sec:phase} is that by \citet{Iwasa2016}.  They considered a hierarchical triple consisting of a star on an orbit around a supermassive black hole (SMBH), with another massive black hole also orbiting the SMBH on a much larger, circular orbit and acting as the perturber of the star-SMBH `binary'.  Their \S C provides a brief explanation of the phase space behaviour as a parameter they call $\gamma$, which is equivalent to our $\epsGR/3$, is varied.  Since the LK problem has $\Gamma =1 > 1/5$, their Figure 2 is qualitatively the same as our Figure \ref{fig:HStar_Contours_Gamma_pt5}.  




\subsection{Approximations and limitations}
\label{sec:limitations}


To derive the Hamiltonian \eqref{H1Star} we truncated the perturbing tidal potential at the quadrupole level.  This is justified if the semimajor axis of the binary is much smaller than the typical outer orbital radius. Next order corrections to the perturbing potential --- so called octupole terms --- are routinely accounted for in LK studies \citep{Naoz2013_Kocsis,Will2017}.  In Appendix E of Paper I we provide the octupole correction to \eqref{H1Star} for arbitrary $\Gamma$. When octupole-order effects are important, the maximum eccentricity can actually be \textit{increased} by GR precession  \citep{Ford2000,Naoz2013_Kocsis,Antonini2014}. However for the applications we have in mind, e.g. a compact object binary of $a\sim 10\mathrm{AU}$ orbiting a stellar cluster at $\sim 1\mathrm{pc}$, octupole corrections are negligible. 

We also employed the test particle approximation, which is valid if the outer orbit contains much more angular momentum than the inner orbit.  One can relax the test particle approximation: in particular, this is often necessary for weakly-hierarchical triples. \citet{Anderson2017} made a detailed study of the `inclination window' that allows fixed points to exist in the (quadrupole) LK phase space for different $\epsGR$, as one varies the ratio of inner to outer orbital angular momenta.  We recover their results in the test particle limit valid for our applications. We also assumed the validity of the DA approximation, the smallness of short-timescale fluctuations (`singly-averaged effects'), etc., all of which are liable to break down at very high eccentricity.  For a full discussion of these issues see Papers I and II. 

Finally, several of the results derived at very high eccentricity (\S\ref{sec:High_ecc_behaviour}) are rather delicate when $\Gamma$ is close to $\pm 1/5$ or when the binary's phase space trajectory is close to a separatrix. These are not major caveats; for instance, in a given stellar cluster potential only a small fraction of binaries will have $\Gamma$ values close enough to $1/5$ to be affected (Paper I). 



\section{Summary}
\label{sec:summary}


In this paper we completed our investigation of doubly-averaged (test-particle quadrupole) cluster tide-driven binary dynamics in the presence of 1PN general relativistic pericentre precession.  Throughout, we parameterised the strength of GR precession relative to tides using the dimensionless number $\epsGR$ (equation (\ref{eq:epsGR})). We can summarise our results as follows:

\begin{itemize}

\item We investigated the effect of non-zero $\epsGR$ on phase space morphology.  For values of $\epsGR$ much less than a critical value $\epsstr$, bifurcations in the dynamics happen at $\Gamma = \pm 1/5, 0$, so that we must consider four $\Gamma$ regimes separately. We found that for $\Gamma \leq 1/5$ a non-zero $\epsGR$ can lead to entirely new phase space morphologies, including (previously undiscovered) fixed points located at $\omega = 0, \pm \pi$.

\item We presented general recipes for computing the locations of fixed points in the phase portrait, for determining whether a given phase space trajectory librates or circulates, and for finding its maximum eccentricity, for arbitrary $\epsGR$.

\item We considered how the maximum eccentricity reached by an initially circular binary is affected by GR precession.  For $\Gamma > 1/5$ the intuitive picture holds that a larger $\epsGR$ leads to a lower maximum eccentricity, but this is not always the case for $\Gamma \leq 1/5$.

\item We delineated four distinct regimes of secular evolution with GR precession depending on the value of $\epsGR$ --- `strong GR', `moderate GR', `weak GR', and 'very weak GR' --- and provided physical justification for transitions between them.  

\item We also studied secular evolution with GR precession in the limit of very high eccentricity. We determined the GR-induced modifications to the minimum angular momentum $\jmin$ achieved by the binary and the time dependence of $j(t)$ near the eccentricity peak, which can be rather non-trivial.  

\item We also provided an approximate analytic description for the evolution of other orbital elements --- pericentre and nodal angles --- near the eccentricity peak, accounting for the GR precession.

\end{itemize}

In upcoming work we will apply the results of this paper to understand the
long-term evolution of compact object binaries due to GW emission, leading to their mergers and the production of LIGO/Virgo GW sources.  Furthermore, these results will inform future studies on the effect of short-timescale fluctuations (`singly-averaged effects') on binaries undergoing cluster tide-driven secular evolution, as well as the population synthesis calculations of merger rates.

\section*{Acknowledgements}

We thank the anonymous referee for several insightful comments on the manuscript. CH is funded by a Science and Technology Facilities Council (STFC) studentship. R.R.R. acknowledges financial support through the STFC grant ST/T00049X/1, NASA grant 15-XRP15-2-0139, and John N. Bahcall Fellowship.

\section*{Data availability}
No new data were generated or analysed in support of this research.



\bibliographystyle{mnras}
\bibliography{Bibliography}


\appendix


\section{Mathematical details of phase space behaviour for $\Gamma > 0$}
\label{sec:mathematical_details_Gamma_positive}


In this Appendix we provide some mathematical details for the results quoted in \S\ref{sec:phase}.


\subsection{Fixed points at \texorpdfstring{$\omega = \pm \pi/2$}{ompmpiby2}}
   \label{sec:fixed_points_omega_piby2}


Putting $\omega = \pm\pi/2$ and $\md \omega/\md t = 0$ into equation (\ref{eom1}) gives us the following quartic equation for $j$ values of the fixed points, which we will call $j_\mathrm{f,\pi/2}$:
\begin{align}
    j_\mathrm{f,\pi/2}\left(j_\mathrm{f,\pi/2}^3 - \frac{\epsGR}{6(1+5\Gamma)} \right)  = \frac{10\Gamma\Theta}{1+5\Gamma}.
    \label{eqn:FPs_at_omega_piby2}
\end{align}
For any $\Gamma > 0$ the right hand side of \eqref{eqn:FPs_at_omega_piby2} is obviously positive. Thus, for there to be a positive (not necessarily physical) solution to equation \eqref{eqn:depressed_quartic} a necessary but insufficient requirement is that
\begin{align}
    j_\mathrm{f,\pi/2} \geq \left[ \frac{\epsGR}{6(1+5\Gamma)}\right]^{1/3},
    \label{eqn:piby_necessary_criterion}
\end{align}
which, since $j<1$, in turn means that $\epsGR$ must necessarily be $\leq 6(1+5\Gamma)$. By differentiating \eqref{eqn:FPs_at_omega_piby2} it is then easy to show that
\begin{align}
\label{eqn:djfdTheta}
    &\left(\frac{\partial j_\mathrm{f,\pi/2}}{\partial \Theta}\right)_{\epsGR}  = \frac{10\Gamma}{1+5\Gamma}\left(4j_\mathrm{f,\pi/2}^3 - \frac{\epsGR}{6(1+5\Gamma)} \right)^{-1}> 0, 
\\
    \label{eqn:djfdepsGR}
    &\left(\frac{\partial j_\mathrm{f,\pi/2}}{\partial \epsGR}\right)_{\Theta} = \frac{j_\mathrm{f,\pi/2}}{6(1+5\Gamma)}\left(4j_\mathrm{f,\pi/2}^3 - \frac{\epsGR}{6(1+5\Gamma)} \right)^{-1} > 0.
\end{align}
In other words, for $\Gamma>0$ the fixed points at $(\omega,j) = (\pm\pi/2,j_{\mathrm{f},\pi/2})$ always get pushed to lower eccentricity when we increase $\epsGR$ or $\Theta$ (see Figures \ref{fig:HStar_Contours_Gamma_pt5}, \ref{fig:HStar_Contours_Gamma_pt1} \& \ref{fig:HStar_Theta_pt001_logplot}).

The criteria for these $\omega = \pm\pi/2$ fixed points to exist can be found by demanding that the condition \eqref{eq:AMconstr} is obeyed, i.e. that $\sqrt{\Theta} < j_\mathrm{f,\pi/2} < 1$. Let us begin by fixing $\Gamma$ and $\Theta$; then, owing to the monotonic behaviour of $j_\mathrm{f,\pi/2}(\epsGR)$ (equation \eqref{eqn:djfdepsGR}), we simply look for the $\epsGR$ values that correspond to $j_{\mathrm{f},\pi/2}=\sqrt{\Theta}$ and $j_{\mathrm{f},\pi/2}=1$. Doing so, we arrive straightforwardly at the condition \eqref{eqn:epsGR_constraint_fp_piby2} on $\epsGR$.

Next we wish to instead fix $\Gamma$ and $\epsGR$ and look for the resulting condition on $\Theta$ that allows the fixed points to exist.  To begin with, we look for the critical $\Theta$ values for which $j_\mathrm{f,\pi/2}=1$ and $j_\mathrm{f,\pi/2}=\sqrt{\Theta}$. The former is $\Theta_1$, the expression for which is given in equation \eqref{eqn:Theta_1}, and the latter is $\Theta_2$ which is determined implicitly through the equation
\begin{align}
\Theta_2^{1/2} \left(\Theta_2 - 
    \frac{10\Gamma}{1+5\Gamma} \right) = \frac{\epsGR}{6(1+5\Gamma)}.
    \label{eqn:Theta_crit}
\end{align}
For $\Gamma > 0$ this equation can have meaningful ($0 \leq \Theta_2  \leq 1$) solutions only if $10\Gamma/(1+5\Gamma) \leq 1$, i.e. if $\Gamma \leq 1/5$. For $\Gamma>1/5$ the value of $\Theta_2$ has no physical significance. Next, to determine the proper constraint on $\Theta$ we begin by setting $\Theta=0$ in (\ref{eqn:FPs_at_omega_piby2}) --- we see that the fixed point exists and has value $j_\mathrm{f,\pi/2} = (\epsGR/[6(1+5\Gamma)])^{1/3}$, so that $j_\mathrm{f,\pi/2} >\sqrt{\Theta}$ at $\Theta=0$. 
As we increase $\Theta$, there are two possibilities. The first is that $j_\mathrm{f,\pi/2}$ increases steeply enough that it reaches unity (at $\Theta=\Theta_1$) before it intersects $\sqrt{\Theta}$. In this case the constraint on $\Theta$ for fixed points to exist is $\Theta < \Theta_1$. The second scenario is that $j_\mathrm{f,\pi/2}$ intersects $\sqrt{\Theta}$ (at $\Theta_2$) before it reaches unity. Then for the inequality $\sqrt{\Theta}< j_\mathrm{f,\pi/2}$ to be satisfied one needs $\Theta < \Theta_2$. Of course, since $\Theta_2$ is only physically meaningful for $\Gamma \leq 1/5$, the second scenario can only occur in that $\Gamma$ regime. This reasoning leads us to the constraint \eqref{eqn:Theta_constraint_fp_piby2} for fixed points to exist at $\omega = \pm\pi/2$. In the limit $\epsGR \to 0$ this constraint reduces to the non-GR constraint \eqref{eq:Theta_constr}.

In summary, for $\Gamma>0$, fixed points at $\omega = \pm \pi/2$ exist if the constraints on both $\Theta$ and $\epsGR$ are satisfied simultaneously; thus they exist in the sub-volume of $(\Gamma,\Theta,\epsGR)$ space bounded by the inequalities \eqref{eqn:epsGR_constraint_fp_piby2}, \eqref{eqn:Theta_constraint_fp_piby2}. We can use this information to understand Figure \ref{fig:Theta_max} in more detail. Recall that in this figure we plotted $\Theta_\mathrm{max}(\Gamma)$, namely the maximum value of $\Theta$ for which fixed points exist at $\omega = \pm\pi/2$ for a given $\epsGR$ (equation \eqref{eqn:Theta_constraint_fp_piby2}). We now seek to understand separately the behaviour for $\Gamma > 1/5$ and $0<\Gamma \leq 1/5$.

For $\Gamma > 1/5$, the lines in Figure \ref{fig:Theta_max} correspond to $\Theta_\mathrm{max}=\Theta_1$ (equation \eqref{eqn:Theta_1}). Then $\partial \Theta_1/\partial \Gamma = \Gamma^{-2}(\epsGR-6)/60$, so that $\Theta_1$ increases (decreases) monotonically with $\Gamma$ for $\epsGR>6$ ($\epsGR<6$).  For the special value $\epsGR=6$ we have $\Theta_1 = 1/2 = $ const., hence the straight horizontal brown line in Figure \ref{fig:Theta_max}.

On the other hand, for $0<\Gamma \leq 1/5$ we have $\Theta_\mathrm{max} = \mathrm{min}[\Theta_1,\Theta_2]$.  By equating $\Theta_1 = \Theta_2$ in equations \eqref{eqn:Theta_1}, \eqref{eqn:Theta_crit} it is straightforward to show that $\Theta_2$ becomes smaller than $\Theta_1$ when $\Gamma$ is reduced below a critical value  $\Gamma_\mathrm{crit}=(6-\epsGR)/30$, and that this happens at $\Theta_1=\Theta_2=1$. This is reflected in Figure \ref{fig:Theta_max} --- as we decrease $\Gamma$ starting from $1/5$, the red ($\epsGR=0$), yellow ($\epsGR=3$) and green ($\epsGR=5$) lines transition from solid ($\Theta_1$) to dotted ($\Theta_2$) at the points $(1/5, 1)$, $(1/15, 1)$ and $(1/30, 1)$ respectively.  For $\epsGR>6$ (blue, pink and black lines in Figure \ref{fig:Theta_max}) we have $\Gamma_\mathrm{crit} < 0$, so this transition never occurs for positive $\Gamma$. Finally, for the special value $\epsGR=6$ we have from equation \eqref{eqn:Theta_crit} that $\Theta_2 = (1-5\Gamma)^{-1} > 1$, which is obviously greater than $\Theta_1=1/2$, so $\Theta_\mathrm{max}=\Theta_1$.  Hence the brown horizontal line in Figure \ref{fig:Theta_max} extends all the way to $\Gamma \to 0$.

\subsection{Fixed points at $\omega = 0$}
\label{sec:fixed_points_omega_0}

For $\Gamma \leq 1/5$ we found (e.g. Figure \ref{fig:HStar_Contours_Gamma_pt1}) that hitherto undiscovered fixed points could arise at $\omega = 0$.
To find the eccentricity of these fixed points we plug $\omega = 0$ into $\md \omega/\md t =0$ using equation \eqref{eom1}.  The result is a cubic equation for $j$ with no quadratic or linear terms. The solution is $j=j_\mathrm{f,0}$ with $j_\mathrm{f,0}$ given in equation \eqref{eqn:jf0}, which is physically meaningful only for $0 < \Gamma \leq 1/5$ (i.e. fixed points at $\omega=0$ do not exist for $\Gamma > 1/5$).
The determinant of the Hessian matrix of $H^*(\omega,j)$ evaluated at the point $(0,j_\mathrm{f,0})$ is equal to
\begin{align}
    -180\epsGR\Gamma \times \frac{(j_\mathrm{f,0}^2-\Theta)(1-j_\mathrm{f,0}^2)}{j_\mathrm{f,0}^5}.
    \label{eqn:hessian}
\end{align}
Clearly for $0 < \Gamma \leq 1/5$ the determinant \eqref{eqn:hessian} is negative whenever the fixed point $j_\mathrm{f,0}$ exists, so $(\omega,e) = (0,e_\mathrm{f,0})$ is necessarily a saddle point in the phase portrait, consistent with Figures \ref{fig:HStar_Contours_Gamma_pt1}b,c,h.

\subsection{Does a given orbit librate or circulate?}
\label{sec:lib_or_circ}

Here we show how to determine whether a phase space trajectory is librating or circulating, given $\Gamma, \Theta, \epsGR$ and the initial phase space coordinates $(\omega_0, e_0)$. For $\Gamma>0$, librating orbits cannot cross $\omega = 0$ and so any trajectory that passes through $\omega=0$ must be circulating\footnote{This general statement does not hold for $\Gamma \leq 0$ --- see Appendix \ref{sec:Gamma_negative}.}. Therefore we can figure out whether an orbit librates or circulates by determining whether it crosses $\omega=0$.
Plugging $\omega =0$ into $H^*(\omega,j)$ gives us a depressed cubic polynomial:
\begin{align}
    j^3 -j_0^2 j + q = 0,
    \label{eqn:depressed_cubic}
\end{align}
where
\begin{align}
    q &\equiv \frac{\epsGR}{3(1-5\Gamma)},
    \label{eqn:q}
\end{align}
and we used the definition \eqref{eqn:j0}, \eqref{eqn:def_D} of $j_0^2$, which need not be positive.
We call the real roots of this polynomial $j(\omega = 0)$.  
In the limit $\epsGR=0$ we have $q=0$ and so we find $j(\omega=0)=j_0$, recovering the expression for $j(\omega=0)$ from equation (14) of Paper II.
  For $\epsGR\neq 0$, the nature of the roots of  \eqref{eqn:depressed_cubic} depends on the sign of the discriminant 
\begin{align}
    \Delta \equiv 4j_0^6 -27q^2.
\end{align} 
We can evaluate $\Delta$ given $(\omega_0,e_0,\Theta,\Gamma,\epsGR)$.
There are then a few different cases to consider:
\begin{itemize}
    \item 
If $\Delta < 0$, equation \eqref{eqn:depressed_cubic} has one real root, which may or may not be physical. If $j_0^2 > 0$ then this root can be written as
\begin{align}
  j(\omega=0)=&  -2\frac{\vert q \vert}{q}\sqrt{\frac{j_0^2}{3}} \nn \\ & \times \cosh\left( \frac{1}{3}\mathrm{arccosh}\left[\frac{3 \vert q\vert}{2j_0^2}\sqrt{\frac{3 }{j_0^2}}\right]\right),
\end{align}
whereas for $j_0^2 < 0$ it is given by
\begin{align}
  j(\omega=0)=  -2\sqrt{\frac{-j_0^2}{3}}\sinh\left( \frac{1}{3}\mathrm{arcsinh}\left[\frac{-3q}{2j_0^2}\sqrt{\frac{-3}{j_0^2}}\right]\right).
\end{align}
Once $j(\omega=0)$ has been determined, the orbit circulates if $\sqrt{\Theta} < j(\omega=0) <1$, and librates otherwise.
\\
\item
If $\Delta>0$ (which necessarily requires $j_0^2 > 0$) there are three distinct real roots, and they can be expressed as
\begin{align}
    j(\omega=0) = 2\sqrt{\frac{j_0^2}{3}}\cos \left( \frac{1}{3} \cos^{-1} \left[\frac{-3q}{2j_0^2} \sqrt{\frac{3}{j_0^2}}\right] - \frac{2\pi k}{3} \right),
    \label{eqn:cosine_solution}
\end{align}
for $k=0,1, 2$. From the theory of polynomial equations we also know that that the product of the three real roots of \eqref{eqn:depressed_cubic} is $-q = \epsGR/[3(5\Gamma-1)]$ and their sum is $0$. 

For $\Gamma>1/5$ this implies that two roots (namely $k=1,2$) must be negative and one ($k=0$) positive. Thus the orbit circulates if the $k=0$ solution lies in $(\sqrt{\Theta},1)$, and librates otherwise.

For $0 < \Gamma \leq1/5$, one root ($k=2$) must be negative and the other two ($k=0,1$) positive. If either or both of the two positive roots lies in $(\sqrt{\Theta},1)$ then the orbit circulates. If neither of them do then it librates. The case of both positive roots lying in $(\sqrt{\Theta},1)$ corresponds to two coexisting families of circulating orbits that share values of $H^*$, one above $e_\mathrm{f,0}$ and one below, as in Figure \ref{fig:HStar_Contours_Gamma_pt1}b,c,h. 
To determine the family of circulating orbits to which the trajectory belongs we compare its initial eccentricity $e_0$ with that of the saddle point $e_\mathrm{f,0}$. If $e_0 > e_\mathrm{f,0}$ then the orbit circulates in the family `above' the saddle point, and vice versa. 

\end{itemize}


\section{High-eccentricity behaviour for orbits whose eccentricity maxima are found at \texorpdfstring{$\omega = 0$}{om0}}
\label{sec:high_e_omega_0}


When GR is switched off, the only binaries whose eccentricity is maximised at $\omega=0$ are those on circulating phase space trajectories in the regime $0 < \Gamma \leq 1/5$ (e.g. Figure \ref{fig:HStar_Theta_pt001_logplot}f; see Paper II for a thorough discussion). The minimum $j$ in this case is $j_\mathrm{min}=j_0$ (Paper II), which is given in equation \eqref{eqn:j0}. For this to correspond to very high eccentricity one needs $D\approx 1$, and the orbit must sit very close to the separatrix between librating and circulating orbits, which can be hard to achieve in practice.

Nevertheless, suppose $j_0 \sim \Theta^{1/2}\ll 1$ for $\epsGR=0$;  then for $e_\mathrm{max}$ not to be changed radically when we do include GR, a necessary but insufficient condition is \eqref{eqn:epsGR_ll_1_minus_5_Gamma}. Finding the minimum $j$ at $\omega=0$ requires that we set the final square bracket in \eqref{eq:djdtGR} to zero, which is the same as solving the depressed cubic \eqref{eqn:depressed_cubic}. In \S\ref{sec:lib_or_circ} we explained how to determine the appropriate explicit solution to \eqref{eqn:depressed_cubic} for arbitrary initial conditions.  In particular, we note that the $\epsGR=0$ solution $j_\mathrm{min}=j_0$ corresponds exactly to equation \eqref{eqn:cosine_solution} with $k=0$. However, the general solutions for $\epsGR\neq 0$ are not very enlightening. We can make some analytical progress if we further assume that
 \begin{align}
     \epsGR \ll \vert 1-5\Gamma \vert j_0^3.
     \label{eqn:j0_expand_cond}
 \end{align}
If \eqref{eqn:j0_expand_cond} is true, then the first order solution for finite $\epsGR$ is
\begin{align}
    \jmin \approx j_0\left[ 1 - \frac{\epsGR}{6(1-5\Gamma)j_0^{3}}\right] = j_0\left[ 1 - \left(\frac{j_{\mathrm{f},0}}{j_0}\right)^3\right].
    \label{eqn:j0_perturbative}
\end{align}
In other words, since $j_0^2 \sim \Theta \ll 1$ by construction, $\jmin$ starts to substantially deviate from $j_0$ when the saddle points appear at $\omega=0$ --- see equation (\ref{eqn:fp_omega_0_inequality}) and \S\ref{sec:fixed_points_0}. Note that the condition \eqref{eqn:j0_expand_cond} is very stringent and requires that the binary be deep in the very weak GR regime (\S\ref{sect:corollary}), so \eqref{eqn:j0_perturbative} may not be useful in practice.  


\section{Analytic solution for orbital elements at high eccentricity}
\label{sec:analytic}


In this Appendix we present an analytic solution to the DA equations of motion for all orbital elements in the limit of high eccentricity, assuming $\Gamma>0$.  To do this we will make the following four assumptions:
\begin{itemize}
    \item (I) $j=j_\mathrm{min} \ll 1$ is realised at $\omega=\pm\pi/2$,
    \item (II) Weak or very weak GR, i.e. $\epsGR \ll \epsweak$,
    \item (III) $\sigma \ll 1$ (equation \eqref{eqn:dimensionless_number_2}),
    \item (IV) $j^4/\Theta \ll 1$.
\end{itemize}
Assumptions (I)-(III) are familiar from \S\ref{sec:weak_GR} (and of course if we set $\epsGR=0$ then (II) and (III) are satisfied automatically).  Taken together, assumptions (I)-(III) imply, in particular, that $j(t)$ takes the form \eqref{eq:t_min} with $j_\mathrm{min}=j_-$, $j_1 = j_+$, $j_2= j_0$, which can be seen by expanding the weak GR equation \eqref{eqn:djdt_weak_GR} for $\sigma \ll 1$.

However, assumption (IV) is new. It is equivalent to the requirement that
\begin{align}
 \frac{j}{\cos^2 i_\mathrm{min}} \ll 1, \,\,\,\,\,\,\,\,\, \mathrm{where} \,\,\,\,\,\,\,\,\,
    \cos^2 i_\mathrm{min} \equiv \Theta/(1-e^2_\mathrm{max})
    \label{eqn:j4_over_Theta_condition}
\end{align}
is the cosine of the binary's minimum inclination.
Assumption (IV) is nearly always satisfied at high eccentricities since we normally have\footnote{Indeed, equation \eqref{eqn:jf}, which is valid in the very weak GR regime, tells us that $j_\mathrm{f}^4 \sim \Theta \ll 1$ and we know that  $j_\mathrm{f}$ then provides an upper bound on $j_\mathrm{min}$ for $\Gamma > 1/5$.} $j^2 \sim \Theta$. The additional assumption (IV) allows us to take the solution for $j(t)$ from \eqref{eq:t_min} that we got using assumptions (I)-(III) and simplify the expression for $t_\mathrm{min}$.  The result is:
\begin{align} 
 \label{eqn:j_DA_analytic}
    j(t)&= j_\mathrm{min} \sqrt{1+\left(\frac{t}{t'_\mathrm{min}}\right)^2}, 
\end{align} 
where\footnote{To see this we take the explicit expressions for $j_1=j_+$ and $j_2=j_0$ from \eqref{eqn:jpm}-\eqref{eqn:def_D} and simplify them using assumption (I). Plugging the simplified expressions into \eqref{eq:t_min} and expanding the result using assumption (IV) we recover equation \eqref{eqn:tmin_DAsoln}.}
\begin{align}
&t'_\mathrm{min} \equiv  \frac{j_\mathrm{min}^3}{60\Gamma\sqrt{\Theta(j_\mathrm{min}^2-\Theta)}}\frac{L}{C}
\label{eqn:tmin_DAsoln}.
\end{align}
We note that $t'_\mathrm{min}$ diverges as $j_\mathrm{min} \to \Theta$, that is as $e_\mathrm{max} \to e_\mathrm{lim}$.  This is as expected from e.g. Figure \ref{fig:HStar_Theta_pt001_logplot}a, since trajectories that approach $e_\mathrm{lim}$ become ever `flatter' in the vicinity of $\omega=\pm\pi/2$, i.e. less and less sharply peaked around their eccentricity maxima, so the fraction of a secular period they spend in the vicinity of $e_\mathrm{max}$ increases.

\begin{figure}
\centering
\includegraphics[width=0.9\linewidth,trim={1.8cm 1.2cm 0.2cm 1.3cm},clip]{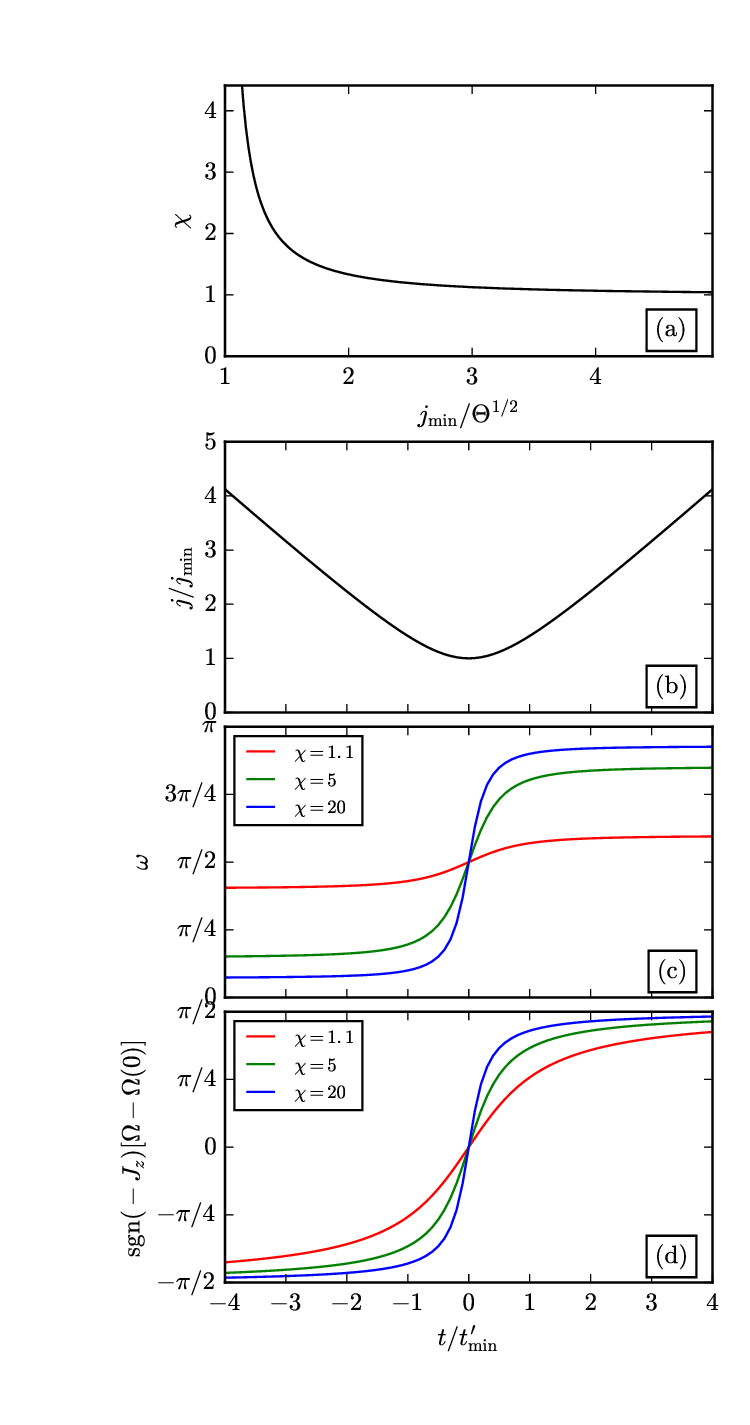}
\caption{Analytic solution to the DA equations of motion without GR at high eccentricity, for binaries that achieve maximum eccentricity at $\omega = \pm\pi/2$. The solution depends on the parameter $\chi$ (equation \eqref{eqn:chi}) which is plotted as a function of $j_\mathrm{min}/\Theta^{1/2}$ in panel (a).  Panels (b)-(d) show the evolution of $j(t), \omega(t)$ and $\Omega(t)$ respectively, where we have taken the maximum eccentricity to coincide with $t=0$.}
\label{fig:DA_solution_column}
\end{figure}

Next we obtain the solution for $\omega(t)$.  First, using the conservation of $H^*$ (equation \eqref{eqn:DA_Hamiltonian}) and assumptions (I) and (IV) we easily get an expression for $\cos^2\omega(j)$ without stipulating any particular form of $j(t)$:
\begin{align}
\cos^2\omega = \frac{\Theta}{j^2-\Theta} \left[ \frac{j^2}{j_\mathrm{min}^2} - 1 + \frac{\epsGR j_\mathrm{min}}{30\Gamma\Theta} \left( \frac{j^2}{j^2_\mathrm{min}} - \frac{j}{j_\mathrm{min}}\right) \right].
\label{eqn:cossquaredomega_DA}
\end{align}
Now plugging in the particular form \eqref{eqn:j_DA_analytic} for $j(t)$ we find the following explicit solution for\footnote{
We have included the $\mathrm{sgn}(t)$ factor in \eqref{eqn:omega_DA_analytic} because for $\Gamma>0$ the pericentre angle $\omega$ must increase towards $\pi/2$ as $j$ decreases to $j_\mathrm{min}$ (at $t=0$), and continue to increase as $j$ increases away from $j_\mathrm{min}$.} $\omega(t)$: 
\begin{align} 
    \omega(t) &= \frac{\pi}{2} + \mathrm{sgn}(t)\cos^{-1}\left(\sqrt{\frac{1+(t/t'_\mathrm{min})^2 - \epsGR P(t)}{1+(t/t'_\mathrm{min})^2\chi}}\right),
    \label{eqn:omega_DA_improved}
\end{align} 
where
\begin{align}
\chi &\equiv \frac{ j_\mathrm{min}^2}{j_\mathrm{min}^2 - \Theta} = \left(1-\frac{\Theta}{j_\mathrm{min}^2}\right)^{-1} = \frac{1}{\sin^2 i_\mathrm{min}},
\label{eqn:chi}
\end{align}
and
\begin{align}
 P(t) &\equiv \frac{\chi}{30\Gamma j_\mathrm{min}}\left[ 1+(t/t_\mathrm{min}')^2 - \sqrt{1+(t/t_\mathrm{min}')^2} \right].
 \label{eqn:omega_soln_auxiliary_fn}
\end{align}
is a dimensionless function of time.
In Figure \ref{fig:DA_solution_column}a we show how $\chi$ varies as a function of $j_\mathrm{min}/\Theta^{1/2} \geq 1$.  We see that $\chi > 1$ always and that typical values of $\chi$ are $\sim$ a few.

Finally we can get the solution for $\Omega(t)$ by using assumption (I) in equation \eqref{eqn:dOmegadt_DA}, plugging in the solutions \eqref{eqn:j_DA_analytic} and \eqref{eqn:omega_DA_improved} for $j(t)$ and $\omega(t)$ respectively, and integrating in time. The result is
 \begin{align}
\Omega(t) = \Omega(0) &+ \mathrm{sgn}(j_z) \Big\{- \tan^{-1} \Big(\sqrt{\chi} \frac{t}{t'_\mathrm{min}}\Big) \nn \\ &+ \frac{\epsGR}{30\Gamma} \frac{\chi}{j_\mathrm{min}} \Big[ 
\tan^{-1} \Big(\sqrt{\chi}\frac{t}{t'_\mathrm{min}}\Big)
\nn \\
&-
\sqrt{\frac{\chi}{\chi-1}}
\tan^{-1} \Big(\sqrt{\chi-1}\frac{t/t'_\mathrm{min}}{\sqrt{1+(t/t'_\mathrm{min})^2}}\Big)
\Big]\Big\},
\label{eqn:Omega_DA_improved}
\end{align}
where we introduced $j_z \equiv J_z/L = j\cos i$.  

In equation \eqref{eqn:Omega_DA_improved} the value of $\Omega(0)$ is an arbitrary constant to be prescribed. Otherwise, equations \eqref{eqn:j_DA_analytic}-\eqref{eqn:tmin_DAsoln}, \eqref{eqn:omega_DA_improved}-\eqref{eqn:Omega_DA_improved},  and the equation $j_z(t) = j_z(0)$ provide a complete, explicit description of the DA dynamics in the high-$e$ limit whenever assumptions (I)-(IV) are satisfied. 

We note that $\omega, \Omega$ make finite `swings' across the maximum eccentricity peak. Indeed, equation \eqref{eqn:omega_DA_improved} tells us that $\omega$ takes asymptotic values
\begin{align}  
\omega(t\to \pm \infty) = \frac{\pi}{2} 
\pm \cos^{-1}
\left(
\sqrt{ \frac{1}{\chi}-\frac{\epsGR}{30\Gamma j_\mathrm{min}} }
\right),
\label{eq:om_jump}
\end{align}
giving a total swing of magnitude $\vert \Delta \omega \vert = 2 \cos^{-1} \sqrt{[\chi^{-1} - \epsGR/(30\Gamma j_\mathrm{min})]}$.
Clearly the larger $\epsGR$, the bigger is this swing\footnote{Of course this value becomes ill-defined when $\epsGR > 30\Gamma j_\mathrm{min}/\chi \sim \epsweak/\chi$.  For typical values of $\chi \sim 1$ this is never an issue in the weak GR regime $\epsGR \ll \epsweak$.}, which makes sense since GR promotes fast apsidal precession. Similarly from \eqref{eqn:Omega_DA_improved} we find:
\begin{align}  \Omega(t\to \pm \infty) &= \Omega(0) \mp \mathrm{sgn}(j_z)
\Big\{
\frac{\pi}{2} \nn \\ &- \frac{\epsGR \chi}{30\Gamma j_\mathrm{min}}\left[\frac{\pi}{2} - \sqrt{\frac{\chi}{\chi-1}}\tan^{-1}(\sqrt{\chi-1})\right]
\Big\}.
\end{align}
Thus, the size of the swing in $\Omega$ across the eccentricity peak $\vert \Delta \Omega \vert = \pi - \mathcal{O}(\epsGR)$ is \textit{reduced} by GR effects.

\subsection{Analytic solution in the LK limit}


To apply the analytic solution to the LK case of hierarchical triples, let the tertiary perturber have mass $\mathcal{M}$ and the outer orbit have semimajor axis $a_\mathrm{g}$ and eccentricity $e_\mathrm{g}$. Then we set $\Gamma=1$, evaluate $\epsGR$ using equations \eqref{eqn:A_LK}, \eqref{eq:epsGR} and take $t'_\mathrm{min}$ equal to
\begin{align}
t'_\mathrm{min, LK}  
\equiv \frac{4a_\mathrm{g}^3(1-e_\mathrm{g}^2)^{3/2}\sqrt{m_1+m_2}}{15 G^{3/2}\mathcal{M}a^{3/2}}\frac{\chi^{3/2}}{\sqrt{\chi-1}}.
\end{align}
(To derive this formula we have used the results of Paper I, Appendix B).



\subsection{Simplified analytic solution in the limit $\epsGR=0$}


One can get a simplified version of the analytic solution if one takes the non-GR limit. For $\epsGR = 0$ the solution for $j(t)$ takes the same form \eqref{eqn:j_DA_analytic}, while \eqref{eqn:omega_DA_improved} and \eqref{eqn:Omega_DA_improved} simplify to
\begin{align} 
    \omega(t) &= \frac{\pi}{2} + \mathrm{sgn}(t)\cos^{-1}\left(\sqrt{\frac{1+(t/t'_\mathrm{min})^2}{1+(t/t'_\mathrm{min})^2\chi}}\right),
    \label{eqn:omega_DA_analytic}
\\
        \Omega(t) &= \Omega(0) - \mathrm{sgn}(j_z) \tan^{-1}\left(\frac{\sqrt{\chi} \,t}{t'_\mathrm{min}}\right).
        \label{eqn:Omega_DA_analytic}
\end{align} 
In Figure \ref{fig:DA_solution_column} we show the characteristic behaviour of this non-GR solution.  
In panel (b) we plot $j/j_\mathrm{min}$ as a function of $t/t_\mathrm{min}'$: obviously $j(t)$ is quadratic in $t$ for $t\lesssim t_\mathrm{min}'$ and linear for $t\gtrsim t_\mathrm{min}'$. Panels (c) and (d) demonstrate how the solutions for $\omega(t)$ and $\Omega(t)$ look for various $\chi$. We note that both angles evolve very rapidly during the interval $-1 \lesssim t/t_\mathrm{min}' \lesssim 1$ and rather slowly otherwise, particularly for $\chi \gg 1$. We see also that the behaviour of $\omega$ is quite strongly dependent on $\chi$; it completes a swing 
$\vert \Delta \omega \vert = 2\cos^{-1}(\chi^{-1/2})$ 
as $t$ runs from  $-\infty$ to $+\infty$. The evolution of $\Omega$ depends somewhat less strongly on $\chi$, and its asymptotic value is independent of $\chi$, so that that the total swing in $\Omega$ across the eccentricity peak is always
$\vert \Delta \Omega \vert = \pi. $
Of course these swings in $\omega$ and $\Omega$ are not completely correct because we expect our analytic formulae to break down once $e$ differs significantly from unity (\S\ref{sec:analytic_validity}).
\begin{figure*}
\begin{minipage}{.48\textwidth}
  \includegraphics[width=.99\linewidth]{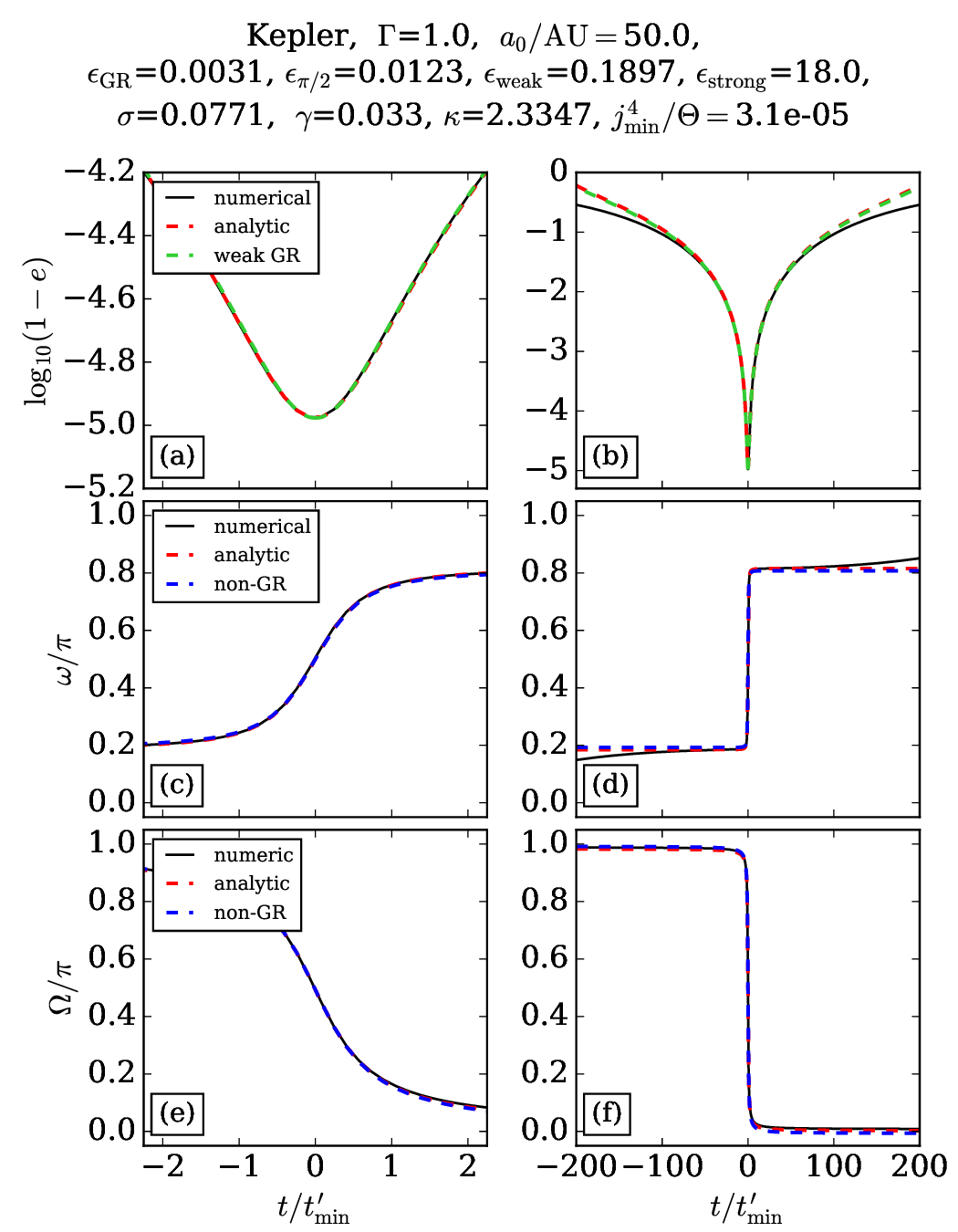}
  \caption{Comparing analytic results at high eccentricity with exact numerical integration (see \S\ref{sec:analytic_validity} for details). In this example $\epsGR < \epspibytwo$ so the binary is in the very weak GR regime. Note also that $\sigma \ll 1$.}
  \label{fig:Kep_a50}
\end{minipage}%
\hspace{0.02\textwidth}
\begin{minipage}{.48\textwidth}
  \includegraphics[width=.99\linewidth]{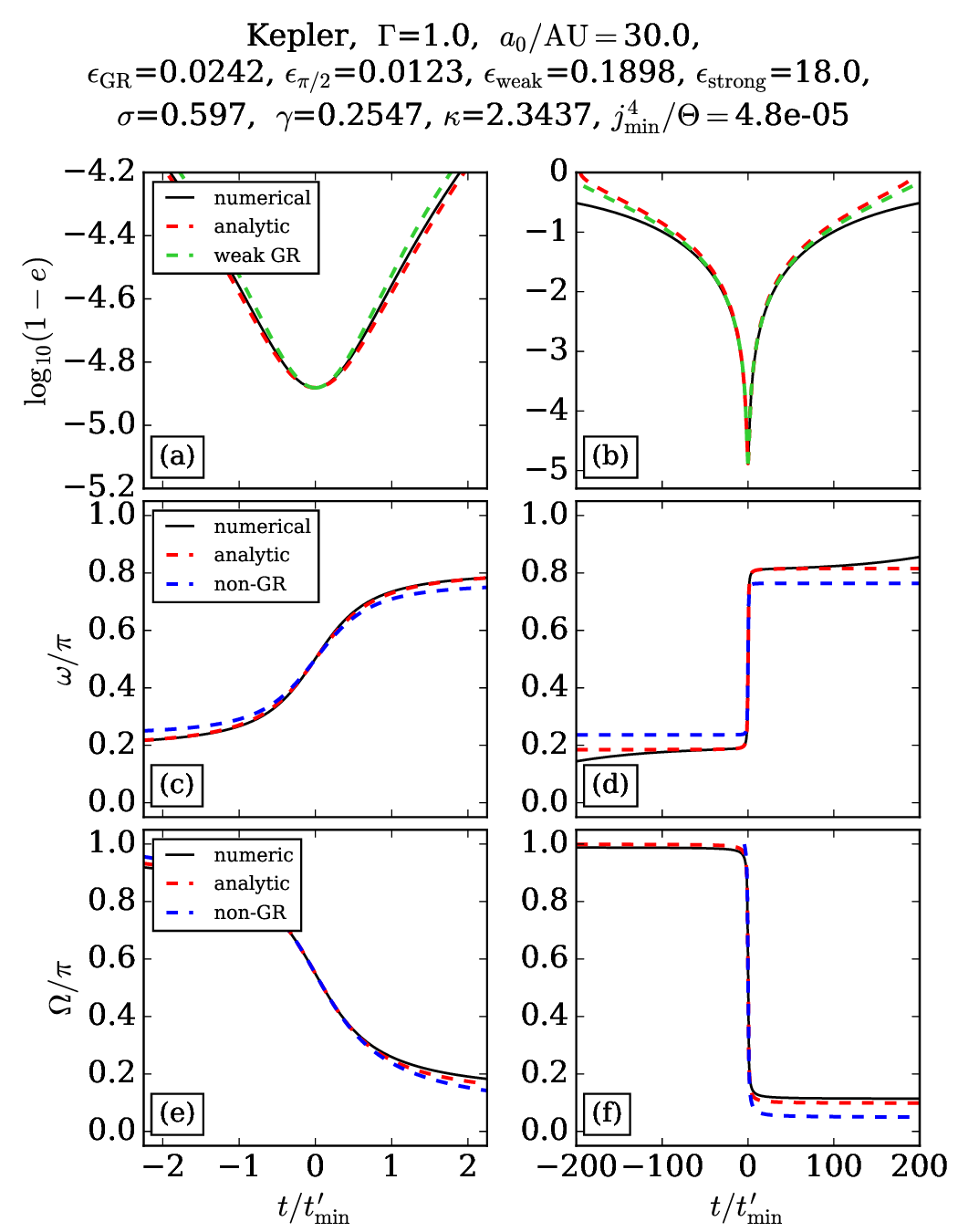}
  \caption{As in Figure \ref{fig:Kep_a50} except we take $a_0 = 30\mathrm{AU},$ so the binary is in the weak GR regime, $\epspibytwo < \epsGR < \epsweak$, and the value of $\sigma$ is now approaching unity.}
  \label{fig:Kep_a30}
\end{minipage}
\begin{minipage}{.48\textwidth}
  \includegraphics[width=.99\linewidth]{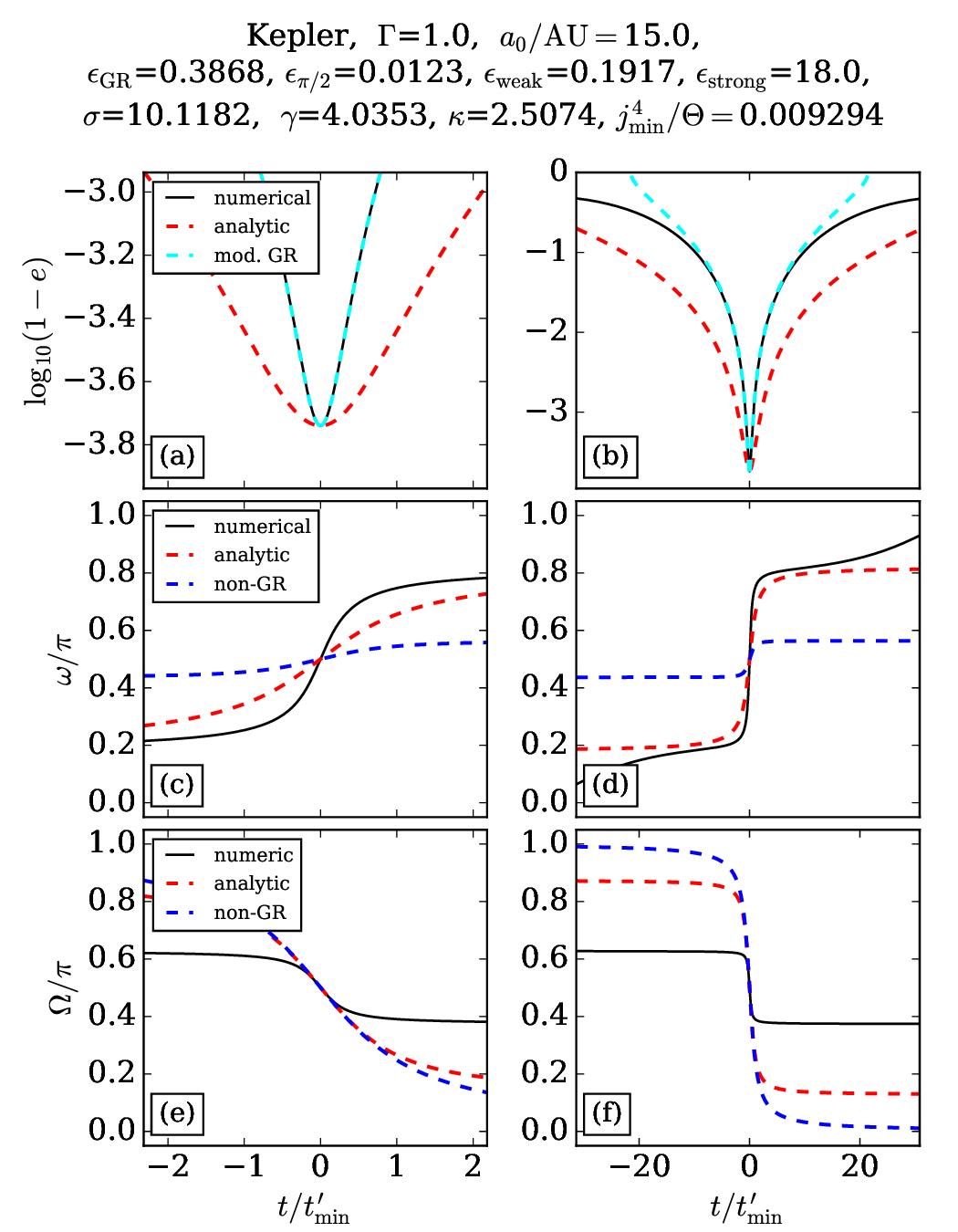}
  \caption{As in Figure \ref{fig:Kep_a50} except we take $a_0 = 15\mathrm{AU},$ so the binary is in the moderate GR regime, $\epsweak< \epsGR < \epsstr$. Note that $\sigma$ is significantly larger than unity in this case.}
  \label{fig:Kep_a15}
\end{minipage}
\hspace{0.02\textwidth}
\begin{minipage}{.48\textwidth}
  \includegraphics[width=.99\linewidth]{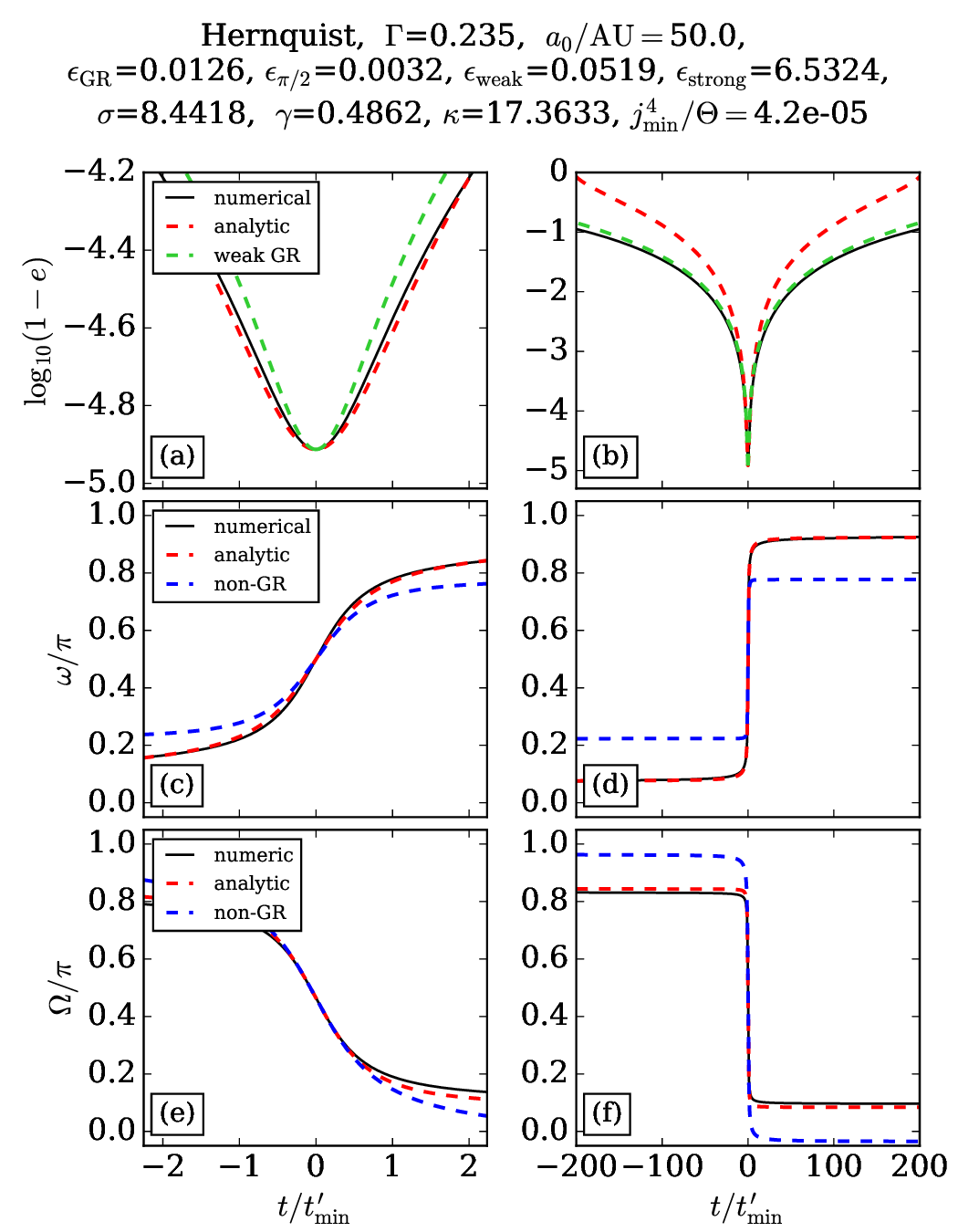}
  \caption{As in Figure \ref{fig:Kep_a50} except we replace the Kepler potential with the Hernquist potential with scale radius $1\mathrm{pc}$.  This results in $\Gamma = 0.235$ so $\sigma \gg 1$ even though the binary is in the weak GR regime.}
  \label{fig:Hern_a50}
\end{minipage}
\end{figure*}
\newpage
\begin{figure*}
\centering
\begin{minipage}{.47\textwidth}
  \centering
  \includegraphics[width=.99\linewidth]{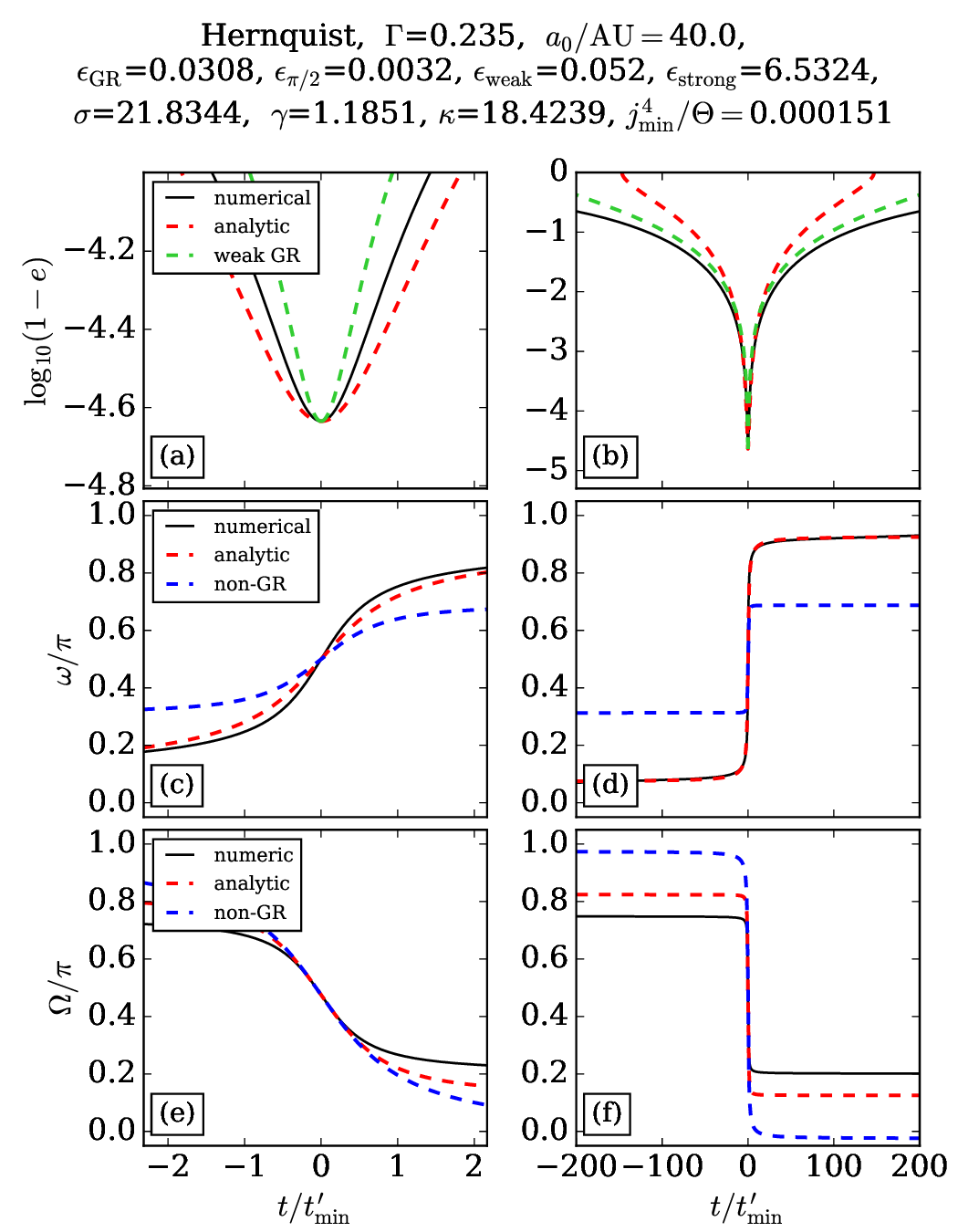}
  \caption{As in Figure \ref{fig:Hern_a50} except we take $a_0 = 40\mathrm{AU}$. Again the binary is in the weak GR regime, $\epspibytwo < \epsGR < \epsweak$, but is approaching the moderate GR regime.}
  \label{fig:Hern_a40}
\end{minipage}
\hspace{0.02\textwidth}
\begin{minipage}{.47\textwidth}
  \centering
  \includegraphics[width=.99\linewidth]{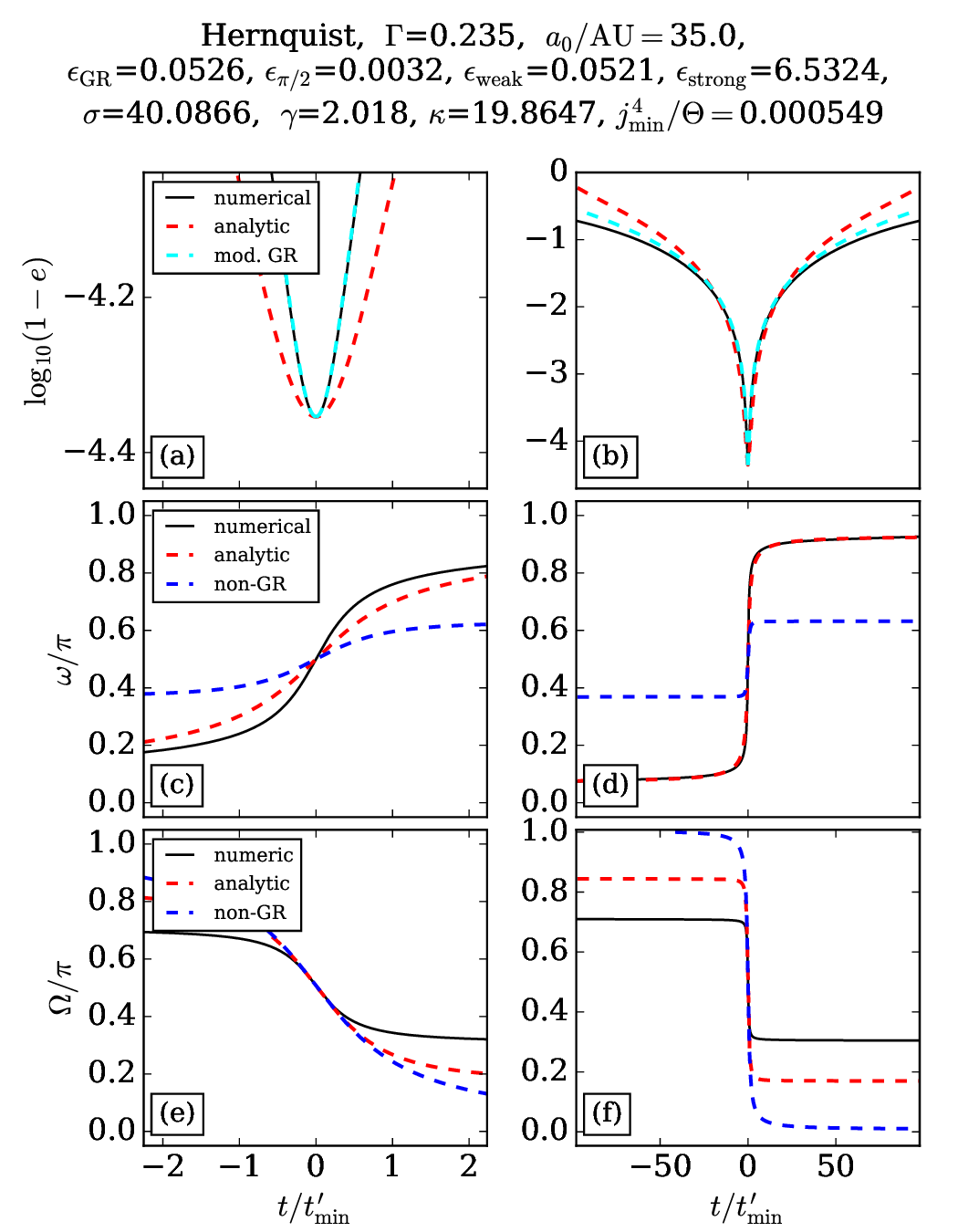}
  \caption{As in Figure \ref{fig:Hern_a50} except we take $a_0 = 35\mathrm{AU}$. In this case the binary is (just) in the moderate GR regime $\epsweak < \epsGR < \epsstr$.}
  \label{fig:Hern_a35}
\end{minipage}
\end{figure*}
\subsection{Validity of the analytic solution}
\label{sec:analytic_validity}

In this section we test the accuracy of the analytical solution \eqref{eqn:j_DA_analytic}, \eqref{eqn:omega_DA_improved}, \eqref{eqn:Omega_DA_improved} and the simplified solution \eqref{eqn:omega_DA_analytic},  \eqref{eqn:Omega_DA_analytic} derived in the non-GR limit, against direct numerical integration of the DA equations of motion \eqref{eom1}, \eqref{eom2}, \eqref{eqn:dOmegadt_DA}, in different dynamical regimes.


\subsubsection{Three examples with $\Gamma =1$}

First we consider some examples in the LK case of $\Gamma=1$. Precisely, we consider a binary with component masses $m_1=m_2=1M_\odot$ orbiting a point mass $\mathcal{M} = 10^5M_\odot$. For the outer orbit we choose a pericentre distance  $\rp = 0.4$pc and an apocentre distance $\ra = 0.6$pc.  The outer orbit is then an ellipse with semimajor axis $a_\mathrm{g} = 0.5$pc and eccentricity $e_\mathrm{g}=0.2$.  For the inner binary orbit we take the initial conditions $e_0=0.5$, $i_0=89.75^\circ$ (so that $\Theta = 1.4\times 10^{-5}$), $\omega_0 = 0^\circ$. When we integrate the equations of motion we will shift the time coordinate so that maximum eccentricity is achieved at $t=0$; in each example we choose a value of $\Omega_0$ so that $\Omega(0) \approx \pi/2$. All that remains is to specify the initial semimajor axis $a_0$.

In Figure \ref{fig:Kep_a50} we take $a_0 = 50$AU.  Then in each panel we plot the result of the direct numerical integration with a black line and we show the analytic solution \eqref{eqn:j_DA_analytic}, \eqref{eqn:omega_DA_improved}, \eqref{eqn:Omega_DA_improved} with a dashed red line. Panel (a) shows the evolution of $\log_{10}(1-e)$ as a function of time $t$ over a short time interval $-2.2 \leq t/t_\mathrm{min}' \leq 2.2$ centred on the eccentricity peak. Panel (b) shows the same solution zoomed out over a much longer time interval $-200 \leq t/t_\mathrm{min}' \leq 200$.  Analogously, panels (c) and (d) show the numerical and analytical solutions for the apsidal angle $\omega(t)$ over these same time intervals, while panels (e) and (f) show the evolution of the nodal angle $\Omega(t)$.  Finally, in panels (c)-(f) we plot blue dashed lines which correspond to the simple non-GR form of the analytic solution, namely equations \eqref{eqn:omega_DA_analytic}, \eqref{eqn:Omega_DA_analytic}, though to evaluate it we still use the GR-modified value of $j_\mathrm{min}$.  (There are also green dashed lines in panels (a), (b) --- see \S\ref{sec:weak_GR}).  At the top of the figure we show the values of various key quantities that allow us to check the validity of the assumptions (I)-(IV). 

Overall, in Figure \ref{fig:Kep_a50} the analytic solution provides an excellent fit to the exact numerical integration. Errors are only noticeable once $e$ falls below $\sim 0.9$ (panels (b) and (d)). This good agreement reflects the fact that $\sigma, j_\mathrm{min}^4/\Theta \ll 1$ and $\epsGR \ll \epsweak$, meaning that all assumptions (I)-(IV) are fulfilled. Moreover, we see that the full analytic solution (red dashed lines) and non-GR solution (blue dashed lines) overlap almost exactly in panels (c)-(f).
This is unsurprising because the binary actually sits in the {\it very weak} GR regime $\epsGR < \epspibytwo$, meaning GR effects are negligible (\S\ref{sect:corollary}).

In Figure \ref{fig:Kep_a30}
we use all the same system parameters as in Figure \ref{fig:Kep_a50} except we set $a_0=30\mathrm{AU}$.  This increases $\epsGR$ and puts the binary in the {\it weak} GR regime $\epspibytwo < \epsGR < \epsweak$.  The fact that $\epsGR$ is no longer smaller than $\epspibytwo$ is responsible for the disagreement between the analytic and non-GR solutions in panels (c)-(f).  Nevertheless, the analytic solution still matches the numerical one very well for $e\gtrsim 0.9$, although not quite as well as in Figure \ref{fig:Kep_a50}, owing to the fact that $\sigma$ is now comparable to unity (breaking assumption (III)).

Next, in Figure \ref{fig:Kep_a15} we again run the same experiment but this time with $a_0=15$AU.  This puts the binary in the {\it moderate} GR regime, $\epsweak < \epsGR < \epsstr$, which violates assumption (II). Additionally we have $\sigma \gg 1$, violating assumption (III). We see solutions \eqref{eqn:j_DA_analytic}, \eqref{eqn:omega_DA_improved}, \eqref{eqn:Omega_DA_improved} largely fail to capture the high-eccentricity behaviour even over a very short timescale. At the same time, we note that the moderate GR solution (\ref{eq:j_implicit_modGR}) captures the $j(t)$ behaviour extremely well in this case.


\subsubsection{Three examples with $\Gamma =0.235$}

Next we consider some examples with a different value of $\Gamma$. To achieve this we replace the Kepler potential with a Hernquist potential $\Phi(r) = -G\mathcal{M}(b+r)^{-1}$, where the total mass $\mathcal{M}=10^5M_\odot$ and the scale radius $b=1$pc.  (The outer orbit still has $\rp=0.4$pc and $\ra=0.6$pc, but will now fill a 2D annulus rather than forming a closed ellipse --- see Paper I). As a result we find $\Gamma = 0.235$.  Also in this case both $\sigma$ and $\kappa$ attain large values, putting our analytical solutions to a demanding test.

In Figure \ref{fig:Hern_a50} we integrate exactly the same system as in Figure \ref{fig:Kep_a50} except for this replacement of the potential --- in particular, we again take $a_0 = 50$AU. We see that this puts the binary in the weak GR regime $\epspibytwo < \epsGR < \epsweak$ (as in Figure \ref{fig:Kep_a30}), but that $\sigma$ is much larger than unity (unlike in Figure \ref{fig:Kep_a30}).  One consequence of this is that the analytic approximation to $\log_{10}(1-e)$ fails rather early on, with significant errors by the time $e$ falls below $0.99$ (Figure \ref{fig:Hern_a50}b). Despite this, the analytic approximations to $\omega(t)$ and $\Omega(t)$ are still excellent (panels (c)-(f)).  This is because $\omega$ and $\Omega$ are sensitive only to the eccentricity behaviour at the very peak --- they change very rapidly over the interval $-2.2 < t/t_\mathrm{min}' < 2.2$ (panels (c) and (e)), but are almost constant the rest of the time. Thus, as long as $j(t)$ is captured well near the very peak eccentricity, as it is in panel (a), the analytic solutions for $\omega$, $\Omega$ work well despite assumption (II) being broken.

In Figure \ref{fig:Hern_a40} we investigate the same system except with $a_0$ reduced to $40$AU. The binary is still in the weak GR regime but only just so, violating assumption (II).  It also has $\sigma \gg 1$ like it did in Figure \ref{fig:Hern_a50}, violating assumption (III).  We see that the analytic fit to $\log_{10}(1-e)$ is quite poor even at the very peak (panel (a)). Interestingly though, the evolution of $\omega$ (panel (d)) is reproduced rather accurately, highlighting how sensitive $\omega(t)$ is to the value of peak eccentricity $j_{\rm min}$ (see equation (\ref{eq:om_jump})), and how insensitive it is to anything else. However, the evolution of $\Omega(t)$ is not reproduced very well. The same conclusions hold for Figure \ref{fig:Hern_a35}, in which we have reduced the semimajor axis further to $a_0=35$AU, putting the binary squarely in the moderate GR regime (so that both assumptions (II) and (III) are broken).

\subsubsection{Conclusions}

While assumptions (I) and (IV) are almost always good provided we consider binaries that reach very high eccentricity ($1-e \ll 0.1$), assumptions (II) and (III) are liable to fail in some regimes.

We have seen that for $\log_{10}(1-e)$ to be accurately reproduced by the analytical solution \eqref{eqn:j_DA_analytic} for $e\gtrsim 0.9$, all four assumptions (I)-(IV) must be valid.  

However, the analytic solution \eqref{eqn:Omega_DA_improved} for $\Omega$ can be very accurate even for $\sigma \gg 1$ (violating assumption (III)) provided the behaviour of $j(t)$ in the close vicinity of $j_\mathrm{min}$ is reproduced reasonably well.

What is more, the solution \eqref{eqn:omega_DA_improved} for $\omega(t)$, and the swing $\Delta\omega$ in particular, can be very accurate even if the system is in the moderate GR regime, invalidating both assumptions (III) and (IV).  This is because $\omega$ is extremely sensitive to the behaviour of $j$ around absolute peak eccentricity and largely insensitive to $j$ otherwise.

Lastly, if all assumptions (I)-(IV) are valid and we additionally have $\epsGR \lesssim \epspibytwo$, then one can employ the simpler non-GR form of the solution for $\omega, \Omega$ (equations \eqref{eqn:omega_DA_analytic}, \eqref{eqn:Omega_DA_analytic}).  


  \section{Phase space behaviour and maximum eccentricity in \texorpdfstring{$\Gamma \leq 0$}{GamNeg} regimes}
\label{sec:Gamma_negative}

     \begin{figure*}
\centering
\includegraphics[width=0.97\linewidth,trim={1.2cm 8.5cm 0cm 0.38cm},clip]{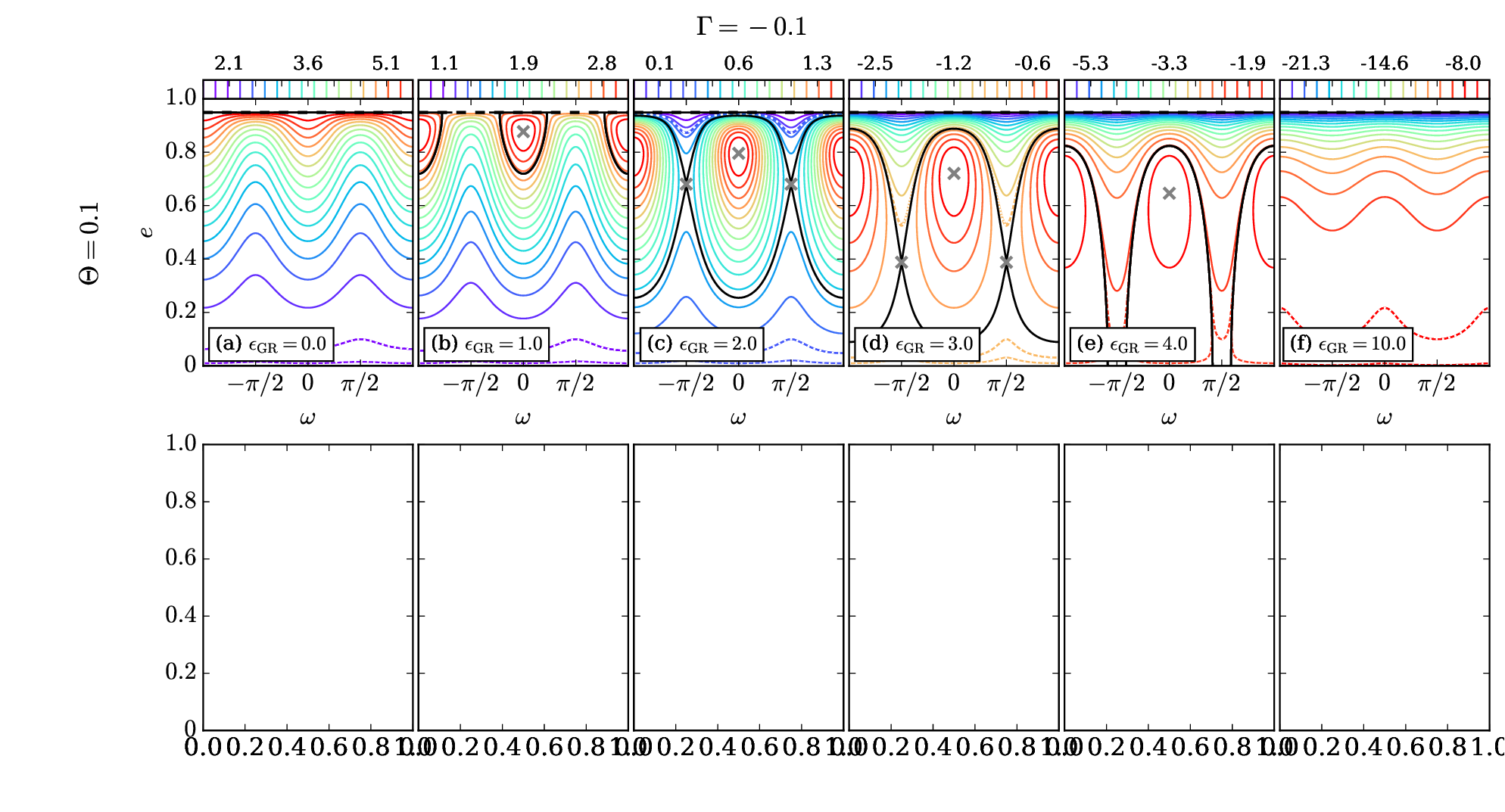}
\caption{As in the top row of Figure \ref{fig:HStar_Contours_Gamma_pt1} except we take $\Gamma=-0.1$ and use some new $\epsGR$ values.}
\label{fig:HStar_Contours_Gamma_minuspt1}
\end{figure*}
   \begin{figure*}
\includegraphics[width=0.99\linewidth,trim={0.9cm 0.3cm 0cm 
0cm},clip]{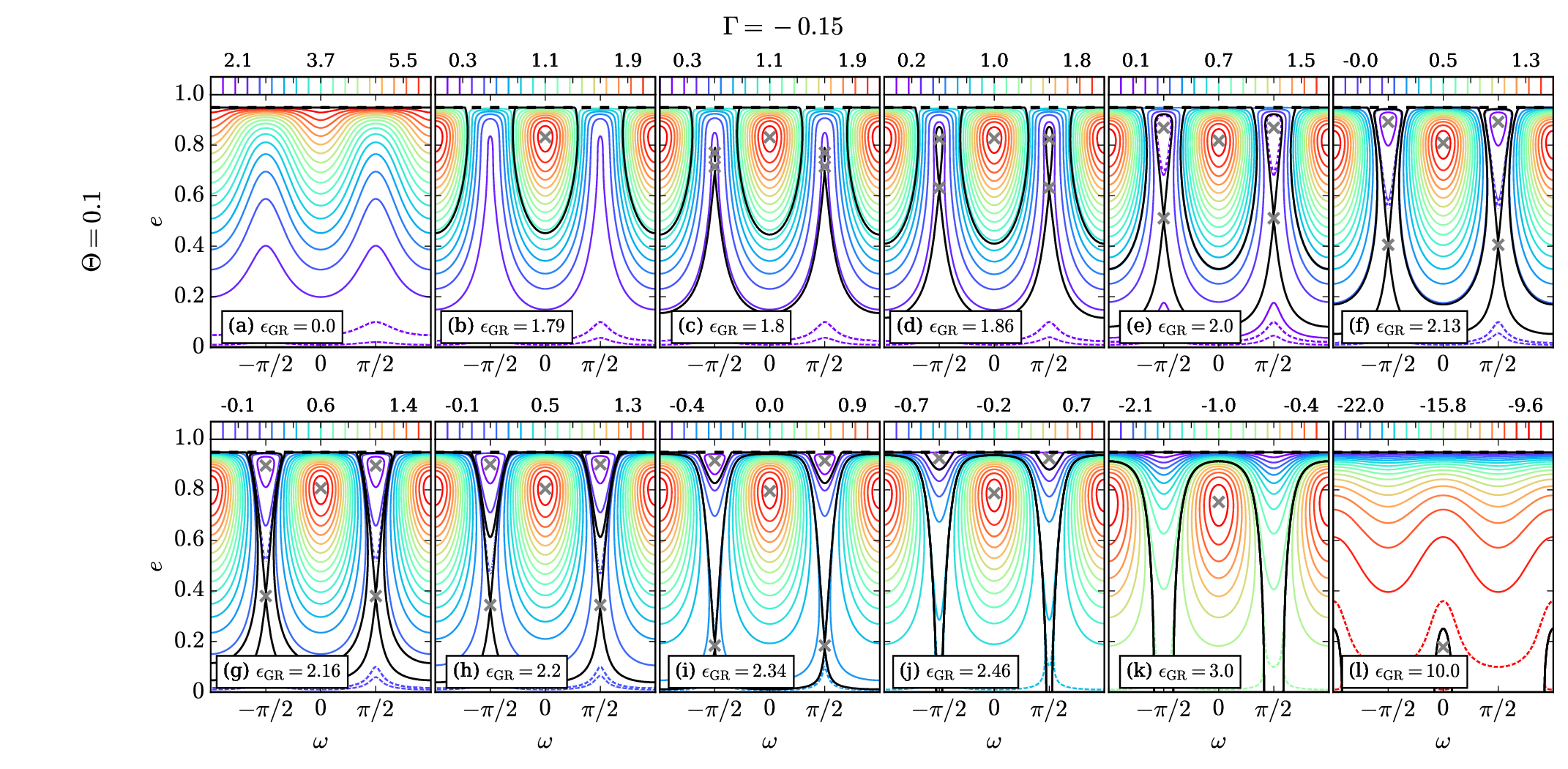}
\caption{As in Figure \ref{fig:HStar_Contours_Gamma_minuspt1} except for $\Gamma=-0.15$, and for twelve values of $\epsGR$.}
\label{fig:HStar_Contours_Gamma_minuspt15}
\end{figure*}
   \begin{figure*}
\centering
\includegraphics[width=0.99\linewidth,trim={0.9cm 8.33cm 0cm 
0.38cm},clip]{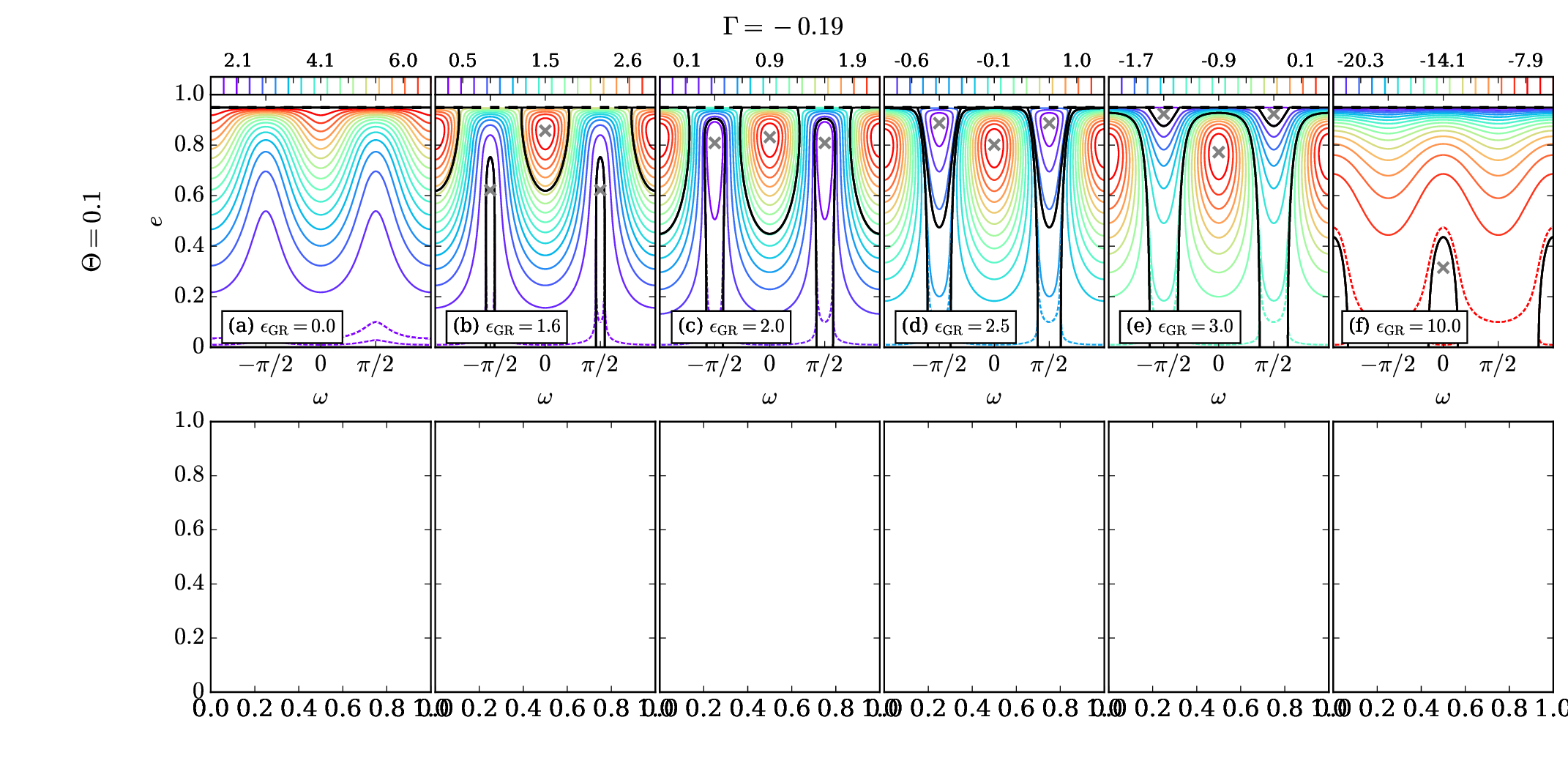}
\caption{As in Figure \ref{fig:HStar_Contours_Gamma_minuspt1} except for $\Gamma=-0.19$ and some different $\epsGR$ values.}
\label{fig:HStar_Contours_Gamma_minuspt19}
\end{figure*}
   \begin{figure*}
\centering
\includegraphics[width=0.99\linewidth,trim={0.9cm 8.33cm 0cm 0.38cm},clip]{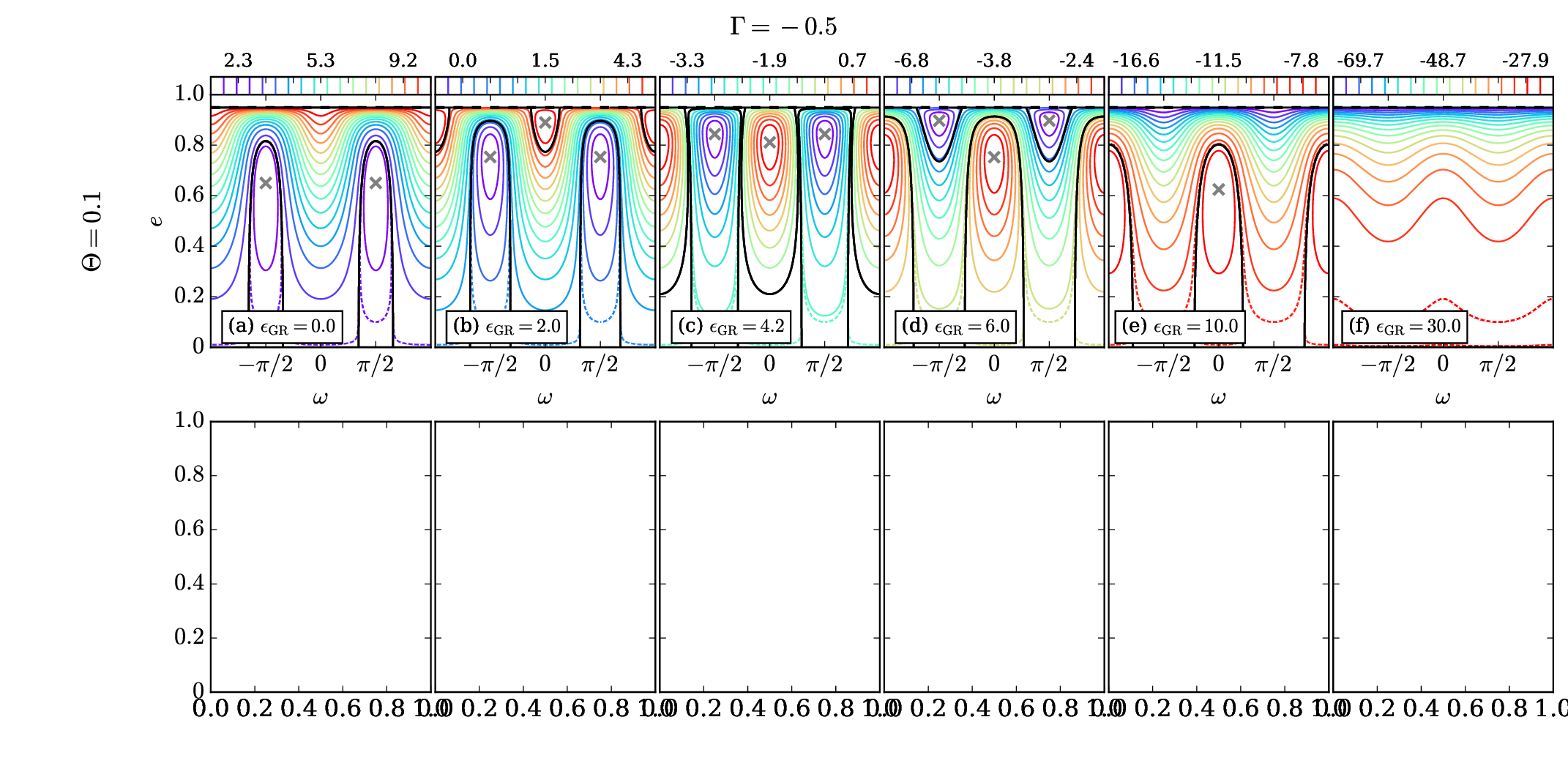}
\caption{As in Figure \ref{fig:HStar_Contours_Gamma_minuspt1} except for $\Gamma=-0.5$, and some new $\epsGR$ values.}
\label{fig:HStar_Contours_Gamma_minuspt5}
\end{figure*}
   \begin{figure*}
\centering
\includegraphics[width=0.87\linewidth]{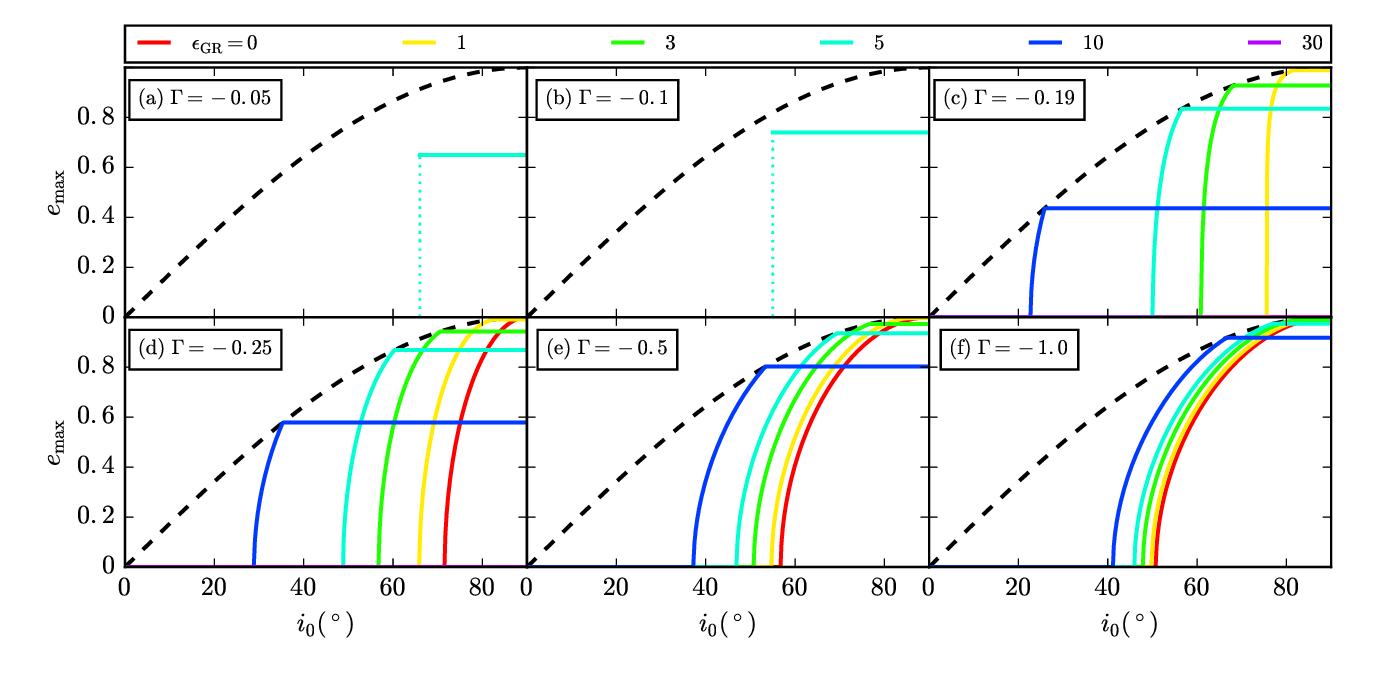}
\caption{As in Figure \ref{fig:emax_initially_circular} except for $-1/5 < \Gamma \leq 0$ (panels (a)-(c)) and $\Gamma \leq -1/5$ (panels (d)-(f)). The vertical dotted lines in panels (a) and (b) show the critical inclination $i_0 = \cos^{-1} \sqrt{\Theta_1(\Gamma, \epsGR)}$ for $\epsGR=5$ --- see \S\ref{sec:max_ecc_circular_negative_Gamma}.}
\label{fig:emax_initially_circular_negative_Gamma}
\end{figure*}

In this Appendix we discuss the dynamical behaviour that arises in negative $\Gamma$ regimes. This behaviour can be significantly more complicated than for positive $\Gamma$.
  In what follows we offer an overview of the phase space dynamics for $-1/5 < \Gamma \leq 0$ (in \S\ref{sec:morphology_Gamma_Regime_III}) and $\Gamma \leq -1/5$ (in \S\ref{sec:morphology_Gamma_Regime_IV}).
   Lastly we consider the eccentricity maxima of binaries with negative $\Gamma$ (\S\ref{sec:lib_or_circ_negative_Gamma}), focusing mainly on initially near-circular orbits. 
  
   
   \subsection{Phase space behaviour in the case \texorpdfstring{$-1/5 < \Gamma \leq 0$}{GamIII}}
   \label{sec:morphology_Gamma_Regime_III}

Unlike for $\Gamma > 0$, the dynamical behaviour in the regime $-1/5 < \Gamma \leq 0$ cannot be understood using only one value of $\Gamma$ as an example. Thus, we consider three values. In Figures \ref{fig:HStar_Contours_Gamma_minuspt1}, \ref{fig:HStar_Contours_Gamma_minuspt15} and \ref{fig:HStar_Contours_Gamma_minuspt19} we plot contours of constant $H^*$ in the $(\omega,e)$ phase space for $\Theta = 0.1$, taking $\Gamma=-0.1$, $\Gamma = -0.15$ and $\Gamma = -0.19$ respectively. The manually-added dashed contours are the same as in Figure \ref{fig:HStar_Contours_Gamma_pt1}. We now discuss these three figures in turn.

First we discuss Figure \ref{fig:HStar_Contours_Gamma_minuspt1} ($\Gamma=-0.1$).  From Paper II we know that when $\epsGR=0$, fixed points never exist in the phase space for $-1/5 < \Gamma \leq 0$. Thus all phase space trajectories circulate and their maximum eccentricity is found at $\omega= \pm\pi/2$, as in panel (a). Now we consider finite $\epsGR$. In panel (b), namely for $\epsGR = 1.0$, we see that fixed points have appeared at $\omega=0, \pm\pi$, which we will refer to simply as $\omega = 0$ from now on. These fixed points are not saddle points like they were for $0< \Gamma \leq 1/5$ (Figure \ref{fig:HStar_Contours_Gamma_pt1}); instead they are maxima of $H^*$ and host a region of librating orbits that is connected to $e_\mathrm{lim}$.  

As we increase $\epsGR$ further we see that these fixed points move down the page to lower eccentricity, and their associated librating islands become larger in area.  At some threshold value of $\epsGR$ the librating islands become disconnected from the line $e=e_\mathrm{lim}$, coinciding with the appearance of new saddle points at $\omega=\pm\pi/2, e=e_\mathrm{lim}$.  As we increase $\epsGR$ beyond this threshold the fixed point at $\omega=0$ continues to move down the page (panels (c) and (d)), as do the saddle points at $\omega = \pm \pi/2$, and a new family of high-$e$ circulating orbits runs over the top of the librating islands, reminiscent of what we saw for $0<\Gamma \leq 1/5$ in Figure \ref{fig:HStar_Contours_Gamma_pt1}.  Partitioning the different librating islands and circulating regions in panels (c) and (d) are separatrices that cross at the saddle points. Continuing to increase $\epsGR$ forces both kinds of fixed point to move to lower eccentricities.  The saddle points move fastest and disappear first; in panel (e), the fixed point at $\omega=0$ remains but the saddle points at $\omega = \pm\pi/2$ have disappeared through $e=0$. Increasing $\epsGR$ even further still, the $\omega = 0$ fixed point reaches $e=0$ and then disappears.  This leaves a phase space filled with circulating trajectories (panel (f)), which is similar to the $\epsGR =0$ case shown in panel (a) except that the maximum eccentricities are now found at $\omega=0$ rather than $\omega = \pm \pi/2$, and the locations of the maxima and minima of $H^*$ are reversed (see the colourbars).

Moving on to Figure \ref{fig:HStar_Contours_Gamma_minuspt15} ($\Gamma = -0.15$), we find a completely different picture of rather impressive dynamical diversity. In this figure we have to use twelve panels to fully illustrate the complex phase space behaviour.  To begin with, panels (a) and (b) in Figure \ref{fig:HStar_Contours_Gamma_minuspt15} have the same morphology as Figures \ref{fig:HStar_Contours_Gamma_minuspt1}a,b. However, panel (c) is very different from Figure \ref{fig:HStar_Contours_Gamma_minuspt1}c. This time, at some threshold value of $\epsGR$ a \textit{pair} of fixed points emerges from a single point at $\omega=\pi/2$, $e=e_\mathrm{f,\pi/2}$,  and the same thing happens at $\omega=-\pi/2$. An increase in $\epsGR$ nudges these fixed points apart in their eccentricity values (panel (d)): one of them moves up the page and the other moves down. In each pair, the fixed point with higher $e$ is a minimum of $H^*$ and hosts a region of librating orbits. The fixed point with lower $e$ is a saddle point, and sits on the separatrix that surrounds the upper point's librating region. In addition we still have the usual fixed point and accompanying librating island at $\omega=0$. As a result, we now find two families of circulating trajectories. One runs close to $e=0$ under the separatrices passing through the saddle points. The other runs above these separatrices, but below the separatrices surrounding the librating islands centred on $\omega=0, \pm\pi$. This second type of circulating trajectory reaches high eccentricity by running above the upper fixed points at $\omega=\pm \pi/2$. Quite remarkably, these circulating trajectories also exhibit {\it non-monotonic} behaviour of $\omega(t)$, i.e. $\dot{\omega}$ is $>0$ at some times and $<0$ at others, despite the trajectory being a circulating one.

Increasing $\epsGR$ further, the upper fixed point (minimum) at $\omega = \pm\pi/2$ continues to move up the page, while the lower fixed point (saddle) moves down (panels (e)-(g)). Meanwhile the $\omega = 0$ fixed points also move down the page, albeit much more slowly. Eventually the librating region surrounding the upper fixed point at $\omega = \pm\pi/2$ becomes connected to $e=e_\mathrm{lim}$. Simultaneously, the saddle point at $\omega=\pm\pi/2$ and its associated separatrices merge with the separatrices surrounding the $\omega=0$ librating regions (see the transition from panel (g) to panel (h)). Accompanying this transition is the change in the nature of the second family of finite eccentricity circulating orbits described above --- they now run above (below) the saddle points at $\omega=0, \pm\pi$ ($\omega=\pm \pi/2$). As $\epsGR$ continues to increase the pair of fixed points at $\omega=\pm\pi/2$ continue to move apart in eccentricity, until eventually the lower one disappears at $e = 0$ (panel (j)) followed by the upper one at $e=e_\mathrm{lim}$ (panel (k)).  In panels (k) and (i) we retain only the fixed points at $\omega = 0$, with qualitatively the same overall phase space behaviour as in Figure \ref{fig:HStar_Contours_Gamma_minuspt1}e. The $\omega=0$ fixed points also disappear once $\epsGR$ becomes sufficiently large.

Figure \ref{fig:HStar_Contours_Gamma_minuspt19} ($\Gamma=-0.19$) shows yet again a different qualitative behaviour.  Like in Figures \ref{fig:HStar_Contours_Gamma_minuspt1}, \ref{fig:HStar_Contours_Gamma_minuspt15}, fixed points emerge at $\omega =0$ followed by additional fixed points at $\omega=\pm\pi/2$, $e=e_\mathrm{f,\pi/2}$ (panel(b)).  However, this time the $\omega=\pm\pi/2$ fixed points do not come in pairs like they did in Figure \ref{fig:HStar_Contours_Gamma_minuspt15}.  Instead they are minima of $H^*$ and are surrounded by a librating island that stretches to $e=0$ (though $e_\mathrm{f,\pi/2} \neq 0$ for any $\epsGR$). These fixed points move up the page as we increase $\epsGR$ (panel(c)) until they become connected to $e= e_\mathrm{lim}$ (panel (d)). At this stage circulating trajectories exhibit a transition similar to that in Figure \ref{fig:HStar_Contours_Gamma_minuspt15}. Thereafter we have qualitatively the same behaviour as in Figure \ref{fig:HStar_Contours_Gamma_minuspt15}j. 
\\
\\
As these three examples demonstrate, the qualitative dynamical behaviour in the regime $-1/5 < \Gamma \leq 0$ is highly complex. It is also very difficult to analyse mathematically. The simplest place to start is with the fixed points at $\omega=0,$ $j=j_\mathrm{f,0}$. The formulae describing these fixed points can be carried over from \S\ref{sec:morphology_Gamma_Regime_II} and Appendix \ref{sec:mathematical_details_Gamma_positive}: the value of $j_\mathrm{f,0}$ is still determined by equation \eqref{eqn:jf0} and the fixed points exist provided equation \eqref{eqn:fp_omega_0_inequality} is true.  The key difference for negative $\Gamma$ compared to positive $\Gamma$ is that the determinant of the Hessian matrix of $H^*$ evaluated at the fixed points, namely the expression \eqref{eqn:hessian}, is now manifestly positive rather than negative. Thus the fixed points at $\omega=0,$ $j=j_\mathrm{f,0}$ are now true extrema (more precisely, maxima) of $H^*$ and host a librating island, which is reflected in Figures \ref{fig:HStar_Contours_Gamma_minuspt1}-\ref{fig:HStar_Contours_Gamma_minuspt19}.

Understanding the fixed points at $\omega = \pm\pi/2$,  $j=j_\mathrm{f,\pi/2}$ is much harder.  Just like for $\Gamma>0$, to find $j_\mathrm{f,\pi/2}$ we must solve the depressed quartic \eqref{eqn:FPs_at_omega_piby2}. From this equation we can once again derive a necessary but insufficient condition for fixed points to exist at $\omega = \pm\pi/2$; however since $10\Gamma/(1+5\Gamma) < 0$ in this $\Gamma$ regime, rather than the upper bound $\epsGR < 6(1+5\Gamma)$ that we found for $\Gamma > 0$ (\S\ref{sec:morphology_Gamma_Regime_I}) we instead get a lower bound, $\epsGR > 6(1+5\Gamma)\Theta^{3/2}$. Unfortunately it is not easy to write down analogues of the sufficient conditions \eqref{eqn:epsGR_constraint_fp_piby2}-\eqref{eqn:Theta_constraint_fp_piby2}\footnote{The difficulty arises because the signs of $\partial j_\mathrm{f,\pi/2}/\partial \Theta$ and $\partial j_\mathrm{f,\pi/2}/\partial \epsGR$ (expressions for which are given in \eqref{eqn:djfdTheta}-\eqref{eqn:djfdepsGR}) are not fixed in this $\Gamma$ regime, so we cannot look for e.g. the bounding values of $\epsGR$ that give $j=\sqrt{\Theta}, 1$.}.Indeed, as we saw in Figure \ref{fig:HStar_Contours_Gamma_minuspt15}, for $-1/5<\Gamma \leq 0$ fixed points can arise in pairs at $\omega = \pi/2$ (with another, separate pair at $\omega = -\pi/2$), corresponding to there being two physical solutions to the quartic \eqref{eqn:FPs_at_omega_piby2}.

Finally, even the nature of the $\omega = \pm\pi/2$ fixed points is a non-trivial issue. The determinant of the Hessian of $H^*(\omega,j)$ evaluated at $(\pm\pi/2,j_\mathrm{f,\pi/2})$ is given by
 \begin{align}
    [3(1+5\Gamma) j_\mathrm{f,\pi/2}^4 + 10\Gamma\Theta]  \frac{360\Gamma(j_\mathrm{f,\pi/2}^2-\Theta)(1-j_\mathrm{f,\pi/2}^2)}{j_\mathrm{f,\pi/2}^6},
    \label{eqn:hessian_negative_Gamma}
\end{align}
where we eliminated $\epsGR$ using equation \eqref{eqn:FPs_at_omega_piby2}.
For negative $\Gamma$, the sign of \eqref{eqn:hessian_negative_Gamma} depends on the sign of the first bracket.  If $[3(1+5\Gamma) j_\mathrm{f,\pi/2}^4 + 10\Gamma\Theta] > 0$ then the fixed point at $\omega=\pm\pi/2$ is a saddle point; otherwise it is a true extremum (in fact a minimum). This puts an implicit constraint on $\epsGR$ (since $j_\mathrm{f,\pi/2}$ depends on $\epsGR$) when determining the nature of the fixed poitns. That constraint is responsible for the fact that even for a fixed $\Theta=0.1$, the $\omega = \pm\pi/2$ fixed points are saddle points in Figure \ref{fig:HStar_Contours_Gamma_minuspt1}, minima in Figure  \ref{fig:HStar_Contours_Gamma_minuspt19}, and both are present in Figure \ref{fig:HStar_Contours_Gamma_minuspt15}.

 


      \subsection{Phase space behaviour in the case \texorpdfstring{$\Gamma \leq -1/5$}{GamIV}}
   \label{sec:morphology_Gamma_Regime_IV}

We now turn to the final $\Gamma$ regime, $\Gamma \leq -1/5$, which luckily is not as complicated as $0 < \Gamma \leq 1/5$. We need only illustrate it with a single example, namely Figure \ref{fig:HStar_Contours_Gamma_minuspt5}, which is for $\Theta = 0.1$ and $\Gamma = -0.5$. 

For $\epsGR=0$ (panel (a)) the phase portrait looks almost identical to those typical of $\Gamma > 1/5$ (e.g. Figure \ref{fig:HStar_Contours_Gamma_pt5}a). However, as we noted in Paper II, despite their similarities the dynamical regimes $\Gamma > 1/5$ and $\Gamma \leq -1/5$ are significantly different.  In particular, the phase space trajectories in each regime are traversed in opposite directions (see the arrows in Figures 4a and 7d of Paper II). One consequence of this is that for $\Gamma \leq -1/5$, increasing $\epsGR$ always pushes the fixed points at $\omega = \pi/2$ \textit{up} the page to higher eccentricity --- see panels (b) and (c) of Figure \ref{fig:HStar_Contours_Gamma_minuspt5}. This behaviour is easy to reconstruct mathematically. Since $10\Gamma/(1+5\Gamma) > 0$ in this $\Gamma$ regime,  equation \eqref{eqn:FPs_at_omega_piby2} tells us that for fixed points at $\omega = \pm \pi/2$ to exist necessarily requires $j_{\mathrm{f},\pi/2}^3 > \epsGR/[6(1+5\Gamma)]$. Then it is easy to show (c.f. equations \eqref{eqn:djfdTheta}-\eqref{eqn:djfdepsGR}) that
\begin{align}
    \left(\frac{\partial j_\mathrm{f,\pi/2}}{\partial \Theta}\right)_{\epsGR} > 0, \,\,\,\,\,\,\,\,\,\,\, \mathrm{and} \,\,\,\,\,\,\,\,\,
    \left(\frac{\partial j_\mathrm{f,\pi/2}}{\partial \epsGR}\right)_{\Theta} < 0.
    \label{eqn:partial_derivatives_negative_Gamma}
\end{align}
In other words, increasing $\Theta$ decreases the eccentricity of the fixed points at $\omega=\pm\pi/2$ should they exist (same as $\Gamma > 0$), but increasing $\epsGR$ increases their eccentricity (opposite to $\Gamma > 0$). 
The condition on $\epsGR$ for the existence of these fixed points is the same as \eqref{eqn:epsGR_constraint_fp_piby2} but reversing the inequalities, i.e. replacing each `$<$' with `$>$'. The condition on $\Theta$ is the same as that given for $0 < \Gamma \leq 1/5$ in equation \eqref{eqn:Theta_constraint_fp_piby2}. Additionally, the fixed points at $\omega=\pm\pi/2$ are always true extrema (minima) of $H^*$ in this $\Gamma$ regime since the expression \eqref{eqn:hessian_negative_Gamma} is always positive.

Meanwhile, the fixed points at $\omega=0$ follow exactly the same rules as for $-1/5 < \Gamma \leq 0$, appearing at $\omega=0, e=e_\mathrm{lim}$ when $\epsGR$ reaches the critical value $\epsGR = 6(1-5\Gamma)\Theta^{3/2}$ and then working their way down the page towards $e=0$ as $\epsGR$ is increased, disappearing for $\epsGR > 6(1-5\Gamma)$ (equation \eqref{eqn:fp_omega_0_inequality}). The only difference is that these fixed points are maxima of $H^*$, not saddle points, which follows from the fact that the quantity \eqref{eqn:hessian} is positive for $\Gamma < 0$.

\subsection{Orbit families and maximum eccentricity for $\Gamma \leq 0$ regimes}
\label{sec:lib_or_circ_negative_Gamma}

Owing to the highly complex phase space morphology, working out a trajectory's orbital family analytically is often a very tedious job for negative $\Gamma$ values.  The same goes for finding a binary's maximum eccentricity: in practice it is best simply to take a brute-force approach by solving the cubic and quartic equations \eqref{eqn:depressed_cubic}, \eqref{eqn:depressed_quartic} numerically to get all seven possible roots, and then declaring $\jmin$ to be the real root closest to but smaller than the initial $j$ value. We will not pursue any further technical details here.


\subsubsection{Maximum eccentricity for initially near-circular binaries}
\label{sec:max_ecc_circular_negative_Gamma}

With this brute-force approach it is straightforward to calculate $\emax$ for a given $i_0,\Gamma$ and $\epsGR$ for initially near-circular binaries when $\Gamma \leq 0$ (c.f. \S\ref{sec:max_ecc_circular}).
In Figure \ref{fig:emax_initially_circular_negative_Gamma} we show $\emax(i_0)$ for various $\epsGR$ values.  In each panel we use a different negative value of $\Gamma$ (c.f. Figure \ref{fig:emax_initially_circular}).  

In the top row of Figure \ref{fig:emax_initially_circular_negative_Gamma} (panels (a)-(c)) we explore the regime $-1/5 < \Gamma \leq 0$.
To understand panels (a) and (b) it is worth looking back at Figures \ref{fig:HStar_Contours_Gamma_minuspt1} and \ref{fig:HStar_Contours_Gamma_minuspt15} and asking what we expect of the behaviour of initially near-circular orbits. We expect from Figures \ref{fig:HStar_Contours_Gamma_minuspt1}a,b,c,d and \ref{fig:HStar_Contours_Gamma_minuspt15}a-i that for low enough $\epsGR$ the maximum eccentricity will be zero. This immediately explains, for instance, why there is no red line (corresponding to $\epsGR=0$) in panels (a) and (b) of Figure \ref{fig:emax_initially_circular_negative_Gamma}. However, when $\epsGR$ takes a value such that (I) the fixed point exists at $\omega = 0$ and (II) the librating region that this fixed point hosts is connected to $e=0$, then the eccentricity of initially circular orbits is maximised at $\omega = 0$ and is nonzero (Figure \ref{fig:HStar_Contours_Gamma_minuspt1}e and Figure \ref{fig:HStar_Contours_Gamma_minuspt15}j,k,l). 

We know that (I) is true if and only if $\epsGR$ satisfies \eqref{eqn:fp_omega_0_inequality}. We also know that for (II) to be true the fixed points at $\omega = \pm\pi/2$ must have disappeared below $e=0$. By examining equations \eqref{eqn:FPs_at_omega_piby2} and \eqref{eqn:djfdepsGR} in the limit of $j_{\mathrm{f},\pi/2} \approx 1$ and small $\Theta$, we find that (II) becomes true when $\epsGR$ is increased beyond the threshold value $6(1+5\Gamma)$. Putting these constraints together and using the fact that $-1/5 < \Gamma \leq 0$, we find that in the limit $\Theta \to 0$ a necessary condition for both (I) and (II) to be true is $6(1+5\Gamma) < \epsGR < 6(1-5\Gamma)$, which is the same as \eqref{eqn:piby2_but_no_saddle} if we replace `$<$' with `$>$'.  For $\Gamma = -0.05$ this gives $4.5 < \epsGR < 7.5$, which is why there is only a cyan line in Figure \ref{fig:emax_initially_circular_negative_Gamma}a.  Similarly, for $\Gamma = -0.1$ we get $3 < \epsGR < 9$, hence the solo cyan line in Figure \ref{fig:emax_initially_circular_negative_Gamma}b.

This necessary constraint on $\epsGR$ was derived for $\Theta \to 0$, i.e. $i_0\to 90^\circ$.  To find the necessary constraint on $i_0$ for (I) and (II) to be true, we need a constraint on $\Theta$ (equivalent to $\cos^2 i_0$ for $e_0\approx 0$). By considering equations \eqref{eqn:fp_omega_0_inequality}, \eqref{eqn:FPs_at_omega_piby2} and \eqref{eqn:djfdTheta} for $j_{\mathrm{f},\pi/2}\approx 1$ and $\Gamma$ not too close to\footnote{Values of $\Gamma$ close to $-1/5$ are more complicated, essentially because the sign of the right hand side of \eqref{eqn:djfdTheta} is liable to change in this regime even for $j_{\mathrm{f},\pi/2} \approx 1$.  This is the case in particular for $\Gamma = -0.19$, which is why the behaviour in Figure \ref{fig:emax_initially_circular_negative_Gamma}c is different from the other $-1/5 < \Gamma \leq 0$ examples in Figure \ref{fig:emax_initially_circular_negative_Gamma}a,b.} $-1/5$, we can show that (I) and (II) are true provided $\Theta < \Theta_1(\Gamma,\epsGR)$. So initially near-circular binaries whose $\Gamma$ values produce phase portraits like in Figures \ref{fig:HStar_Contours_Gamma_minuspt1}, \ref{fig:HStar_Contours_Gamma_minuspt15} can achieve a finite $e_\mathrm{max}$ only if they have $i_0$ greater than the critical value $\cos^{-1} \sqrt{\Theta_1(\Gamma,\epsGR)}$. For panels (a) and (b) of Figure \ref{fig:emax_initially_circular_negative_Gamma} these values are $i_0 = 66^\circ$ and $i_0=55^\circ$ respectively, which we show with vertical dotted lines.
Plugging $H^*,\Theta$ from \eqref{eqn:H_Theta_circular} into the depressed cubic \eqref{eqn:depressed_cubic}, we find that the corresponding minimum $j$ value is:
\begin{align}
\label{eqn:circular_omega_0_solution}
   \jmin = \frac{1}{2}\left( -1 +
    \sqrt{1+\frac{4\epsGR}{3(1-5\Gamma)}} \, \right),
\end{align}
which is independent of $i_0$. In panels (a) and (b) of Figure \ref{fig:emax_initially_circular_negative_Gamma}, the straight horizontal cyan lines for $\epsGR = 5$ correspond to the solution \eqref{eqn:circular_omega_0_solution}.

Panels (c)-(f) of Figure \ref{fig:emax_initially_circular_negative_Gamma} all share a similar morphology, so we will consider them together.  In each panel, for a fixed $\epsGR$ a finite $\emax$ arises at some critical value of $i_0$, increases as a function of $i_0$ until it reaches $\emax = e_\mathrm{lim}$, and then is constant for all larger values of $i_0$ up to $90^\circ$. Note that on the non-constant parts of these curves we have essentially the opposite of the intuitive $\Gamma > 1/5$ result: for a fixed initial inclination, a larger $\epsGR$ leads to a \textit{larger} $\emax$. The behaviour we see in these panels is consistent with what we expect from the $\Gamma = -0.19$ example given in Figure \ref{fig:HStar_Contours_Gamma_minuspt19} and the $\Gamma \leq -1/5$ example we studied in Figure \ref{fig:HStar_Contours_Gamma_minuspt5}. In those figures, the fixed points that emerge at $\omega = \pm\pi/2, e=e_{\mathrm{f},\pi/2}$ host regions of librating orbits that are connected to $e= 0$. In each case, the maximum eccentricity of initially circular orbits is determined by eccentricity of the separatrix at the point $\omega=\pm\pi/2$. As we increase $\epsGR$, the value of $e_{\mathrm{f},\pi/2}$ is increased, pushing the separatrix to higher $e$, and so the maximum eccentricity of initially circular orbits grows. Eventually, however, $e_{\mathrm{f},\pi/2}$ is increased so much that the separatrix reaches $e=e_\mathrm{lim}$ (dashed black line) --- see the transition between Figure \ref{fig:HStar_Contours_Gamma_minuspt5}c and \ref{fig:HStar_Contours_Gamma_minuspt5}d.  At the same time, the librating islands that are hosted by fixed points at $\omega = 0$ become connected to $e=0$. After that the maximum eccentricity is given by \eqref{eqn:circular_omega_0_solution} and is independent of $i_0$ --- hence the straight horizontal lines in Figure \ref{fig:emax_initially_circular_negative_Gamma}c-f. The main qualitative difference between panel (c) and panels (d)-(f) is that panel (d) exhibits no red line, i.e. no solution for $\epsGR=0$. This is because in the regime $-1/5<\Gamma \leq 0$ a finite $\epsGR$ is always required for any fixed points to exist (Figures \ref{fig:HStar_Contours_Gamma_minuspt1}-\ref{fig:HStar_Contours_Gamma_minuspt19}).

Finally we may briefly compare Figure \ref{fig:emax_initially_circular_negative_Gamma} with Figure \ref{fig:emax_initially_circular}. Consider what happens if we fix $\epsGR$ and increase $i_0$ from zero. In Figure \ref{fig:emax_initially_circular_negative_Gamma} ($\Gamma\leq 0$), the larger is $\epsGR$, the lower $i_0$ is required to achieve a non-zero $\emax$ (provided $\epsGR$ is not so large that no eccentricity excitation is possible). On the contrary, in Figure \ref{fig:emax_initially_circular} ($\Gamma >0$) the most favourable situation for eccentricity excitation is always to have $\epsGR$ as small as possible: the larger $\epsGR$, the larger $i_0$ is required to get a non-zero maximum eccentricity. Though the two regimes differ in this respect, they are similar in that for binaries with $i_0 \approx 90^\circ$ a larger $\epsGR$ always leads to a smaller maximum eccentricity (again provided $\epsGR$ is such that eccentricity excitation is possible).

\bsp	
\label{lastpage}

\end{document}